\documentclass[reqno,12pt]{article}
\usepackage{amsfonts,amssymb,
amsmath,amscd, bm, mathrsfs, %bbold,
mathrsfs,delarray,subfigure}
\usepackage{hyperref,cite}
\usepackage[all]{xy}%\usepackage{xypic}
\usepackage[dvips]{color}
\usepackage[dvips]{graphicx}
\DeclareMathOperator{\Ad}{Ad}

\DeclareMathOperator{\id}{id}
\DeclareMathOperator{\Id}{Id}

\DeclareMathOperator{\Conn}{Conn} 

\DeclareMathOperator{\Map}{Map}

\DeclareMathOperator{\Gau}{Gau}

\DeclareMathOperator{\rank}{rank}

\DeclareMathOperator{\Cyc}{Cyc}

\numberwithin{equation}{subsection} 
\numberwithin{subsection}{section} 

\font\sansserif=cmss12
\font\scriptsansserif=cmss12 at 7 truept
\font\scriptscriptsansserif=cmss10 at 5 truept
\font\euler=eusm10 at 12.8 truept
\font\scripteuler=eusm7
\font\scriptscripteuler=eusm5 
\newcommand{\ul}[1]{{\underline{#1}}{}}

\CompileMatrices

\newtheorem{defi}{%\indent 
Definition}[subsection]

\newtheorem{prop}{%\indent 
Proposition}[subsection]
\newtheorem{lemma}{%\indent 
Lemma}[subsection]
\newtheorem{notation}{%\indent 
Notation}[subsection]
\newtheorem{remark}{%\indent 
Remark}[subsection]

\begin{document}

\hrule\vskip.5cm
\hbox to 14.5 truecm{October 2014 \hfil DIFA 14}
%\hbox to 14.5 truecm{Version 2 \hfil}
\vskip.5cm\hrule
\vskip.7cm
%\begin{large}
\centerline{\textcolor{blue}{\bf A NEW FORMULATION OF  HIGHER PARALLEL }}   
\centerline{\textcolor{blue}{\bf TRANSPORT  IN HIGHER GAUGE THEORY}}   
%\end{large}
\vskip.2cm
\centerline{by}
\vskip.2cm
\centerline{\bf Emanuele Soncini and Roberto Zucchini}
\centerline{\it Dipartimento di Fisica ed Astronomia, Universit\`a di Bologna}
\centerline{\it V. Irnerio 46, I-40126 Bologna, Italy}
\centerline{\it I.N.F.N., sezione di Bologna, Italy}
\centerline{\it E--mail: emanuele.soncini@studio.unibo.it, zucchinir@bo.infn.it}
\vskip.7cm
\hrule
\vskip.6cm
\centerline{\bf Abstract} 
\par\noindent
In this technical paper, we present a new formulation of higher parallel transport in strict
higher gauge theory required for the rigorous construction of Wilson lines and
surfaces. Our approach is based on an original notion
of  Lie crossed module cocycle and cocycle $1$-- and $2$--gauge 
transformation with a non standard double category theoretic
interpretation.  We show its equivalence to earlier formulations. 

\vfil\eject

\tableofcontents

\vfil\eject

%%%%%%%%%%%%%%%%%%%%%%%%%%%%%%%%%%%  highgau1.tex

\section{\normalsize \textcolor{blue}{Introduction}}\label{sec:intro}

\vfil
\hspace{.5cm} 
{\it Higher gauge theory} is a generalisation of ordinary gauge theory 
describing the dynamics of charged objects of any dimension.
%, particles as well as strings and branes. 
It has found application in string theory in the study of 
$D$-- and $M$--branes \cite{Polchinski:1998rr,Becker:2007zj,Johnson:2003gi} 
as well as loop quantum gravity and, in particular, spin foam models
\cite{Baez:1999sr,Rovelli:2004tv}. 
See ref. \cite{Baez:2010ya} for an easy to read, 
up--to--date exposition of this subject and extensive referencing. 

\vfil
From a mathematical point of view, higher gauge theory intersects various branches
of modern mathematics. On the algebraic side, it is related to the theory of higher
algebraic structures, such as higher categories, higher groups \cite{Baez5,Baez:2003fs} 
and strong homotopy Lie or $L_\infty$ algebras \cite{Lada:1992wc,Lada:1994mn}; on the 
geometrical one, it leads naturally to higher geometrical structures such as gerbes 
both in the Abelian and non Abelian variant \cite{Brylinski:1993ab,Breen:2001ie}.
A recent treatment of these matters with a physical outlook can be found 
in \cite{Schreiber2011,Gruetzmann:2014ica}.  

\vfil
Though higher gauge theory has a long story that can be traced back to the
inception of supergravity theory, in its modern form it has been formulated relatively recently 
in the seminal papers by Baez \cite{Baez:2002jn} and Baez and Schreiber \cite{Baez:2004in,Baez:2005qu}. 
A pivotal role is attributed to the analysis of higher parallel transport. 
Many papers have been written about the precise and rigorous definition of 
parallel transport. We have in mind in particular for the influence they had 
on our work the papers by Schreiber and Waldorf \cite{Schrei:2009,Schrei:2011,Schrei:2008}
and Martins and Picken \cite{Martins:2007,Martins:2008,Martins:2009}. Recent contributions include 
\cite{Morton:2013dla} and \cite{Abad:2014a,Abad:2014b}.

\vfil
In this paper, we propose a new formulation of parallel transport in strict higher gauge 
theory. We do not claim any new results but we only offer a new perspective from which to
view old ones, which hopefully may provide new insight. 
Our interest in this subject has been prompted by our recent formulation 
of semistrict higher gauge theory aimed to higher Chern--Simons theory, 
in which we circumvent the difficulties related to 
the integration of the underlying semistrict Lie $2$--algebra to a semistrict $2$--group,
when possible, by relying on the automorphism $2$--group of the Lie $2$--algebra, which is 
always strict \cite{Zucchini:2011aa,Soncini:2014ara}. (See also 
\cite{Jurco:2014mva} for an alternative approach.) In a companion paper, we plan to 
study the issue of higher holonomy and invariant traces on the same lines \cite{SZ:2015}. 

Our formulation is based on an original notion
of  Lie crossed module cocycle and cocycle $1$-- and $2$--gauge 
transformation with a non standard double category theoretic
interpretation.  (See \cite{Martins:2007,Martins:2008} and
\cite{Morton:2013dla} for related  approaches.)

%\vfil\eject

\subsection{\normalsize \textcolor{blue}{The scope and the plan of this paper}}\label{subsec:scope}

\hspace{.5cm} 
In this introductory subsection, we want to convey an intuitive idea of our formulation of 
higher parallel transport theory by reviewing first the cocycle approach to the ordinary theory 
and then outlining the higher generalization of it we propose. 
Here, we have no pretension of full mathematical rigor. Everything we say below holds in the smooth category.

Let $G$ be a Lie group. A $G$--cocycle is a map $f:\mathbb{R}^2\rightarrow G$ obeying
\begin{equation}
f(x'',x')f(x',x)=f(x'',x). 
\label{scope1}
\end{equation}
A $G$--connection $a$ is just a $\mathfrak{g}$--valued $1$--form on $\mathbb{R}$. 
$G$--cocycles are in one--to--one correspondence with $G$--connections. The $G$--connection $a_f$ 
corresponding to a $G$--cocycle $f$ is defined by
\begin{equation}
a_{fx}(x)=-d_{x'}f(x',x)f(x',x)^{-1}\big|_{x'=x}.
\label{scope2}
\end{equation}
The $G$--cocycle $f_a$ corresponding to a $G$--connection $a$ is given by $f_a(x,x_0)=u_{x_0}(x)$,
where $u_{x_0}$ is the unique solution of the differential problem 
\begin{equation}
d_xu_{x_0}(x)u_{x_0}(x)^{-1}=-a_x(x), \qquad u_{x_0}(x_0)=1_G.  
\label{scope3}
\end{equation}
A $G$--gauge transformation is simply a mapping $\varkappa:\mathbb{R}\rightarrow G$. 
$G$--gauge transformations act on $G$--cocycles and $G$--connections. The gauge transform of
a cocycle $f$ by a gauge transformation $\varkappa$ is  
\begin{equation}
{}^\varkappa f(x',x)=\varkappa(x')f(x',x)\varkappa(x)^{-1}.
\label{scope4}
\end{equation}
The gauge transform of a connection $a$ by a gauge transformation $\varkappa$ is given by the familiar relation  
\begin{equation}
{}^\varkappa a_x(x)=\Ad\varkappa(x)(a_x(x))-d_x \varkappa(x)\varkappa(x)^{-1}.
\label{scope5}
\end{equation}
These actions are furthermore compatible with the cocycle to connection correspondence.
The above has a categorical formulation. Let $\mathbb{GR}$ be the oriented segment groupoid of $\mathbb{R}$,
the familiar groupoid of pairs of elements $\mathbb{R}$, and $BG$ be the delooping of $G$, the one object groupoid
whose morphisms set is $G$. Then, a $G$--cocycle $f$ can be viewed as a functor $f:\mathbb{GR}\rightarrow BG$.
Further, any $G$--gauge transformation $\varkappa$ encodes a natural transformation 
$\varkappa:f\Rightarrow {}^\varkappa f$. 

Parallel transport in a gauge theory with gauge group $G$ on a manifold $M$ can now be defined as follows. 
For simplicity we assume that the background principal $G$--bundle is trivial.
A $G$--connection $\theta$ is then simply a $\mathfrak{g}$--valued $1$--form on $M$.
Given two points $p_0$, $p_1$ of $M$ a curve $\gamma:p_0\rightarrow p_1$ in $M$ with sitting instants 
joining them, the pull--back $\gamma^*\theta$ is a 
$G$--connection in the sense defined in the previous paragraph. With this, there is associated a $G$--cocycle 
$f_{\gamma^*\theta}$. The parallel transport induced by $\theta$ along $\gamma$ is then given by 
\begin{equation}
F_\theta(\gamma)=f_{\gamma^*\theta}(1,0).
\label{scope6}
\end{equation}
A $G$--gauge transformation is just a $G$--valued map $g$ on $M$. It acts on a $G$--connection $\theta$ in the
well--known way,
\begin{equation}
{}^g \theta=\Ad g(\theta)-dgg^{-1}. 
\label{scope7}
\end{equation}
The associated parallel transport transforms correspondingly as 
\begin{equation}
F_{{}^g\theta}(\gamma)=g(p_1)F_\theta(\gamma)g(p_0)^{-1}.
\label{scope8}
\end{equation}
since $g$ yields a $G$--gauge transformation $\gamma^*g$ on $\gamma^*\theta$ in the sense defined in the preceding paragraph. 
From a categorical point of view, it is found that $F_\theta$ defines a functor $F_\theta:(M, P_1M)\rightarrow BG$
from the path groupoid $(M, P_1M)$ of $M$ to $BG$ and that $g$ defines a natural transformation $g:F_\theta\Rightarrow F_{{}^g\theta}$. 

In this paper, we show that the cocycle based formulation of parallel transport of ordinary gauge theory outlined above
admits a non trivial extension to strict higher gauge theory. Let $(G,H)$ be a Lie crossed module. In sect. \ref{sec:highcocy},
we introduce the notion of $(G,H)$--cocycle, a triple of three maps $f:\mathbb{R}^3\rightarrow G$, 
$g:\mathbb{R}^3\rightarrow G$ and $W:\mathbb{R}^4\rightarrow H$ obeying relations extending \eqref{scope1} and a target matching 
condition relating $f$ $g$ and $W$, and recall that of $(G,H)$--connection doublet, a pair of a $\mathfrak{g}$--valued 
$1$--form $a$ and a $\mathfrak{h}$--valued $2$--form $B$ on $\mathbb{R}^2$ satisfying the so--called zero fake curvature 
condition familiar in higher gauge theory. We show then that 
there is a one--to--one correspondence between $(G,H)$--cocycles and $(G,H)$--connection doublets 
analogous to \eqref{scope2}, \eqref{scope3}. 
We introduce next the notion of integral $(G,H)$--$1$--gauge transformation, a triple constituted by three maps 
$\kappa:\mathbb{R}^2\rightarrow G$ and $\varPsi:\mathbb{R}^3\rightarrow H$, $\varPhi:\mathbb{R}^3\rightarrow H$
obeying certain cocycle relations, and of differential $(G,H)$--$1$--gauge transformation, a pair of a $G$--valued map 
$\varkappa$ and a $\mathfrak{h}$--valued $1$--form $\varGamma$ on $\mathbb{R}^2$. We prove then the existence of 
a one--to--one correspondence between integral and differential $(G,H)$--$1$--gauge transformations. 
Integral $(G,H)$--$1$--gauge transformations are next shown to act on $(G,H)$ cocycles by an extension of \eqref{scope5} and, similarly,
differential $(G,H)$--$1$--gauge transformations on $(G,H)$--connection doublets through the usual
higher gauge theoretic prescription generalizing \eqref{scope6} and these actions are found to be compatible with the 
correspondences between cocycles and connection doublets and integral and differential gauge transformations. 
Finally, we introduce the notion of $(G,H)$--$2$--gauge transformation, a single mapping $A:\mathbb{R}^2\rightarrow H$,
and show that $(G,H)$--$2$--gauge transformations act both on integral and differential $(G,H)$--$1$--gauge transformations
in a way that is compatible with the correspondence between the two. 

The above construction has a remarkable double categorical interpretation. The basic ingredients of this are 
the double groupoid $\mathbb{GR}^2$ of oriented rectangles of $\mathbb{R}^2$ and the edge symmetric double groupoid 
$B(G,H)$ canonically associated to the Lie crossed module $(G,H)$. A $(G,H)$--cocycle amounts to a double functor 
$\mathbb{R}^2\rightarrow B(G,H)$. An integral $(G,H)$--$1$--gauge transformation encodes a form of double natural 
transformation between a $(G,H)$--cocycle and its $1$--gauge transform. Finally, a $(G,H)$--$2$--gauge transformation
yields a double modifications between an integral $(G,H)$--$1$--gauge transformation and its $2$--gauge transform. 
The notion of double natural transformation and modification we use are not standard 
and are precisely defined in the appendix. This may be of some interest in category theory. 

In sect. \ref{sec:hiholo}, we rederive higher parallel transport theory originally obtained in the references 
recalled above using higher cocycle theory. 
We consider a strict higher gauge theory with gauge crossed module $(G,H)$ on a manifold $M$ 
for a trivial $(G,H)$ $2$--bundle. A $(G,H)$ connection doublet is a pair of a $\mathfrak{g}$--valued 
$1$--form $\theta$ and a $\mathfrak{h}$--valued $2$--form $\varUpsilon$ on $M$ satisfying 
the zero fake curvature condition. If $\gamma_0$, $\gamma_1$ are curves with the same endpoints and $\varSigma:\gamma_0\Rightarrow \gamma_1$ 
is a surface connecting them, all with sitting instants, then $\varSigma^*\theta$, $\varSigma^*\varUpsilon$ constitute
a $(G,H)$ connection doublet in the sense defined two paragraphs above with which there is associated 
a $(G,H)$--cocycle $f_{\varSigma^*\theta\varSigma^*\varUpsilon}$, $g_{\varSigma^*\theta\varSigma^*\varUpsilon}$, 
$W_{\varSigma^*\theta\varSigma^*\varUpsilon}$. The $1$--parallel transport along the $\gamma_i$ and the $2$--parallel
transport along $\varSigma$ are $F_{\theta,\varUpsilon}(\gamma_i)=f_{\varSigma^*\theta\varSigma^*\varUpsilon|i}(1,0)$
and $F_{\theta,\varUpsilon}(\varSigma)=W_{\varSigma^*\theta\varSigma^*\varUpsilon}(0,1;1,0)$, extending the prescription 
\eqref{scope6}. Next, a $(G,H)$--$1$--gauge transformation is a pair of a $G$--valued map 
$g$ and a $\mathfrak{h}$--valued $1$--form $J$ on $M$. $(G,H)$--$1$--gauge transformations
act on $(G,H)$--connection doublets $\theta$, $\varUpsilon$ according the higher gauge theoretic prescription 
generalizing \eqref{scope7}
and thus on parallel transport. This action comes through the action of the integral $(G,H)$--$1$--transformation 
$\kappa_{\varSigma^*g,\varSigma^*J}$, $\varPsi_{\varSigma^*g,\varSigma^*J}$, $\varPhi_{\varSigma^*g,\varSigma^*J}$
associated to the differential $(G,H)$--$1$--gauge transformation $\varSigma^*g$, $\varSigma^*J$
on the $(G,H)$--cocycle $f_{\varSigma^*\theta\varSigma^*\varUpsilon}$, $g_{\varSigma^*\theta\varSigma^*\varUpsilon}$, 
$W_{\varSigma^*\theta\varSigma^*\varUpsilon}$ and leads to the appropriate  extension of \eqref{scope8}.
Similar considerations hold for $2$--gauge transformations.  

We find that the higher parallel transport operation constructed in this way agrees with that developed
in  earlier literature \cite{Baez:2004in,Baez:2005qu,Schrei:2009,Schrei:2011,Schrei:2008,
Martins:2007,Martins:2008,Martins:2009}. In particular, we recover the remarkable interpretation 
of the higher transport $F_{\theta,\varUpsilon}$ as a $2$--functor $F_{\theta,\varUpsilon}:(M,P_1M,P_2M)\rightarrow B_0(G,H)$
from the path $2$--groupoid $(M,P_1M,P_2M)$ of $M$ to the strict $2$--group $B_0(G,H)$ corresponding to $(G,H)$ 
and of $(G,H)$--$1$-- and $2$--gauge transformation as pseudonatural transformations and modifications, respectively.

\subsection{\normalsize \textcolor{blue}{Outlook}}\label{subsec:look}

\hspace{.5cm} 
Using the results of the present work and restricting to the flat case, we plan to reconsider 
in the companion paper \cite{SZ:2015} the theory of higher holonomy,  
already studied in \cite{Martins:2007,Martins:2008,Martins:2009} and reanalyzed recently 
in a very general setting in \cite{Abad:2014a,Abad:2014b}, and tackle the problem of the proper definition of higher 
holonomy invariants. The quest for the latter 
is particularly important for the applications they may have in a study of $2$--knots in $4$--folds
based on the higher Chern--Simons theory developed in ref. \cite{Soncini:2014ara}.
(See ref. \cite{Cattaneo:2002tk} for a related endeavour.)

\vfil\eject

\section{\normalsize \textcolor{blue}{Lie crossed module cocycle theory}}\label{sec:highcocy}

\hspace{.5cm} In this section, we expound our theory of Lie crossed
module cocycles. Hints of this approach were already present in
refs. \cite{Schrei:2009,Schrei:2011,Schrei:2008}, to which we are
indebted for inspiration. We illustrate our construction stressing its
being an extension of the ordinary Lie group cocycle theory. The
theory of Lie crossed module $1$-- and $2$--gauge transformations is
presented on the same lines. 

The basic algebraic and differential geometric structures on which the following analysis is based are 
those of Lie group, Lie algebra, Lie crossed module and differential Lie crossed module. These are 
reviewed in some detail in the appendix of ref. \cite{Soncini:2014ara}, whose conventions we adopt. 
Below, we use throughout the following notation.

\begin{notation}
With each Lie crossed module $(G,H,t,m)$ there is associated a differential Lie crossed module
$(\mathfrak{g},\mathfrak{h},\dot t,\widehat{m})$ with 
\begin{subequations}
\begin{align}
&\dot t(X)=\frac{dt(C(v))}{dv}\Big|_{v=0},
\vphantom{\Big]}
\label{hiholo1}
\\
&\widehat{m}(x)(X)
=\frac{\partial}{\partial u}\Big(\frac{\partial m(c(u))(C(v))}{\partial v}\Big|_{v=0}\Big)\Big|_{u=0}
\vphantom{\Big]}
\label{hiholo2}
\end{align}
\end{subequations}
for $x\in\mathfrak{g}$, $X\in\mathfrak{h}$, where $c(u)$ is any curve in $G$ 
such that $c(u)\big|_{u=0}=1_G$ and $dc(u)/du\big|_{u=0}=x$ and  $C(v)$ is any curve 
in $H$ such that $C(v)\big|_{v=0}=1_H$ and $dC(v)/dv\big|_{v=0}=X$. 
\end{notation}

\begin{notation}
Each Lie crossed module $(G,H,t,m)$ is characterized by two canonical mappings
$\dot m:G\times\mathfrak{h}\rightarrow \mathfrak{h}$ and 
$Q:\mathfrak{g}\times H\rightarrow \mathfrak{h}$ defined by
\begin{subequations}
\begin{align}
&\dot m(a)(X)=\frac{d}{dv}m(a)(C(v))\Big|_{v=0},
\vphantom{\Big]}
\label{hiholo3}
\\
&Q(x,A)=\frac{d}{du}m(c(u))(A)A^{-1}\Big|_{u=0} 
\vphantom{\Big]}
\label{hiholo4}
\end{align}
\end{subequations}
for $a\in G$, $X\in\mathfrak{h}$, $x\in\mathfrak{g}$, $A\in H$, where $c(u)$ is a curve in $G$ 
such that $c(u)\big|_{u=0}$ $=1_G$ and $dc(u)/du\big|_{u=0}=x$ and $C(v)$ is a curve 
in $H$ such that $C(v)\big|_{v=0}=1_H$ and $dC(v)/dv\big|_{v=0}=X$. 
\end{notation}

%\noindent 
%The following relations are recurrently used,
%\vspace{-.1cm}
%\begin{subequations}
%\label{linftymorph4}
%\begin{align}
%&Q([x,y],A)+[Q(x,A),Q(y,A)]-[x,Q(y,A)]+[y,Q(x,A)]=0,
%\vphantom{\Big]}
%\label{linftymorph4a}
%\\
%&Q(x,AB)=Q(x,A)+AQ(x,B)A^{-1},
%\vphantom{\Big]}
%\label{linftymorph4b}
%\\
%&Q(axa^{-1},A)=\dot m(a)(Q(x,m(a^{-1})(A))).
%\vphantom{\Big]}
%\label{linftymorph4c}
%\end{align}
%\end{subequations}

\vfil\eject

\subsection{\normalsize \textcolor{blue}{Lie crossed module cocycles}}\label{sec:cycle}

\hspace{.5cm} Cocycle theory plays a basic tole in higher holonomy theory and gauge theory. 
We begin by recalling the definition and main properties of Lie group cocycles
and then move to state the definition and study the properties 
of Lie crossed module cocycles.

Let $G$ be a Lie group.

\begin{defi}%{\it Definition}.
A $G$--cocycle is a  map $f\in\Map(\mathbb{R}^2,G)$ 
such that 
\begin{equation}
f(x'',x)=f(x'',x')f(x',x),
\label{cycle1}
\end{equation}
for $x,x',x''\in\mathbb{R}$. We denote the set of $G$--cocycles as $\Cyc(G)$.
\end{defi}

A few basic properties of cocycles follow immediately from the definition.

\begin{prop}
If $f$ is a $G$--cocycle, then
\begin{subequations}
\label{cycle2,3}
\begin{align}
&f(x,x)=1_G,
\vphantom{\Big]}
\label{cycle2}
\\
&f(x,x')=f(x',x)^{-1},
\vphantom{\Big]}
\label{cycle3}
\end{align}
\end{subequations}
for $x,x'\in\mathbb{R}$.
\end{prop}

Lie group cocycles have a categorical interpretation. Though this is well known, we review it here
since it points to and justifies the less known generalization to Lie crossed module cocycle presented below. 

The segment groupoid $\mathbb{GR}$ has one object for each real number $x\in\mathbb{R}$ and 
one arrow for each pair of real numbers $x,x'\in\mathbb{R}$
\begin{equation}
\xymatrix{{\text{\footnotesize $x'$}}&{\text{\footnotesize $x$}}\ar[l]}\!.
\label{cycle4}
\end{equation}
Composition of arrows is carried out by concatenation at their common end. The identity arrows are those with equal ends. 
Inversion of an arrow is performed by exchange of its ends. $\mathbb{GR}$ is evidently isomorphic to the pair groupoid 
of $\mathbb{R}$. 

A Lie group $G$ can be viewed as a one object groupoid $BG$, the delooping of $G$,  
with one arrow for each element of $g\in G$
\begin{equation}
\xymatrix{{\text{\footnotesize $*$}}&{\text{\footnotesize $*$}}\ar[l]_g}\!.
\label{cycle5}
\end{equation}
Composition is given by group multiplication. The identity arrow is that corresponding to the neutral element $1_G$.
Inversion is the same as group inversion.

\begin{prop}
A $G$--cocycle $f$ is equivalent to a smooth functor $\mathbb{GR}\rightarrow BG$
\begin{equation}
\xymatrix{{\text{\footnotesize $x'$}}&{\text{\footnotesize $x$}}\ar[l]}
\quad
\xymatrix{\ar@{|->}[r]&}
\quad
\xymatrix@C=3.5pc{{\text{\footnotesize $*$}}&{\text{\footnotesize $*$}}\ar[l]_{f(x',x)}}\!.
\label{cycle6}
\end{equation}
\end{prop}

\noindent{\it Proof}. The cocycle relation \eqref{cycle1} is a necessary and sufficient
condition for the functoriality of the above mapping. \hfill $\Box$

Every Lie group cocycle yields and can be reconstructed from a Lie valued differential form datum.

\begin{defi} \label{def:r2gconn}
A $G$--connection is a form $a\in\Omega^1(\mathbb{R},\mathfrak{g})$. 
We denote the set of $G$--connections by $\Conn(G)$. 
\end{defi}

The following theorem holds \cite{Schrei:2011}.

\begin{prop} \label{theor:cycle1} There is a canonical 
one--to--one correspondence between the set $\Cyc(G)$ of $G$--cocycles
and that $\Conn(G)$ of $G$--connections. The $G$--connection $a_f$ corresponding to a $G$--cocycle 
$f$ is 
\begin{equation}
a_{fx}(x)=-d_{x'}f(x',x)f(x',x)^{-1}\big|_{x'=x}.
\label{cycle7}
\end{equation}
The $G$--cocycle $f_a$ corresponding to a $G$--connection $a$ is 
\begin{equation}
f_a(x,x_0)=u_{x_0}(x),
\label{cycle8}
\end{equation}
where $u_{x_0}$ is the unique solution
\begin{equation}
d_xu_{x_0}(x)u_{x_0}(x)^{-1}=-a_x(x)
\label{cycle9}
\end{equation}
with $u_{x_0}:\mathbb{R}\rightarrow G$ smooth and satisfying the initial condition 
\begin{equation}
u_{x_0}(x_0)=1_G.
\label{cycle10}
\end{equation}
\end{prop}

\noindent{\it Proof}. If $f$ is a $G$--cocycle, then \eqref{cycle7}
clearly defines a $G$--connection  $a_f$. 
If $a$ is a $G$--connection, then the solution 
$u_{x_0}$ of the differential problem \eqref{cycle9}, \eqref{cycle10}
exists, is unique and smooth in $x_0$. The $G$--valued maps
\begin{subequations}
\label{cyclea1,2}
\begin{align}
&u_1(x)=f_a(x,x_1)f_a(x_1,x_0),
\vphantom{\Big]}
\label{cyclea1}
\\
&u_2(x)=f_a(x,x_0)
\vphantom{\Big]}
\label{cyclea2}
\end{align}
\end{subequations}
solve the differential equation $d_xu(x)u(x)^{-1}=-a_x(x)$ with initial condition
$u(x_1)$ $=f_a(x_1,x_0)$, by \eqref{cycle8}--\eqref{cycle10}. As this differential problem has only one solution,
%By the uniqueness of the solution, 
we have $u_1=u_2$. From \eqref{cyclea1,2}, it follows then 
that $f_a$ obeys the cocycle
condition \eqref{cycle1}. % as required. 
\eqref{cycle9} implies immediately that 
$a_{f_a}=a$. By \eqref{cycle7} and \eqref{cycle2}, $f=f_{a_f}$. The mappings $f\to a_f$ and $a\to f_a$ 
are thus reciprocally inverse. 
\hfill $\Box$

We now present the definition of Lie crossed module cocycle.
Let $(G,H,t,m)$ be a Lie crossed module.

\begin{defi}
A $(G,H)$--cocycle consists of three  mappings
$f\in\Map(\mathbb{R}^2\times\mathbb{R},G)$, $g\in\Map(\mathbb{R}\times\mathbb{R}^2,G)$  
and $W\in\Map(\mathbb{R}^2\times\mathbb{R}^2,H)$ satisfying the target matching condition
\begin{equation}
t(W(x',x;y',y))=g(x;y',y)^{-1}f(x',x;y')^{-1}g(x';y',y)f(x',x;y)
%t(W(x',x;y',y))=g_{|x}(y',y)^{-1}f_{|y'}(x',x)^{-1}g_{|x'}(y',y)f_{|y}(x',x)
\label{cycle15}
\end{equation}
and the relations
\begin{subequations}
\label{cycle11,12,13,14}
\begin{align}
&f_{|y}(x'',x)=f_{|y}(x'',x')f_{|y}(x',x),
\vphantom{\Big]}
\label{cycle11}
\\
&g_{|x}(y'',y)=g_{|x}(y'',y')g_{|x}(y',y),
\vphantom{\Big]}
\label{cycle12}
\\
&W_{|y',y}(x'',x)=W_{|y',y}(x',x)m(f_{|y}(x',x)^{-1})(W_{|y',y}(x'',x')),
\vphantom{\Big]}
\label{cycle13}
\\
&W_{|x',x}(y'',y)=m(g_{|x}(y',y)^{-1})(W_{|x',x}(y'',y'))W_{|x',x}(y',y)
\vphantom{\Big]}
\label{cycle14}
\end{align}
\end{subequations}
\vspace{-7truemm}\eject\noindent
for $x,x',x'',y,y',y''\in\mathbb{R}$. 
We denote the set of $(G,H)$--cocycles as $\Cyc(G,H)$.
\end{defi}
Above, we have set $f_{|y}(x',x)=f(x',x;y)$,
$g_{|x}(y',y)=g(x;y',y)$ and $W_{|y',y}(x',x)=W_{|x',x}(y',y)=W(x',x;y',y)$ for 
convenience. 

Lie crossed module cocycles enjoy a number of properties 
generalizing \eqref{cycle2,3}. 

\begin{prop}
If $(f,g,W)$ is a $(G,H)$--cocycle, then 
\begin{subequations}
\begin{align}
&f_{|y}(x,x)=1_G,
\vphantom{\Big]}
\label{cycle16}
\\
&f_{|y}(x,x')=f_{|y}(x',x)^{-1},
\vphantom{\Big]}
\label{cycle17}
\\
&g_{|x}(y,y)=1_G,
\vphantom{\Big]}
\label{cycle18}
\\
&g_{|x}(y,y')=g_{|x}(y',y)^{-1},
\vphantom{\Big]}
\label{cycle19}
\\
&W_{|y',y}(x,x)=W_{|x',x}(y,y)=1_H,
\vphantom{\Big]}
\label{cycle20}
\\
&W_{|y',y}(x,x')=m(f_{|y}(x',x))(W_{|y',y}(x',x)^{-1}),
\vphantom{\Big]}
\label{cycle21}
\\
&W_{|x',x}(y,y')=m(g_{|x}(y',y))(W_{|x',x}(y',y)^{-1})
\vphantom{\Big]}
\label{cycle22}
\end{align}
\end{subequations}
for $x,x',x'',y,y',y''\in\mathbb{R}$. 
\end{prop}

As we announced above, Lie crossed module cocycles enjoy a categorical interpretation
analogous to and extending that of ordinary Lie group cocycles.
Its statement requires basic notions of double category theory
that are reviewed in app. \ref{sec:dcat} to the benefit of the reader.
%the appendices (cf. apps. \ref{sec:dcdef}--\ref{sec:dccrossed}). 

The rectangle double groupoid $\mathbb{GR}^2$ has one object $(x,y)$ for each 
$x,y\in\mathbb{R}$, one horizontal arrow \hphantom{xxxxxxxxxxxxxxxx}
%\vspace{-.2cm}
\begin{equation}
\vbox{
\xymatrix{{\text{\footnotesize $(x',y)$}}&{\text{\footnotesize $(x,y)$}}\ar[l]}}\!,
\label{cycle23} %dcplane1}
\end{equation}
%\vspace{-.2cm}
for each $x,x',y\in\mathbb{R}$, one vertical arrow 
%\vspace{-.1cm}
\begin{equation}
\xymatrix{{\text{\footnotesize $(x,y')$}}
\\
{\text{\footnotesize $(x,y)$}}\ar[u]}\!,
\label{cycle24} %dcplane2}
\end{equation}
%\vspace{-.3cm}
for each $x,y,y'\in\mathbb{R}$ and one arrow square
%\vspace{-.1cm}
\begin{equation}
\xymatrix{
{\text{\footnotesize $(x',y')$}}  & {\text{\footnotesize $(x,y')$}}\ar[l] \ar@{}[dl]^(.25){}="a"^(.75){}="b" \ar@{=>} "a";"b" %|-{X}
\\                 
{\text{\footnotesize $(x',y)$}} \ar[u] & {\text{\footnotesize $(x,y)_{\vphantom{g}}$}}\ar[u] \ar[l]            
}\!,
\label{cycle25}%dcplane3}
\end{equation}
for each quadruple $x,x',y,y'\in\mathbb{R}$. 
The various operations of composition, identity assignment and inversion of 
arrows and arrow squares are defined in subapp. \ref{sec:dcplane}. Arrow operations are essentially the same 
as those of the segment groupoid. Intuitively, arrow square operations go by concatenation through a common arrow,
identification of opposite arrows and exchange of opposite arrows in either the horizontal or the vertical 
direction. 
 
With a Lie crossed module $(G,H)$ there is canonically associated a double groupoid 
$B(G,H)$ in many ways analogous to the delooping of a Lie group. $B(G,H)$ 
has a single object $*$, one horizontal arrow and one vertical arrow
%\vspace{-.2cm}
\begin{equation}
\vbox{
\xymatrix{{\text{\footnotesize $*$}}&{\text{\footnotesize $*$}}\ar[l]_x}
\vspace{-.75cm}}
\qquad
\xymatrix{{\text{\footnotesize $*$}}
\\
{\text{\footnotesize $*$}}\ar[u]_x}
\label{cycle26} %dccrossed1}
\end{equation}
%\vspace{.2cm}
for each element $x\in G$ and one arrow square 
%\vspace{-.2cm}
\begin{equation}
\xymatrix{
{\text{\footnotesize $*$}}  & {\text{\footnotesize $*$}}\ar[l]_u  \ar@{}[dl]^(.25){}="a"^(.75){}="b" \ar@{=>} "a";"b"_X %|-{X}
\\                 
{\text{\footnotesize $*$}} \ar[u]^v & {\text{\footnotesize $*$}}\ar[u]_x\ar[l]^y             
}
\label{cycle27} %dccrossed3}
\end{equation}
%\vspace{.3cm}
for each $x,y,u,v\in G$ and $X\in H$ satisfying the target matching condition 
\begin{equation}
vy=uxt(X).
\label{cycle28} %dccrossed2}
\end{equation}
The various operations of composition, identity assignment and inversion of 
arrows and arrow squares are defined in subapp. \ref{sec:dccrossed}. 
Arrow operations are essentially the same as those of the delooping $BG$ of $G$. 
Arrow square operations involve the full crossed module structure of $(G,H)$. 
The target matching condition is required for the exchange law to hold.  

\begin{prop}
A $(G,H)$--cocycle $(f,g,W)$ is equivalent to a smooth double functor $\mathbb{GR}^2\rightarrow B(G,H)$
\begin{equation}
\xymatrix{
{\text{\footnotesize $(x',y')$}}  & {\text{\footnotesize $(x,y')$}}\ar[l] \ar@{}[dl]^(.25){}="a"^(.75){}="b" \ar@{=>} "a";"b" %|-{X}
\\                 
{\text{\footnotesize $(x',y)$}} \ar[u] & {\text{\footnotesize $(x,y)$}}\ar[u] \ar[l]            
}
\!\!\!\quad
\hspace{.4cm}
\vbox{
\xymatrix{\ar@{|->}[r]
&
}
\vspace{-.85cm}}
\quad
\xymatrix@C=9pc@R=2.75pc{
{\text{\footnotesize $*$}}  & {\text{\footnotesize $*$}}\ar[l]_{f(x',x;y')}  
\ar@{}[dl]^(.25){}="a"^(.75){}="b" \ar@{=>} "a";"b"_{W(x',x;y',y)\hspace{.9cm}} %|-{X}
\\                 
{\text{\footnotesize $*$}} \ar[u]^{g(x';y',y)} & {\text{\footnotesize $*$}}\ar[u]_{g(x;y',y)}\ar[l]^{f(x',x;y)}            
}
\label{cycle29}
\end{equation}
\end{prop}

\noindent{\it Proof}. Inspection of the double groupoid operations of $\mathbb{GR}^2$, $B(G,H)$
(cf. subapps. \ref{sec:dcplane},\ref{sec:dccrossed})
reveals that the cocycle relations \eqref{cycle11,12,13,14} are 
an equivalent to the double functoriality of the above mapping
(cf. subapp. \ref{sec:dcfnctr}).  %for the definition of double functor.) 
\hfill $\Box$

Analogously to ordinary Lie group cocycles, any Lie crossed module cocycle 
yields and can be reconstructed from differential Lie crossed module 
valued differential form data. 

\begin{defi} \label{def:r2ghconn}
A $(G,H)$--connection doublet is a pair of forms $(a,B)
\in\Omega^1(\mathbb{R}^2$, $\mathfrak{g})\times \Omega^2(\mathbb{R}^2,\mathfrak{h})$ 
satisfying the zero fake curvature condition
\begin{equation}
da+\frac{1}{2}[a,a]-\dot t(B)=0.
\label{cycle30}
\end{equation}
We denote the set of $(G,H)$--connection doublets by $\Conn(G,H)$. 
\end{defi}

The following theorem holds. 
  
\begin{prop} \label{theor:cycle2} There is a canonical one--to--one correspondence between the set 
$\Cyc(G,H)$ of $(G,H)$--cocycles and the set $\Conn(G,H)$ of $(G,H)$--connection doublets. 
The connection doublet $(a_{f,g,W},B_{f,g,W})$ corresponding to a $(G,H)$--cocycle $(f,g,W)$ is 
given by 
\begin{subequations}
\label{cycle31,32}
\begin{align}
&a_{f,g,Wx}(x,y)=-\,\partial _{x'}f(x',x;y)f(x',x;y)^{-1}\big|_{x'=x},
\vphantom{\Big]}
\label{cycle31}
\\
&a_{f,g,Wy}(x,y)=-\,\partial _{y'}g(x;y',y)g(x;y',y)^{-1}\big|_{y'=y},
\nonumber
\vphantom{\Big]}
\\
&B_{f,g,Wxy}(x,y)=-\,\partial_{x'}(\partial_{y'}W(x',x;y',y)W(x',x;y',y)^{-1})\big|_{x'=x,y'=y} % - sign ?
\vphantom{\Big]}
\label{cycle32}
\\
&\hphantom{B_{f,g,Wxy}(x,y)}
=-\,\partial_{y'}(W(x',x;y',y)^{-1}\partial_{x'}W(x',x;y',y))\big|_{x'=x,y'=y}.  % - sign ?
\vphantom{\Big]}
\nonumber
\end{align}
\end{subequations}
The $(G,H)$--cocycle $(f_{a,B},g_{a,B},W_{a,B})$ corresponding to a $(G,H)$--connection doublet
$(a,B)$ is given by 
\begin{subequations}
\label{cycle33,34,35}
\begin{align}
&f_{a,B}(x,x_0;y)=u_{|y,x_0}(x),
\vphantom{\Big]}
\label{cycle33}
\\
&g_{a,B}(x;y,y_0)=v_{|x,y_0}(y),
\vphantom{\Big]}
\label{cycle34}
\\
&W_{a,B}(x,x_0;y,y_0)=E_{|x_0,y_0}(x,y),
\vphantom{\Big]}
\label{cycle35}
\end{align}
\end{subequations}
where $u_{|y,x_0}$, $v_{|x,y_0}$, $E_{|x_0,y_0}$ are the unique solution of the differential problem
\begin{subequations}
\label{cycle36,37,38}
\begin{align}
&\partial_xu_{|y,x_0}(x)u_{|y,x_0}(x)^{-1}=-a_x(x,y),
\vphantom{\Big]}
\label{cycle36}
\\
&\partial_yv_{|x,y_0}(y)v_{|x,y_0}(y)^{-1}=-a_y(x,y),
\vphantom{\Big]}
\label{cycle37}
\\
&\partial_x(\partial_yE_{|x_0,y_0}(x,y)E_{|x_0,y_0}(x,y)^{-1})
\vphantom{\Big]}
\label{cycle38}
\\
&\hspace{3cm}
=-\dot m(v_{|x_0,y_0}(y)^{-1}u_{|y,x_0}(x)^{-1})(B_{xy}(x,y)) ~~\text{or}~~
\nonumber
\vphantom{\Big]}
\\
&\partial_y(E_{|x_0,y_0}(x,y)^{-1}\partial_xE_{|x_0,y_0}(x,y))
\vphantom{\Big]}
\nonumber
\\
&\hspace{3cm}
=-\dot m(u_{|y_0,x_0}(x)^{-1}v_{|x,y_0}(y)^{-1})(B_{xy}(x,y))
\nonumber
\vphantom{\Big]}
\end{align}
\end{subequations}
with $u_{|-,x_0},v_{|-,y_0}:\mathbb{R}^2\rightarrow G$ and 
$E_{|x_0,y_0}:\mathbb{R}^2\rightarrow H$ smooth and satisfying the initial conditions \hphantom{xxxxxxxxxxxxxx}
\begin{subequations}
\label{cycle39,40,41}
\begin{align}
&u_{|y,x_0}(x_0)=1_G,
\vphantom{\Big]}
\label{cycle39}
\\
&v_{|x,y_0}(y_0)=1_G,
\vphantom{\Big]}
\label{cycle40}
\\
&E_{|x_0,y_0}(x_0,y)=E_{|x_0,y_0}(x,y_0)=1_H
\vphantom{\Big]}
\label{cycle41}
\end{align}
\end{subequations}
(cf. eq. \eqref{hiholo3}). 
The two forms of the differential problem \eqref{cycle38} 
with the initial condition \eqref{cycle41} are equivalent:
any solution of one is automatically solution of the other.
\end{prop}

\noindent{\it Proof}. 
If $(f,g,W)$ is a $(G,H)$--cocycle, then \eqref{cycle31}, \eqref{cycle32}
clearly define a $\mathfrak{g}$--valued $1$--form $a_{f,g,W}$ and a 
$\mathfrak{h}$--valued $2$--form $B_{f,g,W}$ on $\mathbb{R}^2$. 
The identity of the two expressions of $B_{f,g,W}$ follows from the relation \pagebreak 
\begin{align}
&\partial_{x'}(\partial_{y'}W(x',x;y',y)W(x',x;y',y)^{-1})
\vphantom{\Big]}
\label{cycleb1}
\\
&\hspace{1.5cm}
=\Ad W(x',x;y',y)(\partial_{y'}(W(x',x;y',y)^{-1}\partial_{x'}W(x',x;y',y)))   %W(x',x;y',y)^{-1}
\vphantom{\Big]}
\nonumber
\end{align}
and \eqref{cycle20}. Using relations \eqref{cycle31}, \eqref{cycle32} %\pagebreak 
and the target matching condition \eqref{cycle15}, we find, 
\begin{align}
&\hspace{-.3cm}\dot t(B_{f,g,W\, xy}(x,y))
\vphantom{\Big]}
\label{}
\\
&=-\,\partial_{x'}(\partial_{y'}t(W(x',x;y',y))t(W(x',x;y',y))^{-1})\big|_{x'=x,y'=y} %
\vphantom{\Big]}
\nonumber
\\
&=-\,\partial_{x'}\big(\partial_{y'}(g(x;y',y)^{-1}f(x',x;y')^{-1}g(x';y',y)f(x',x;y))
\vphantom{\Big]}
\nonumber
\\
&\hspace{1cm}\times f(x',x;y)^{-1}g(x';y',y)^{-1}f(x',x;y')g(x;y',y)\big)\big|_{x'=x,y'=y}
\vphantom{\Big]}
\nonumber
\\
&=-\,\partial_x(\partial _{y'}g(x;y',y)g(x;y',y)^{-1}\big|_{y'=y})
+\partial_y(\partial _{x'}f(x',x;y)f(x',x;y)^{-1}\big|_{x'=x})
\vphantom{\Big]}
\nonumber
\\
&\hspace{1cm}+[\partial _{x'}f(x',x;y)f(x',x;y)^{-1}\big|_{x'=x}, 
\partial _{y'}g(x;y',y)g(x;y',y)^{-1}\big|_{y'=y}]
\vphantom{\Big]}
\nonumber
\\
&=\partial_xa_{f,g,Wy}(x,y)-\partial_ya_{f,g,Wx}(x,y)+[a_{f,g,Wx}(x,y),a_{f,g,Wy}(x,y)]
\vphantom{\Big]}
\nonumber
\end{align}
verifying the zero fake curvature condition \eqref{cycle30}. Thus, 
the pair $(a_{f,g,W},a_{f,g,W})$ is a $(G,H)$--connection doublet. 
This shows the first part of the theorem. 

Proving the second part of the theorem requires some preparatory work.
We assume that $r$, $l$ are $G$--valued maps and $D$ is an $\mathfrak{h}$--valued $2$--form
on $\mathbb{R}^2$ satisfying the differential relations
%with the property that 
\begin{subequations}
\label{cycleb13,14}
\begin{align}
&\partial_x(r(x,y)^{-1}\partial_yr(x,y)-l(x,y)^{-1}\partial_yl(x,y))
\vphantom{\Big]}
\label{cycleb13}
\\
&\hspace{.3cm}+[r(x,y)^{-1}\partial_xr(x,y),r(x,y)^{-1}\partial_yr(x,y)-l(x,y)^{-1}\partial_yl(x,y)]
=\dot t(D_{xy}(x,y)),
\vphantom{\Big]}
\nonumber
\\
&\partial_y(r(x,y)^{-1}\partial_xr(x,y)-l(x,y)^{-1}\partial_xl(x,y))
\vphantom{\Big]}
\label{cycleb14}
\\
&\hspace{.3cm}+[l(x,y)^{-1}\partial_yl(x,y),r(x,y)^{-1}\partial_xr(x,y)-l(x,y)^{-1}\partial_xl(x,y)]
=\dot t(D_{xy}(x,y))
\vphantom{\Big]}
\nonumber
\end{align}
\end{subequations}
and the initial conditions
\begin{equation}
r(x_0,y)l(x_0,y)^{-1}=r(x,y_0)l(x,y_0)^{-1}=1_G.
\label{cycleb8}
\end{equation}

The differential problem 
\begin{align}
&\partial_x(\partial_yR(x,y)R(x,y)^{-1})=\dot m(r(x,y))(D_{xy}(x,y)),
\vphantom{\Big]}
\label{cycleb2}
\\
&R(x_0,y)=R(x,y_0)=1_H
\vphantom{\Big]}
\label{cycleb3}
\end{align}
with $R$ a smooth $H$--valued map on $\mathbb{R}^2$ has a unique solution, 
since it is equivalent to the differential problem 
\begin{align}
&\partial_yR(x,y)R(x,y)^{-1}=\int_{x_0}^x d\xi\,\dot m(r(\xi,y))(D_{xy}(\xi,y)),
\vphantom{\Big]}
\label{cycleb4}
\\
&R(x,y_0)=1_H,
\vphantom{\Big]}
\label{cycleb5}
\end{align}
which does. Similarly, the differential problem 
\begin{align}
&\partial_y(L(x,y)^{-1}\partial_xL(x,y))=\dot m(l(x,y))(D_{xy}(x,y)),
\vphantom{\Big]}
\label{cycleb6}
\\
&L(x_0,y)=L(x,y_0)=1_H
\vphantom{\Big]}
\label{cycleb7}
\end{align}
with $L$ a smooth $H$--valued map on $\mathbb{R}^2$ has a unique solution by being equivalent to the problem 
\begin{align}
&L(x,y)^{-1}\partial_xL(x,y)=\int_{y_0}^y d\eta\,\dot m(l(x,\eta))(D_{xy}(x,\eta)),
\vphantom{\Big]}
\label{}
\\
&L(x_0,y)=1_H. \hspace{6.3cm}
\vphantom{\Big]}
\label{}
\end{align}

Suppose that $Q$ is an $H$--valued map on $\mathbb{R}^2$ such that 
\begin{equation}
t(Q(x,y))=r(x,y)l(x,y)^{-1}.
\label{cycleb9}
\end{equation}
Then,  $R(x,y)=Q(x,y)$ solves the differential problem \eqref{cycleb2}, \eqref{cycleb3}
if and only if $L(x,y)=Q(x,y)$ does that \eqref{cycleb6}, \eqref{cycleb7}, by the relation 
\begin{equation}
\partial_x(\partial_yQ(x,y)Q(x,y)^{-1})
=\Ad Q(x,y)(\partial_y(Q(x,y)^{-1}\partial_xQ(x,y)))    %Q(x,y)^{-1}
\label{cycleb10}
\end{equation}
and the Peiffer identity. 

The auxiliary differential problem 
\begin{align}
&\partial_x(\partial_y\rho(x,y)\rho(x,y)^{-1})=\Ad r(x,y))(\dot t(D_{xy}(x,y))), 
\vphantom{\Big]}
\label{cycleb11}
\\
&\rho(x_0,y)=\rho(x,y_0)=1_G
\vphantom{\Big]}
\label{cycleb12}
\end{align}
with $\rho$ a smooth $G$--valued map on $\mathbb{R}^2$ has a unique solution, by
a reasoning completely analogous to that indicated two paragraphs above. Similarly, the
auxiliary differential problem 
\begin{align}
&\partial_y(\lambda(x,y)^{-1}\partial_x\lambda(x,y))=\Ad l(x,y))(\dot t(D_{xy}(x,y))), 
\vphantom{\Big]}
\label{cycleb15}
\\
&\lambda(x_0,y)=\lambda(x,y_0)=1_G
\vphantom{\Big]}
\label{cycleb16}
\end{align}
with $\lambda$ a smooth $G$--valued map on $\mathbb{R}^2$ has a unique solution. 

Suppose that $Q$ is an $H$--valued map on $\mathbb{R}^2$ such that $R(x,y)=Q(x,y)$ solves 
the differential problem \eqref{cycleb2}, \eqref{cycleb3}. Then, $\rho(x,y)=t(Q(x,y))$ solves 
\eqref{cycleb11}, \eqref{cycleb12}. 
Using \eqref{cycleb13} and \eqref{cycleb8}, it is straightforward to verify that 
$\rho(x,y)=r(x,y)l(x,y)^{-1}$ also solves \eqref{cycleb11}, \eqref{cycleb12}.
By uniqueness, it then follows that \eqref{cycleb9} holds. 
Similarly, by using \eqref{cycleb14} and \eqref{cycleb8} and making reference to the problem 
\eqref{cycleb15}, \eqref{cycleb16} instead, one finds that when 
$Q$ is an $H$--valued map on $\mathbb{R}^2$ such that $L(x,y)=Q(x,y)$ solves 
the differential problem \eqref{cycleb6}, \eqref{cycleb7}, then \eqref{cycleb9} holds. 
We conclude that, under the assumptions \eqref{cycleb13,14} and \eqref{cycleb8}, 
the differential problems \eqref{cycleb2}, \eqref{cycleb3} and \eqref{cycleb6}, \eqref{cycleb7} 
have a unique solution and that this solution is the same for both
and obeys \eqref{cycleb9}.

We can now complete the proof of the second part of the theorem. 
Let $(a,B)$ be a $(G,H)$--connection doublet. The solution 
$u_{|y,x_0}$ of the differential problem \eqref{cycle36}, \eqref{cycle39}
exists, is unique and is smooth in $y$ and $x_0$. Similarly, 
the solution $v_{|x,y_0}$ of the differential problem \eqref{cycle37}, \eqref{cycle40}
exists, is unique and is smooth in $x$ and $y_0$. Using 
\eqref{cycle36}, \eqref{cycle37} and \eqref{cycle39}, \eqref{cycle40}
and the zero fake curvature condition \eqref{cycle30}, \pagebreak 
it is straightforward to check that the $G$--valued maps $r$, $l$ and 
the $\mathfrak{h}$--valued $2$--form $D$ on $\mathbb{R}^2$ defined by 
\begin{subequations}
\begin{align}
&r(x,y)=v_{|x_0,y_0}(y)^{-1}u_{|y,x_0}(x)^{-1},
\vphantom{\Big]}
\label{cycleb17}
\\
&l(x,y)=u_{|y_0,x_0}(x)^{-1}v_{|x,y_0}(y)^{-1},
\vphantom{\Big]}
\label{cycleb18}
\\
&D_{xy}(x,y)=-B_{xy}(x,y)
\vphantom{\Big]}
\label{cycleb19}
\end{align}
\end{subequations}
obey relations \eqref{cycleb13}, \eqref{cycleb14} and \eqref{cycleb8}. 
Therefore, by what was shown above, the solution $E_{|x_0,y_0}$
of the twin differential problems \eqref{cycle38}, \eqref{cycle41}
exists, is unique and is smooth in $x_0$, $y_0$ and furthermore 
it is the same for both and satisfies
\begin{equation}
t(E_{|x_0,y_0}(x,y))=v_{|x_0,y_0}(y)^{-1}u_{|y,x_0}(x)^{-1}v_{|x,y_0}(y)u_{|y_0,x_0}(x).
\label{cycleb20}
\end{equation}
Relations \eqref{cycle33}--\eqref{cycle35} define in this way a $G$--valued map
$f_{a,B}$ on $\mathbb{R}^2\times \mathbb{R}$, a $G$--valued map
$g_{a,B}$ on $\mathbb{R}\times \mathbb{R}^2$ and an $H$--valued map
$W$ on $\mathbb{R}^2\times \mathbb{R}^2$ fulfilling the target matching condition 
\eqref{cycle15}. We have now to show that these objects satisfy the cocycle relations 
\eqref{cycle11,12,13,14}. 
Consider the $G$-- and $H$--valued maps
\begin{subequations}
\label{cycleb21,22,23,24,25,26,27,28} 
\begin{align}
&u_1(x)=f_{a,B|y}(x,x_1)f_{a,B|y}(x_1,x_0), 
\vphantom{\Big]}
\label{cycleb21}
\\
&u_2(x)=f_{a,B|y}(x,x_0),
\vphantom{\Big]}
\label{cycleb22}
\\
&v_1(y)=g_{a,B|x}(y,y_1)g_{a,B|x}(y_1,y_0),
\vphantom{\Big]}
\label{cycleb23}
\\
&v_2(y)=g_{a,B|x}(y,y_0),
\vphantom{\Big]}
\label{cycleb24}
\\
&E_1(x,y)=W_{a,B|y,y_0}(x_1,x_0)m(f_{a,B|y_0}(x_1,x_0)^{-1})(W_{a,B|y,y_0}(x,x_1)),
\vphantom{\Big]}
\label{cycleb25}
\\
&E_2(x,y)=W_{a,B|y,y_0}(x,x_0),
\vphantom{\Big]}
\label{cycleb26}
\\
&E_3(x,y)=m(g_{a,B|x_0}(y_1,y_0)^{-1})(W_{a,B|x,x_0}(y,y_1))W_{a,B|x,x_0}(y_1,y_0),
\vphantom{\Big]}
\label{cycleb27}
\\
&E_4(x,y)=W_{a,B|x,x_0}(y,y_0).
\vphantom{\Big]}
\label{cycleb28}
\end{align}
\end{subequations}  
By \eqref{cycle33}, \eqref{cycle36}, \eqref{cycle39}, 
$u_1$, $u_2$ both solve the differential equation $d_xu(x)u(x)^{-1}=-a_x(x,y)$
with initial condition $u(x_1)=f_{a,B|y}(x_1,x_0)$.
By the uniqueness of the solution of this differential problem, $u_1=u_2$. 
By \eqref{cycleb21}, \eqref{cycleb22}, then, $f_{a,B|y}$ fulfills the cocycle
condition \eqref{cycle11} as required. Similarly, by \eqref{cycle34}, \eqref{cycle37}, \eqref{cycle40}, 
$v_1$, $v_2$ both solve the differential equation 
$d_yv(y)v(y)^{-1}=-a_y(x,y)$ with initial condition $v(y_1)=g_{a,B|x}(y_1,y_0)$, 
so that $v_1=v_2$. By \eqref{cycleb23}, \eqref{cycleb24}, then, $g_{a,B|x}$ fulfills the cocycle
condition \eqref{cycle12}. By \eqref{cycle35}, \eqref{cycle38}, \eqref{cycle41}, 
$E_1$, $E_2$ both solve the differential equation 
\begin{equation}
E(x,y)^{-1}\partial_xE(x,y)=-\int_{y_0}^yd\eta\,\dot m(f_{a,B|y_0}(x,x_0)^{-1}g_{a,B|x}(\eta,y_0)^{-1})
(B_{xy}(x,\eta))
\nonumber
\end{equation}
with initial condition $E(x_1,y)=W_{a,B|y,y_0}(x_1,x_0)$. Again by the uniqueness of the solution of this
differential problem, we have $E_1=E_2$, from which through \eqref{cycleb25}, \eqref{cycleb26}
it follows that $W_{a,B}$ obeys the cocycle condition \eqref{cycle13}. By considering instead the equation
\begin{equation}
\partial_yE(x,y)E(x,y)^{-1}=-\int_{x_0}^xd\xi\,\dot m(g_{a,B|x_0}(y,y_0)^{-1}f_{a,B|y}(\xi,x_0)^{-1})
(B_{xy}(\xi,y))
\nonumber
\end{equation}
one finds that $E_3=E_4$. from which through \eqref{cycleb27}, \eqref{cycleb28} 
it follows that $W_{a,B}$ also obeys the condition \eqref{cycle14}. 

To conclude the proof of the theorem, we have to show that the mappings
$(f,g,W)\to(a_{f,g,W},B_{f,g,W})$ and $(a,B)\to(f_{a,B}, g_{a,B}, W_{a,B})$ are reciprocally
inverse. For a given doublet $(a,B)$, 
inserting the \eqref{cycle33,34,35} into the \eqref{cycle31,32} and using 
\eqref{cycle36,37,38}, \eqref{cycle39,40,41}, it is immediately verified that 
$a_{f_{a,B}, g_{a,B}, W_{a,B}}=a$, $B_{f_{a,B}, g_{a,B}, W_{a,B}}$ $=B$. For a given cocycle $(f,g,W)$, 
from the \eqref{cycle31,32}, 
using the cocycle relations \eqref{cycle11,12,13,14}, it is relatively straightforward 
to check that $u_{|y,x_0}(x)=f(x,x_0;y)$, $v_{|x,y_0}(y)=g(x;y,y_0)$
and $E_{|x_0,y_0}(x,y)=W(x,x_0;y,y_0)$ solve the differential problem 
\eqref{cycle36,37,38}, \eqref{cycle39,40,41} with $a=a_{f,g,W}$, $B=B_{f,g,W}$, 
so that $f_{a_{f,g,W},B_{f,g,W}}=f$, $g_{a_{f,g,W},B_{f,g,W}}=g$, $W_{a_{f,g,W},B_{f,g,W}}=W$.
The claim is so shown. \hfill $\Box$ 

We have so achieved our first goal, the formulation of a
Lie crossed module cocycle theory naturally relating to higher gauge
theory.

\vfil\eject

\subsection{\normalsize \textcolor{blue}{Lie crossed module $1$--gauge transformations}}\label{sec:gauge}

\hspace{.5cm} In ordinary as in higher gauge theory, parallel
transport must be gauge 
covariant. It is important therefore to have the appropriate notion of gauge transformation for cocycles.
We review first gauge transformation of ordinary group cocycles and then we define gauge transformation 
of crossed module cocycles. 

Let $G$ be a Lie group. 

\begin{defi}
A $G$--gauge transformation is a map $\varkappa\in\Map(\mathbb{R},G)$. 
The $G$-- gauge transformations form a set $\Gau(G)$. 
\end{defi}

The following proposition is basic. 

\begin{prop} For any $G$--cocycle $f$ and any $G$--gauge transformation $\varkappa$, the mapping 
${}^\varkappa f\in\Map(\mathbb{R},G)$ defined by the expression 
\begin{equation}
{}^\varkappa f(x',x)=\varkappa(x')f(x',x)\varkappa(x)^{-1}.
\label{gauge1}
\end{equation}
is also a $G$--cocycle, the gauge transform %${}^\varkappa f$
 of $f$ by $\varkappa$.
\end{prop} 

\noindent
{\it Proof}. It is readily checked that ${}^\varkappa f$ obeys the cocycle relation \eqref{cycle1}.
\hfill $\Box$  

As we showed in subsect. \ref{sec:cycle}, every Lie group cocycle represents secretly a smooth functor form the segment
groupoid  to the delooping groupoid of the Lie group. In the same spirit, every gauge transformation 
defines a natural transformation between a Lie group cocycle and its gauge transform. 
%Though this is well--known, 
%we recall it to prepare the ground for %the treatment of the higher case.

\begin{prop}
If $f$ is $G$--cocycle and $\varkappa$ is a $G$--gauge transformation, then $\varkappa$ yields
a natural transformation $\varkappa:f\Rightarrow {}^\varkappa f$ of the functors
$f,{}^\varkappa f:\mathbb{GR}\rightarrow BG$. 
\end{prop}

\noindent{\it Proof}. %In a categorical approach, 
By \eqref{gauge1}, a gauge transformation $\varkappa$ amounts to a mapping
\vspace{-.2cm}
\begin{equation}
\vbox{
\xymatrix{{\text{\footnotesize $x$}}\hspace{.3cm}\ar@{|->}[r]
&
}
\vspace{-.75cm}}
\xymatrix{
{\text{\footnotesize $*$}}
\\
{\text{\footnotesize $*$}}\ar[u]_{\varkappa(x)}
}
\label{gauge2}
\end{equation}
of the objects of $\mathbb{GR}$ to the arrows of $BG$ such that for each arrow
\begin{equation}
\xymatrix{ {\text{\footnotesize $y$}}&{\text{\footnotesize $x$}}\ar[l]
}
\label{gauge3}
\end{equation}
of $\mathbb{GR}$, the diagram of $BG$ \hphantom{xxxxxxxxxxxx}
\begin{equation}
\xymatrix@C=3.5pc{
{\text{\footnotesize $*$}}  & {\text{\footnotesize $*$}}\ar[l]_{{}^\varkappa f(x',x)}  
\\                 
{\text{\footnotesize $*$}} \ar[u]^{\varkappa(y)} & {\text{\footnotesize $*$}}\ar[u]_{\varkappa(x)}\ar[l]^{f(x',x)}             
}
\label{gauge4}
\end{equation}
commutes. This is precisely the statement that $\varkappa$ is a natural transformation
$f\Rightarrow {}^\varkappa f$ of the functors $f,{}^\varkappa f:\mathbb{GR}\rightarrow BG$. 
\hfill $\Box$

By prop. \ref{theor:cycle1}, there is one--to--one correspondence between $G$--cocycles $f$  
and $G$--connections $a$. Hence,
the action of a $G$--gauge transformation $\varkappa$ on $f$ must translate into one
on the form $a_f$. %Indeed, the following classic theorem holds. 

\begin{prop} \label{theor:gauge3} Let $f$ be a $G$ cocycle and  
$\varkappa$ be a gauge transformation. 
Then, the form $a_{{}^\varkappa f}$ associated with the gauge transformed cocycle ${}^\varkappa f$ is 
\begin{equation}
a_{{}^\varkappa f}=\Ad \varkappa(a_f)-d\varkappa\varkappa^{-1}.
\label{gauge30}
\end{equation}
\end{prop}

\noindent{\it Proof}. This follows readily from inserting \eqref{gauge1}
into \eqref{cycle7}. See  also  ref. \cite{Schrei:2011}. 
\hfill $\Box$

An action of  $G$--gauge transformations on $G$--connections is so yielded. 

\begin{defi} 
Let $a$ be a $G$--connection. For a $G$--gauge transformation $\varkappa$, 
\begin{equation}
{}^\varkappa a=\Ad \varkappa (a)-d\varkappa\varkappa^{-1}.
\vphantom{\Big]}
\label{gauge31}
\end{equation}
\end{defi}

We now extend the above to a Lie crossed module $(G,H,t,m)$. 

\begin{defi} Let $(f,g,W)$ be a $(G,H)$--cocycle. 
An $(f,g,W)$--$1$--gauge transformation, or an integral $(G,H)$--$1$--gauge transformation when $(f,g,W)$
is understood, consists of three maps $\kappa\in\Map(\mathbb{R}\times\mathbb{R},G)$,
$\varPsi\in\Map(\mathbb{R}^2\times \mathbb{R},H)$, $\varPhi\in\Map(\mathbb{R}\times \mathbb{R}^2,H)$
satisfying the relations 
\begin{subequations}
\label{gauge5,6}
\begin{align}
&\varPsi_{|y}(x'',x)=\varPsi_{|y}(x',x)m(f_{|y}(x',x)^{-1})(\varPsi_{|y}(x'',x')),
\vphantom{\Big]}
\label{gauge5}
\\
&\varPhi_{|x}(y'',y)=\varPhi_{|x}(y',y)m(g_{|x}(y',y)^{-1})(\varPhi_{|x}(y'',y')),
\vphantom{\Big]}
\label{gauge6}
\end{align}
\end{subequations}
where we have set  %$\kappa_{|x}(y)=\kappa_{|y}(x)=\kappa(x;y)$, 
$\varPsi_{|y}(x',x)=\varPsi(x',x;y)$ and $\varPhi_{|x}(y',y)=\varPhi(x;y',y)$
for clarity. 
The $(f,g,W)$--$1$--gauge transformations form a set $\Gau_{1\,f,g,W}(G,H)$.
\end{defi}

The following properties of crossed module cocycles are immediately proven.

\begin{prop}
If $(f,g,W)$ is a $(G,H)$--cocycle and $(\kappa,\varPsi,\varPhi)$ is an $(f,g,W)$ --$1$--gauge transformation, then 
\begin{subequations}
\label{gauge7,8,9,10}
\begin{align}
&\varPsi_{|y}(x,x)=1_H,
\vphantom{\Big]}
\label{gauge7}
\\
&\varPsi_{|y}(x,x')=m(f_{|y}(x',x))(\varPsi_{|y}(x',x)^{-1}),
\vphantom{\Big]}
\label{gauge8}
\\
&\varPhi_{|x}(y,y)=1_H,
\vphantom{\Big]}
\label{gauge9}
\\
&\varPhi_{|x}(y,y')=m(g_{|x}(y',y))(\varPhi_{|x}(y',y)^{-1})
\vphantom{\Big]}
\label{gauge10}
\end{align}
\end{subequations}
for $x,x',x'',y,y',y''\in\mathbb{R}$. 
\end{prop}

Just as ordinary gauge transformations act on group cocycles
$1$--gauge transformations act on crossed module cocycles.  

\begin{prop}
Let $(f,g,W)$ be a $(G,H)$--cocycle and $(\kappa,\varPsi,\varPhi)$ be an $(f,g,W)$ --gauge transformation. Then,  
the mappings ${}^{\kappa,\varPsi,\varPhi}f\in\Map(\mathbb{R}^2\times \mathbb{R},G)$,
${}^{\kappa,\varPsi,\varPhi}g\in\Map(\mathbb{R}\times \mathbb{R}^2,G)$
and ${}^{\kappa,\varPsi,\varPhi}W\in\Map(\mathbb{R}^2\times\mathbb{R}^2,H)$
defined by the expressions 
\begin{subequations}
\label{gauge11,12,13}
\begin{align}
&{}^{\kappa,\varPsi,\varPhi}f_{|y}(x',x)
=\kappa_{|y}(x')f_{|y}(x',x)t(\varPsi_{|y}(x',x))^{-1}\kappa_{|y}(x)^{-1},
\vphantom{\Big]}
\label{gauge11}
\\
&{}^{\kappa,\varPsi,\varPhi}g_{|x}(y',y)
=\kappa_{|x}(y')g_{|x}(y',y)t(\varPhi_{|x}(y',y))^{-1}\kappa_{|x}(y)^{-1},
\vphantom{\Big]}
\label{gauge12}
\\
&{}^{\kappa,\varPsi,\varPhi}W(x',x;y',y)
=m(\kappa(x;y))\big(\varPhi_{|x}(y',y)m(g_{|x}(y',y)^{-1})(\varPsi_{|y'}(x',x))
\vphantom{\Big]}
\label{gauge13}
\\
%&\hspace{.5cm}
%=m(\kappa(x;y))\big(\varPhi_{|x}(y',y)m(g_{|x}(y',y)^{-1})(\varPsi_{|y'}(x',x))
%\nonumber
%\vphantom{\Big]}
%\\
&\hspace{2.5cm}\times W(x',x;y',y)
m(f_{|y}(x',x)^{-1})(\varPhi_{|x'}(y',y))^{-1}\varPsi_{|y}(x',x)^{-1}\big),
\nonumber
\vphantom{\Big]}
\end{align}
\end{subequations}
where we have set  $\kappa_{|x}(y)=\kappa_{|y}(x)=\kappa(x;y)$ for clarity, 
constitute a $(G,H)$--cocycle 
$({}^{\kappa,\varPsi,\varPhi}f$, ${}^{\kappa,\varPsi,\varPhi}g,{}^{\kappa,\varPsi,\varPhi}W)$, 
the gauge transform of $(f,g,W)$ by $(\kappa,\varPsi,\varPhi)$. 
\end{prop}

\noindent{\it Proof}. Exploiting  the \eqref{gauge5,6}, one checks  
that $({}^{\kappa,\varPsi,\varPhi}f,{}^{\kappa,\varPsi,\varPhi}g,{}^{\kappa,\varPsi,\varPhi}W)$
satisfies the target matching condition \eqref{cycle15} and the cocycle relations 
\eqref{cycle11,12,13,14} whenever $(f,g,W)$  does. 
\hfill $\Box $

As we showed in subsect. \ref{sec:cycle}, every Lie crossed module cocycle represents secretly a smooth functor from 
the rectangle double groupoid to the delooping double groupoid of the Lie crossed module. Analogously to the ordinary case, 
every $1$--gauge transformation defines a double natural transformation between a Lie crossed module cocycle 
and its gauge transform. %We remark to the reader however that 
The notion of double natural transformation we use, however, is not the customary one %usually stated %in the literature 
and presupposes that the target category is edge symmetric and folded 
(cf. subapps. \ref{sec:dcedge}, \ref{sec:dcedsym}, \ref{sec:dccrossed}).

\begin{prop}
If $(f,g,W)$ is $(G,H)$--cocycle and $(\kappa,\varPsi,\varPhi)$ is a $(f,g,W)$--$1$--gauge transformation, 
then $(\kappa,\varPsi,\varPhi)$ is equivalent to a double natural transformation 
$(f,g,W)\Rightarrow ({}^{\kappa,\varPsi,\varPhi}f,{}^{\kappa,\varPsi,\varPhi}g,{}^{\kappa,\varPsi,\varPhi}W)$ 
of the double functors $(f,g,W),({}^{\kappa,\varPsi,\varPhi}f$, ${}^{\kappa,\varPsi,\varPhi}g,{}^{\kappa,\varPsi,\varPhi}W)
:\mathbb{GR}^2\rightarrow B(G,H)$. 
\end{prop}

\noindent{\it Proof}. The data of a $(f,g,W)$--$1$--gauge transformation $(\kappa,\varPsi,\varPhi)$  are equivalent to 
a mapping of the set of object of $\mathbb{GR}^2$ into the set of vertical arrows of $B(G,H)$,
\begin{equation}
\vbox{
\xymatrix{{\text{\footnotesize $(x,y)$}}\hspace{.3cm}\ar@{|->}[r]
&
}
\vspace{-.8cm}}
\xymatrix{
{\text{\footnotesize $*$}}
\\
{\text{\footnotesize $*$}}
\ar[u]_{\kappa(x;y)}
}
%\vspace{.2cm}
\label{gauge14} %dcnatr1}
\end{equation}
and two compatible functors from the horizontal and vertical arrow groupoids of $\mathbb{GR}^2$ into 
the horizontal truncation groupoid $B(G,H)_h$ of $B(G,H)$
\hskip11mm\noindent
\begin{equation}
\vbox{
\hbox{\xymatrix{
{\text{\footnotesize $(x',y)$}}&{\text{\footnotesize $(x,y)$}}\ar[l]
}
\quad\xymatrix{\ar@{|->}[r]
&
}\quad
}
\vspace{-.8cm}}
\xymatrix@C=6pc{
{\text{\footnotesize $*$}}  & {\text{\footnotesize $*$}}\ar[l]_{{}^{\kappa,\varPsi,\varPhi}f(x',x;y)}  
\ar@{}[dl]^(.25){}="a"^(.75){}="b" \ar@{=>} "a";"b"_{\varPsi(x',x;y)\hspace{.5cm}} %|-{X}
\\                 
{\text{\footnotesize $*$}} \ar[u]^{\kappa(x';y)} & {\text{\footnotesize $*$}}\ar[u]_{\kappa(x;y)}\ar[l]^{f(x',x;y)}             
}
\label{gauge15} %dcnatr2}
\end{equation}
\begin{equation}
\xymatrix{
{\text{\footnotesize $(x,y')$}}
\\
{\text{\footnotesize $(x,y)$}}\ar[u]
}
\quad\vbox{
\xymatrix{\ar@{|->}[r]
&
}
\vspace{-.67cm}}\quad
\xymatrix@C=6pc{
{\text{\footnotesize $*$}}  & {\text{\footnotesize $*$}}\ar[l]_{{}^{\kappa,\varPsi,\varPhi}g(x;y',y)}  
\ar@{}[dl]^(.25){}="a"^(.75){}="b" \ar@{=>} "a";"b"_{\varPhi(x;y',y)\hspace{.5cm}} %|-{X}
\\                 
{\text{\footnotesize $*$}} \ar[u]^{\kappa(x;y')} & {\text{\footnotesize $*$}}\ar[u]_{\kappa(x;y)\,.}\ar[l]^{g(x;y',y)}             
}
\nonumber
\end{equation}
\vskip5mm\noindent
(cf. eqs. \eqref{dcnatr1}, \eqref{dcnatr2}). 
The fulfillment of the target matching condition \eqref{dccrossed2} is guaranteed by relations
\eqref{gauge11}, \eqref{gauge12}. The functoriality of the mappings \eqref{gauge15}
is equivalent to relations \eqref{gauge5}, \eqref{gauge6} and the ensuing relations
\eqref{gauge7}-\eqref{gauge10}. \eqref{gauge14}, \eqref{gauge15} 
 are precisely the data required for a double natural transformation
from the first to the second of the double functors
$(f,g,W), ({}^{\kappa,\varPsi,\varPhi}f$, ${}^{\kappa,\varPsi,\varPhi}g,{}^{\kappa,\varPsi,\varPhi}W)
:\mathbb{GR}^2\rightarrow B(G,H)$.
The only thing left to check is the double naturality condition \eqref{dcnatr5}. 
Using the expressions of the operations of the double groupoid $B(G,H)$ of subapp.
\ref{sec:dccrossed}, it is easily checked that
this is equivalent to relation \eqref{gauge13} written in the form
\hskip1.4cm\noindent
\begin{align}
&\varPhi(x;y',y)m(g(x;y',y)^{-1})(\varPsi(x',x;y'))W(x',x;y',y)
\vphantom{\Big]}
\label{gauge16}
\\
&\hspace{.5cm}=m(\kappa(x;y)^{-1})({}^{\kappa,\varPsi,\varPhi}W(x',x;y',y))
\varPsi(x',x;y)m(f(x',x;y)^{-1})(\varPhi(x';y',y)).
\nonumber
\vphantom{\Big]}
\end{align}
%\vskip-5mm\eject\noindent
Intuitively, the double naturality condition can be interpreted as the requirement that
the cube diagram of $B(G,H)$ %\vfil\noindent
\vskip8mm\noindent
\begin{equation}
\vbox{
\hbox{
\hspace{1.3cm}\xymatrix{\ar@{.>}'[dr]^{\hspace{-.3cm}\varPsi(x',x;y')\vphantom{\ul{\ul{\ul{\ul{g}}}}}}[ddrr]&&
\\
&&
\\
&&
\\
&&}
}
\vspace{-3.cm}
\hbox{
\xymatrix@C=1.3pc@R=2pc{\\
&&&
\\
\ar@{.>}[rrr]^{\varPhi(x';y',y)\hspace{1cm}}&&&
\\
&&&
}
\hspace{-1.7cm}
\xymatrix@C=2.5pc@R=2.5pc{
{\text{\footnotesize $*$}}  && {\text{\footnotesize $*$}} \ar[ll]_{{}^{\kappa,\varPsi,\varPhi}f(x',x;y')} 
\ar@{}[dl]^(.25){}="a"^(.75){}="b" \ar@{} "a";"b"|-{{}^{\kappa,\varPsi,\varPhi}W(x',x;y',y)}
\\
& {\text{\footnotesize $*$}} \ar[ul]^{{}^{\kappa,\varPsi,\varPhi}g(x';y',y)\!\!\!\!} 
&& {\text{\footnotesize $*$}}\ar[ul]_{\!\!\!\!\!{}^{\kappa,\varPsi,\varPhi}g(x;y',y)} 
\ar[ll]^{{}^{\kappa,\varPsi,\varPhi}f(x',x;y)} 
\\
{\text{\footnotesize $*$}} \ar[uu]^{\kappa(x',y')} &&  {\text{\footnotesize $*$}} 
\ar'[l]_{f(x',x;y')}[ll] \ar'[u]_{\kappa(x;y')}[uu] 
\ar@{}[dl]^(.25){}="a"^(.75){}="b" \ar@{} "a";"b"|-{W(x',x;y',y)}
\\
&  {\text{\footnotesize $*$}} \ar[ul]^{g(x';y',y)} \ar[uu]^{\kappa(x';y)\vphantom{\ul{\ul{\ul{\ul{\ul{\ul{g}}}}}}}} 
&& {\text{\footnotesize $*$}} \ar[ul]_{g(x;y',y)} \ar[ll]^{f(x',x;y)} \ar[uu]_{\kappa(x;y)}
}
\hspace{-1.7cm}
\xymatrix@C=1.3pc@R=2pc{\\
&&&
\\
&&&\ar@{.>}[lll]_{\hspace{1cm}\varPhi(x;y',y)}
\\
&&&
}
}
\vspace{-3.cm}
\hbox{
\hspace{4.7cm}
\xymatrix{
&&
\\
&&
\\
&&
\\
&&\ar@{.>}'[ul]^{\vphantom{\Big[}\varPsi(x',x;y)\hspace{-.4cm}}[uull]}
}
}
\label{gauge17} %{dcnatr6}
\end{equation} 
\vfill\eject\noindent
commutes for any arrow square of $\mathbb{GR}^2$,  
%\vspace{-.1cm}
\begin{equation}
\xymatrix{
{\text{\footnotesize $(x',y')$}}  & {\text{\footnotesize $(x,y')$}}\ar[l] \ar@{}[dl]^(.25){}="a"^(.75){}="b" \ar@{=>} "a";"b" %|-{X}
\\                 
{\text{\footnotesize $(x',y)$}} \ar[u] & {\text{\footnotesize $(x,y)_{\vphantom{g}}$}}\ar[u] \ar[l]            
}\!,
\label{cycle25/1}%dcplane3} 
\end{equation}
where we have dropped all double arrows in order not to clog the diagram
(cf. eq. \eqref{dcnatr6}). 
The precise meaning of this statement is given by the diagrammatic identity
\eqref{dcnatr5} adapted to the edge symmetric folded groupoid $B(G,H)$. \hfill $\Box$  

In contrast to ordinary gauge transformations, a crossed module $1$--gauge transformation 
yields and can be reconstructed from differential Lie crossed module 
valued differential form data. 

\begin{defi} \label{def:r2dghgau}
A differential $(G, H)$--$1$--gauge transformation is a pair $(\varkappa,\varGamma)\in
\Map(\mathbb{R}^2,G)\times \Omega^1(\mathbb{R}^2,\mathfrak{h})$. 
We denote the set of differential $(G, H)$--$1$--gauge transformation by $\Gau_1(G,H)$.
\end{defi}

The following theorem holds. 

\begin{prop} \label{theor:gauge1} For a fixed $(G,H)$--cocycle $(f,g,W)$, 
there is a canonical one--to--one correspondence between the set  
$\Gau_{1\,f,g,W}(G,H)$ of $(f,g,W)$--$1$--gauge transformations and the set 
$\Gau_1(G,H)$ differential $(G, H)$--$1$--gauge transformations.
The differential $(G, H)$--$1$--gauge transformation
$(\varkappa_{\kappa,\varPsi,\varPhi},\varGamma_{\kappa,\varPsi,\varPhi})$ corresponding to 
a $(f,g,W)$--$1$--gauge transformation $(\kappa,\varPsi,\varPhi)$ is given by 
\begin{subequations}
\label{gauge18,19}
\begin{align}
&\varkappa_{\kappa,\varPsi,\varPhi}(x,y)=\kappa(x;y),
\vphantom{\Big]}
\label{gauge18}
\\
&\varGamma_{\kappa,\varPsi,\varPhi \,x}(x,y)
=-\dot m(\kappa(x;y))(\varPsi(x',x;y)^{-1}\partial_{x'}\varPsi(x',x;y)\big|_{x'=x}),
\vphantom{\Big]}
\label{gauge19}
\\
&\varGamma_{\kappa,\varPsi,\varPhi\,y}(x,y)=
-\dot m(\kappa(x;y))(\varPhi(x;y',y)^{-1}\partial_{y'}\varPhi(x;y',y)\big|_{y'=y})
\nonumber
\vphantom{\Big]}
\end{align}
\end{subequations}
(cf. eq. \eqref{hiholo3}). Conversely, the $(f,g,W)$--$1$--gauge transformation $(\kappa_{\varkappa,\varGamma},
\varPsi_{\varkappa,\varGamma}$, $\varPhi_{\varkappa,\varGamma})$
corresponding to a differential $(G, H)$--$1$--gauge transformation 
$(\varkappa,\varGamma)$ is \vspace{2truemm} \pagebreak 
\begin{subequations}
\label{gauge20,21,21x}
\begin{align}
&\kappa_{\varkappa,\varGamma}(x;y)=\varkappa(x,y),
\vphantom{\Big]}
\label{gauge20}
\\
&\varPsi_{\varkappa,\varGamma}(x,x_0;y)=\varLambda_{|y,x_0}(x),
\vphantom{\Big]}
\label{gauge21}
\\
&\varPhi_{\varkappa,\varGamma}(x;y,y_0)=\varXi_{|x,y_0}(y),
\label{gauge21x}
\vphantom{\Big]}
\end{align}
\end{subequations}
where $\varLambda_{|y,x_0}$, $\varXi_{|x,y_0}$ are the unique solutions of the 
differential problem 
\begin{subequations}
\label{gauge22,23}
\begin{align}
&\varLambda_{|y,x_0}(x)^{-1}\partial_x\varLambda_{|y,x_0}(x)
=-\dot m(f(x,x_0;y)^{-1}\varkappa(x,y)^{-1})(\varGamma_x(x,y)),
\vphantom{\Big]}
\label{gauge22}
\\
&\varXi_{|x,y_0}(y)^{-1}\partial_{|y}\varXi_{|x,y_0}(y)
=-\dot m(g(x;y,y_0)^{-1}\varkappa(x,y)^{-1})(\varGamma_y(x,y))
\vphantom{\Big]}
\label{gauge23}
\end{align}
\end{subequations}
with the initial conditions 
\begin{subequations}
\label{gauge24,25}
\begin{align}
&\varLambda_{|y,x_0}(x_0)=1_H,
\vphantom{\Big]}
\label{gauge24}
\\
&\varXi_{|x,y_0}(y_0)=1_H.
\vphantom{\Big]}
\label{gauge25}
\end{align}
\end{subequations}
\end{prop}

\noindent{\it Proof}. 
If $(\kappa,\varPsi,\varPhi)$ is an $(f,g,W)$--$1$--gauge transformation, then \eqref{gauge18}, 
\eqref{gauge19} clearly define a $G$--valued map $\varkappa_{\kappa,\varPsi,\varPhi}$ and an 
$\mathfrak{h}$--valued $1$--form $\varGamma_{\kappa,\varPsi,\varPhi}$ on $\mathbb{R}^2$,
so a  differential $1$--gauge transformation.
This shows the first part of the theorem. 

Let $(\varkappa,\varGamma)$ be a differential $1$--gauge transformation. The solution 
$\varLambda_{|y,x_0}$ of the differential problem \eqref{gauge22}, \eqref{gauge24}
exists, is unique and is smooth in $y$ and $x_0$. Similarly, 
the solution $\varXi_{|x,y_0}$ of the differential problem \eqref{gauge23}, \eqref{gauge25}
exists, is unique and is smooth in $x$ and $y_0$. Relations \eqref{gauge20}, \eqref{gauge21}
define in this way a $G$--valued map $\kappa_{\varkappa,\varGamma}$ on $\mathbb{R}\times \mathbb{R}$
and two $H$--valued maps $\varPsi_{\varkappa,\varGamma}$ and $\varPhi_{\varkappa,\varGamma}$
on $\mathbb{R}^2\times \mathbb{R}$ and $\mathbb{R}\times \mathbb{R}^2$, respectively. 
We have now to show that the cocycle relations \eqref{gauge5,6} are identically obeyed. 
Consider the $H$--valued maps
\begin{subequations}
\label{gaugeb1,2,3,4}
\begin{align}
&\varLambda_1(x)=\varPsi_{\varkappa,\varGamma|y}(x_1,x_0)m(f_{|y}(x_1,x_0)^{-1})(\varPsi_{\varkappa,\varGamma|y}(x,x_1)),
\vphantom{\Big]}
\label{gaugeb1}
\\
&\varLambda_2(x)=\varPsi_{\varkappa,\varGamma|y}(x,x_0),
\vphantom{\Big]}
\label{gaugeb2}
\\
&\varXi_1(y)=\varPhi_{\varkappa,\varGamma|x}(y_1,y_0)m(g_{|x}(y_1,y_0)^{-1})(\varPhi_{\varkappa,\varGamma|x}(y,y_1)),
\vphantom{\Big]}
\label{gaugeb3}
\\
&\varXi_2(y)=\varPhi_{\varkappa,\varGamma|x}(y,y_0).
\vphantom{\Big]}
\label{gaugeb4}
\end{align}
\end{subequations} 
In virtue of  \eqref{gauge21}, \eqref{gauge22}, \eqref{gauge24}, 
$\varLambda_1$, $\varLambda_2$ are both solution of the differential equation 
\begin{equation}
\varLambda(x)^{-1}d_x\varLambda(x)=-\dot m(f_{|y}(x,x_0)^{-1}\varkappa(x,y)^{-1})(\varGamma_x(x,y))
\nonumber
\end{equation}
with initial condition $\varLambda(x_1)=\varPsi_{\varkappa,\varGamma|y}(x_1,x_0)$.
By the uniqueness of the solution of this differential problem, $\varLambda_1=\varLambda_2$. 
By \eqref{gaugeb1}, \eqref{gaugeb2}, then, $\varPsi_{\varkappa,\varGamma|y}$ fulfills the cocycle
condition \eqref{gauge5} as required. Similarly, by \eqref{gauge21x}, \eqref{gauge23}, \eqref{gauge25}, 
$\varXi_1$, $\varXi_2$ are both solution of the differential equation 
\begin{equation}
\varXi(y)^{-1}d_y\varXi(y)=-\dot m(g_{|x}(y,y_0)^{-1}\varkappa(x,y)^{-1})(\varGamma_y(x,y))
\nonumber
\end{equation}
with initial condition $\varXi(y_1)=\varPhi_{\varkappa,\varGamma|x}(y_1,y_0)$, 
so that $\varXi_1=\varXi_2$. By \eqref{gaugeb3}, \eqref{gaugeb4}, then, $\varPhi_{\varkappa,\varGamma|x}$ fulfills the cocycle
condition \eqref{gauge6}.

To conclude the proof of the theorem, we have to show that the mappings
$(\kappa,\varPsi,\varPhi)\to(\varkappa_{\kappa,\varPsi,\varPhi},\varGamma_{\kappa,\varPsi,\varPhi})$ 
and $(\varkappa,\varGamma)\to(\kappa_{\varkappa,\varGamma}, \varPsi_{\varkappa,\varGamma}, \varPhi_{\varkappa,\varGamma})$ 
are reciprocally inverse. For a given differential $1$--gauge transformation $(\varkappa,\varGamma)$,
inserting the \eqref{gauge20,21,21x} into the \eqref{gauge18,19} and using 
\eqref{gauge22,23}, \eqref{gauge24,25}, it is immediately verified that 
$\varkappa_{\kappa_{\varkappa,\varGamma}, \varPsi_{\varkappa,\varGamma}, \varPhi_{\varkappa,\varGamma}}=\varkappa$, 
$\varGamma_{\kappa_{\varkappa,\varGamma}, \varPsi_{\varkappa,\varGamma}, 
\varPhi_{\varkappa,\varGamma}}$ $=\varGamma$. For a given integral $1$--gauge transformation $(\kappa,\varPsi,\varPhi)$,
from the \eqref{gauge18,19}, 
using the cocycle relations \eqref{gauge5,6}, it is straightforwardly
checked that $\varLambda_{|y,x_0}(x)=\varPsi(x,x_0;y)$, 
$\varXi_{|x,y_0}(y)=\varPhi(x;y,y_0)$ solve the differential problem 
\eqref{gauge22,23}, \eqref{gauge24,25} with $\varkappa=\varkappa_{\kappa,\varPsi,\varPhi}$, 
$\varGamma=\varGamma_{\kappa,\varPsi,\varPhi}$, 
so that $\kappa_{\varkappa_{\kappa,\varPsi,\varPhi},\varGamma_{\kappa,\varPsi,\varPhi}}=\kappa$, 
$\varPsi_{\varkappa_{\kappa,\varPsi,\varPhi},\varGamma_{\kappa,\varPsi,\varPhi}}=\varPsi$, 
$\varPhi_{\varkappa_{\kappa,\varPsi,\varPhi},\varGamma_{\kappa,\varPsi,\varPhi}}=\varPhi$. The claim is so shown. 
\hfill $\Box$   

\begin{remark}
Since $\varkappa$, $\varGamma$ do not obey any conditions, the sets $\Gau_{1\,f,g,W}(G,H)$
with varying cocycle $(f,g,W)$ are all in canonical one--to--one
correspondence.
\end{remark}

By prop. \ref{theor:cycle2}, there exists one--to--one correspondence between $(G,H)$--cocycles $(f,g,W)$  
and connection doublets $(a,B)$. Hence,
the action of a $(f,g,W)$ --$1$--gauge transformation $(\kappa,\varPsi,\varPhi)$ must translate into one
on the associated doublet $(a_{f,g,W},B_{f,g,W})$. 

\begin{prop} \label{theor:gauge2} Let $(f,g,W)$ be a $(G,H)$--cocycle and 
$(\kappa,\varPsi,\varPhi)$ be an $(f,g$, $W)$--$1$--gauge transformation. 
The  $(G,H)$--connection doublet $(a_{{}^{\kappa,\varPsi,\varPhi}f,{}^{\kappa,\varPsi,\varPhi}g,{}^{\kappa,\varPsi,\varPhi}W}$, \linebreak 
$B_{{}^{\kappa,\varPsi,\varPhi}f,{}^{\kappa,\varPsi,\varPhi}g,{}^{\kappa,\varPsi,\varPhi}W})$
associated with the gauge transformed cocycle $({}^{\kappa,\varPsi,\varPhi}f,{}^{\kappa,\varPsi,\varPhi}g$, 
${}^{\kappa,\varPsi,\varPhi}W)$ is then given by the expressions
\begin{subequations}
\label{gauge26,27}
\begin{align}
&a_{{}^{\kappa,\varPsi,\varPhi}f,{}^{\kappa,\varPsi,\varPhi}g,{}^{\kappa,\varPsi,\varPhi}W}=\Ad \varkappa_{\kappa,\varPsi,\varPhi} (a_{f,g,W})
-d\varkappa_{\kappa,\varPsi,\varPhi}\varkappa_{\kappa,\varPsi,\varPhi}{}^{-1}-\dot t(\varGamma_{\kappa,\varPsi,\varPhi}),
\vphantom{\Big]}
\\
&B_{{}^{\kappa,\varPsi,\varPhi}f,{}^{\kappa,\varPsi,\varPhi}g,{}^{\kappa,\varPsi,\varPhi}W}
=\dot m(\varkappa_{\kappa,\varPsi,\varPhi})(B_{f,g,W})-d\varGamma_{\kappa,\varPsi,\varPhi}
-\frac{1}{2}[\varGamma_{\kappa,\varPsi,\varPhi},\varGamma_{\kappa,\varPsi,\varPhi}]
\vphantom{\Big]}
\label{gauge27}
\\
&\hspace{2.1cm}-\widehat{m}(\Ad \varkappa_{\kappa,\varPsi,\varPhi} (a_{f,g,W}) 
-d\varkappa_{\kappa,\varPsi,\varPhi} \varkappa_{\kappa,\varPsi,\varPhi}{}^{-1}
-\dot t(\varGamma_{\kappa,\varPsi,\varPhi}),\varGamma_{\kappa,\varPsi,\varPhi})
\nonumber
\vphantom{\Big]}
\end{align}
\end{subequations}
(cf. eqs. \eqref{hiholo1}--\eqref{hiholo3}). 
\end{prop}

\noindent{\it Proof}. These relations follow from substituting the 
\eqref{gauge11,12,13} into the \eqref{cycle31,32} through 
a relatively straightforward calculation. See also ref. \cite{Schrei:2011}.  
\hfill $\Box$

If we take the $(G,H)$--connection doublets  and the differential 
$(G,H)$--$1$--gauge transformations as basic cocycle and gauge transformation data 
relying on props. \ref{theor:cycle2}, \ref{theor:gauge1}, 
then the \eqref{gauge26,27} define an action of differential $1$--gauge transformations on connection doublets.

\begin{defi} \label{def:gauconn}
Let $(a,B)$ be a $(G,H)$--connection doublet. For a differential 
$(G,H)$--$1$--gauge transformations $(\varkappa,\varGamma)$ let 
\begin{subequations}
\label{gauge28,29}
\begin{align}
&{}^{\varkappa,\varGamma}a=\Ad \varkappa (a)-d\varkappa\varkappa^{-1}-\dot t(\varGamma),
\vphantom{\Big]}
\label{gauge28}
\\
&{}^{\varkappa,\varGamma}B=\dot m(\varkappa)(B)-d\varGamma
-\frac{1}{2}[\varGamma,\varGamma]-\widehat{m}(\Ad \varkappa (a) -d\varkappa\varkappa^{-1}-\dot t(\varGamma),\varGamma).
\vphantom{\Big]}
\label{gauge29}
\end{align}
\end{subequations}
\end{defi}
It can be checked that this gauge transformation is compatible with the zero fake
curvature condition \eqref{cycle30}. 

We have in this way achieved our second goal, the incorporation of
gauge transformation into Lie crossed module cocycle theory
in a manner that naturally relates to gauge invariance in higher gauge
theory.

\vfil\eject

\subsection{\normalsize \textcolor{blue}{Lie crossed module $2$--gauge transformations}}\label{sec:twogau}

\hspace{.5cm} We consider now $2$--gauge transformations, which have no nontrivial counterpart 
in ordinary gauge theory. 

\begin{defi} 
A $(G,H)$--$2$--gauge transformation is a mapping $A\in \Map(\mathbb{R}\times \mathbb{R},H)$.
We denote by $\Gau_2(G,H)$ the set of all $(G,H)$--$2$--gauge transformations.
\end{defi}

$2$--gauge transformations are gauge for gauge transformations: they
act on $1$--gauge transformations.

\begin{prop} \label{prop:twogau0}
Let $(f,g,W)$ be a $(G,H)$--cocycle, $(\kappa,\varPsi,\varPhi)$ be an $(f,g,W)$--$1$--gauge transformation
and $A$ be a $(G,H)$--$2$--gauge transformation. Then, the maps ${}^A\kappa\in\Map(\mathbb{R}\times\mathbb{R},G)$,
${}^A\varPsi\in\Map(\mathbb{R}^2\times \mathbb{R},H)$, ${}^A\varPhi\in\Map(\mathbb{R}\times \mathbb{R}^2,H)$
defined by the expressions
\begin{subequations}
\label{twogay1,2,3}
\begin{align}
&{}^A\kappa(x;y)=\kappa(x;y)t(A(x;y)),
\vphantom{\Big]}
\label{twogau1}
\\
&{}^A\varPsi_{|y}(x',x)
=A_{|y}(x)^{-1}\varPsi_{|y}(x',x)m(f_{|y}(x',x)^{-1})(A_{|y}(x')),
%m(f'{}_{|y}(x',x)^{-1})(A_{|y}(x'))=\varPsi'{}_{|y}(x',x)^{-1}A_{|y}(x)\varPsi_{|y}(x',x)
\vphantom{\Big]}
\label{twogau2}
\\
&{}^A\varPhi_{|x}(y',y)
=A_{|x}(y)^{-1}\varPhi_{|x}(y',y)m(g_{|x}(y',y)^{-1})(A_{|x}(y')),
%m(g'{}_{|x}(y',y)^{-1})(A_{|x}(y'))=\varPhi'{}_{|x}(y',y)^{-1}A_{|x}(y)\varPhi_{|x}(y',y)
\vphantom{\Big]}
\label{twogau3}
\end{align}
\end{subequations}
where we have set $A_{|y}(x)=A_{|x}(y)=A(x;y)$ for clarity, constitute 
an $(f,g,W)$--$1$--gauge transformation $({}^A\kappa,{}^A\varPsi,{}^A\varPhi)$, the 
$2$--gauge transform of $(\kappa,\varPsi,\varPhi)$ by $A$.
\end{prop}

\noindent{\it Proof}. 
Using the defining relations \eqref{twogay1,2,3}, one verifies that $({}^A\kappa,{}^A\varPsi,{}^A\varPhi)$ 
satisfies $1$--gauge cocycle conditions \eqref{gauge5,6} whenever $(\kappa,\varPsi,\varPhi)$ does. 
\hfill $\Box$

$2$--gauge equivalent $1$--gauge transformations yield the the same gauge transform of the underlying 
cocycle. 

\begin{prop} \label{prop:twogau1}
Let $(f,g,W)$ be a $(G,H)$--cocycle, $(\kappa,\varPsi,\varPhi)$ be an $(f,g,W)$--$1$--gauge transformation
and $A$ be a $(G,H)$--$2$--gauge transformation. Then the transformed cocycles 
$({}^{\kappa,\varPsi,\varPhi}f, {}^{\kappa,\varPsi,\varPhi}g, {}^{\kappa,\varPsi,\varPhi}W)$,
$({}^{{}^A\kappa,{}^A\varPsi,{}^A\varPhi}f, {}^{{}^A\kappa,{}^A\varPsi,{}^A\varPhi}g, {}^{{}^A\kappa,{}^A\varPsi,{}^A\varPhi}W)$
are equal.
\end{prop}

\noindent{\it Proof}. This is readily checked by computing 
$({}^{{}^A\kappa,{}^A\varPsi,{}^A\varPhi}f, {}^{{}^A\kappa,{}^A\varPsi,{}^A\varPhi}g, {}^{{}^A\kappa,{}^A\varPsi,{}^A\varPhi}W)$
inserting the expressions \eqref{twogay1,2,3} %of $({}^A\kappa,{}^A\varPsi,{}^A\varPhi)$ 
into the \eqref{gauge11,12,13} and using 
the target matching condition \eqref{cycle15}. \hfill $\Box$

As we proved in subsects. \ref{sec:cycle}, \ref{sec:gauge}, every Lie crossed module cocycle can be regarded as 
a smooth functor form the rectangle double groupoid to the delooping double groupoid of the Lie crossed module 
and any $1$--gauge transformation as a double natural transformation between a Lie crossed module cocycle and its gauge transform. 
In the same spirit, a $2$--gauge transformation can be viewed as a double modification between a $1$--gauge transformation 
and its $2$--gauge transform (cf. subapp. \ref{sec:dcmod}). We warn the reader that our definition of double modification hinges on that 
of double natural transformation (cf. subapp. \ref{sec:dcnatr}), 
which, as we have recalled above,  differs from the one customarily
provided in the literature.

\begin{prop}
If $(f,g,W)$ is $(G,H)$--cocycle, $(\kappa,\varPsi,\varPhi)$ is a $(f,g,W)$--$1$--gauge transformation
and $A$ is a $(G,H)$--$2$--gauge transformation. Then, $A$ is equivalent to a double modification 
$(\kappa,\varPsi,\varPhi)\Rrightarrow ({}^A\kappa,{}^A\varPsi,{}^A\varPhi)$ 
of the double natural transformations $(\kappa,\varPsi,\varPhi)$, $({}^A\kappa,{}^A\varPsi,{}^A\varPhi)$.
\end{prop}

\noindent{\it Proof}. 
The data of a $2$--gauge transformation $A$ are equivalent to 
a mapping of the set of object of $\mathbb{GR}^2$ into the set of arrow squares of $B(G,H)$,
\begin{equation}
\vbox{
\xymatrix{{\text{\footnotesize $(x,y)$}}\hspace{.3cm}\ar@{|->}[r]
&
}
\vspace{-.8cm}}
\xymatrix@C=6pc{
{\text{\footnotesize $*$}}  & {\text{\footnotesize $*$}}\ar[l]_{1_G}  
\ar@{}[dl]^(.25){}="a"^(.75){}="b" \ar@{=>} "a";"b"_{A(x;y)\hspace{.5cm}} %|-{X}
\\                 
{\text{\footnotesize $*$}} \ar[u]^{{}^A\kappa(x;y)} & {\text{\footnotesize $*$}}\ar[u]_{\kappa(x;y)}\ar[l]^{1_G}             
}
\label{twogau4} 
\end{equation}
(cf. eqs. \eqref{dcmod1}). 
The fulfillment of the target matching condition \eqref{dccrossed2} is guaranteed by relation
\eqref{twogau1}. \eqref{twogau4} are precisely the data required for a double modification
from the first to the second of the double natural transformations 
$(\kappa,\varPsi,\varPhi), ({}^A\kappa,{}^A\varPsi,{}^A\varPhi):
(f,g,W)\Rightarrow({}^{\kappa,\varPsi,\varPhi}f,{}^{\kappa,\varPsi,\varPhi}g,{}^{\kappa,\varPsi,\varPhi}W)
=({}^{{}^A\kappa,{}^A\varPsi,{}^A\varPhi}f$, ${}^{{}^A\kappa,{}^A\varPsi,{}^A\varPhi}g, {}^{{}^A\kappa,{}^A\varPsi,{}^A\varPhi}W)$.
The only thing left to check is the double modification conditions \eqref{dcmod3}, \eqref{dcmod5}.
Using the expressions of the operations of the double groupoid $B(G,H)$ of subapp.
\ref{sec:dccrossed}, it is easily checked that
these are equivalent to relations \eqref{gauge13} written in the form 
\begin{subequations}
\label{twogau5,6}
\begin{align}
&A(x;y){}^A\varPsi(x',x;y)
=\varPsi(x',x;y)m(f(x',x;y)^{-1})(A(x';y)),
\vphantom{\Big]}
\label{twogau5}
\\
&A(x;y){}^A\varPhi(x;y',y)
=\varPhi(x;y',y)m(g(x;y',y)^{-1})(A(x;y')).
\label{twogau6}
\vphantom{\Big]}
\end{align}
\end{subequations}
%\vskip-5mm\eject\noindent
Intuitively, the double modification condition can be interpreted as the requirement that, for any horizontal and vertical arrow
of $\mathbb{GR}^2$  
 %\vspace{-.1cm}
\begin{equation}
\vbox{
\xymatrix{
{\text{\footnotesize $(x',y)$}}  & {\text{\footnotesize $(x,y)$}}\ar[l]}
\vspace{-.75cm}
}
\hspace{1.5cm}
\xymatrix{
{\text{\footnotesize $(x,y')$}}  
\\                 
{\text{\footnotesize $(x,y)$}} \ar[u]}
\label{twogau7}
\end{equation}
the cylinder diagrams 
\begin{subequations}
\begin{equation}
\xymatrix@C=1.pc@R=1.1pc{%@=.6pc{
& {\text{\scriptsize $\kappa(x';y)$}}\ar@/_.7pc/[dl]& & &{\text{\scriptsize $\kappa(x;y)$}}\ar@/_.7pc/[dl] 
\ar@{.}[lll]|-{\varPsi(x',x;y)}&
\\
{\text{\footnotesize $*$}}& & \text{$\hphantom{x}$} \ar[ll]_{{}^{\kappa,\varPsi,\varPhi}f(x',x;y)}& {\text{\footnotesize  $*$}} \ar@{-}[l] &&
\\
&& {\text{\footnotesize $*$}} \ar@{-}@/_1.15pc/[uul] \ar@{}[ull]^{A(x';y)}
& &&\ar@{-}@/_1.15pc/[uul]\ar[lll]_>>>>>>>>>>>>>>>>>{f(x',x;y)}{\text{\footnotesize  $*$}} \ar@{}[ull]_{A(x;y)}
\\
&{\text{\scriptsize ${}^A\kappa(x';y)$}} \ar@/^1.2pc/[uul]\ar@{-}@/_.7pc/[ur]&& 
&{\text{\scriptsize ${}^A\kappa(x;y)$}}\ar@/^1.2pc/[uul]|->>>>>{~\text{$\vphantom{\ul{\ul{\ul{\ul{\ul{f}}}}}}$}}
\ar@{-}@/_.7pc/[ur] \ar@{.}[lll]|-{{}^A\varPsi(x',x;y)}&
}
\label{twogau8} %dcmod6}
\end{equation}
\vspace{0mm}
\begin{equation}
\xymatrix@C=1.pc@R=1.1pc{%@=.6pc{
& {\text{\scriptsize $\kappa(x;y')$}}\ar@/_.7pc/[dl]& & &{\text{\scriptsize $\kappa(x;y)$}}\ar@/_.7pc/[dl] 
\ar@{.}[lll]|-{\varPhi(x;y',y)}&
\\
{\text{\footnotesize $*$}}& & \text{$\hphantom{x}$} \ar[ll]_{{}^{\kappa,\varPsi,\varPhi}g(x;y',y)}& {\text{\footnotesize  $*$}} \ar@{-}[l] &&
\\
&& {\text{\footnotesize $*$}} \ar@{-}@/_1.15pc/[uul] \ar@{}[ull]^{A(x;y')}
& &&\ar@{-}@/_1.15pc/[uul]\ar[lll]_>>>>>>>>>>>>>>>>>{g(x;y',y)}{\text{\footnotesize  $*$}} \ar@{}[ull]_{A(x;y)}
\\
&{\text{\scriptsize ${}^A\kappa(x;y')$}} \ar@/^1.2pc/[uul]\ar@{-}@/_.7pc/[ur]&& 
&{\text{\scriptsize ${}^A\kappa(x;y)$}}\ar@/^1.2pc/[uul]|->>>>>{~\text{$\vphantom{\ul{\ul{\ul{\ul{\ul{f}}}}}}$}}
\ar@{-}@/_.7pc/[ur] \ar@{.}[lll]|-{{}^A\varPhi(x;y',y)}&
}
\label{twogau9} %dcmod7}
\end{equation}
\end{subequations}
\vskip6mm\noindent
both commute, where all double arrows have been dropped for clarity and the identity morphisms 
of the modification arrow squares have been collapsed
(cf. eqs. \eqref{dcmod6}, \eqref{dcmod7}). 
The precise meaning of this statement is given by the diagrammatic identities
\eqref{dcmod3}, \eqref{dcmod5} adapted to the edge symmetric folded groupoid $B(G,H)$. 
\hfill $\Box$

%At the level of form data equivalent

By prop. \ref{theor:gauge1}, there exists a one--to--one correspondence between integral $(f,g,W)$--$1$--gauge
transformations $(\kappa,\varPsi,\varPhi)$  and differential $(G,H)$--$1$--gauge transformations 
$(\varkappa,\varGamma)$. So, the action of a $(G,H)$--$2$--gauge transformation $A$ must translate into one
on the data $(\varkappa_{\kappa,\varPsi,\varPhi},\varGamma_{\kappa,\varPsi,\varPhi})$. 

\begin{prop} Let $(f,g,W)$ be a $(G,H)$--cocycle, $(\kappa,\varPsi,\varPhi)$ be an $(f,g,W)$--$1$--gauge transformation
and $A$ a $(G,H)$--$2$--gauge transformation. Then, 
\begin{subequations}
\label{twogau10,11}
\begin{align}
&\varkappa_{{}^A\kappa,{}^A\varPsi,{}^A\varPhi}=t(\tilde A)\varkappa_{\kappa,\varPsi,\varPhi},
\vphantom{\Big]}
\label{twogau10}
\\
&\varGamma_{{}^A\kappa,{}^A\varPsi,{}^A\varPhi}
=\tilde A\varGamma_{\kappa,\varPsi,\varPhi}\tilde A^{-1}-d\tilde A\tilde A^{-1}
-Q(a_{{}^{\kappa,\varPsi,\varPhi}f,{}^{\kappa,\varPsi,\varPhi}g,{}^{\kappa,\varPsi,\varPhi}W},\tilde A)
\vphantom{\Big]}
\label{twogau11}
\end{align}
\end{subequations}
(cf. eq. \eqref{hiholo4}), where we have set \hphantom{xxxxxxxxxxxxx}
\begin{equation}
\tilde A=m(\kappa)(A)
\label{twogau12}
\end{equation}
with $\tilde A$ viewed as an element of $\Map(\mathbb{R}^2,H)$.
\end{prop}

\noindent{\it Proof}. These relations follow from substituting the 
\eqref{twogay1,2,3} into the \eqref{gauge18,19} through 
a relatively straightforward calculation. See also ref. \cite{Schrei:2011}.  \hfill $\Box$

If we take  the 
$(G,H)$--connection doublets and the differential 
$(G,H)$--$1$--gauge transformations as basic cocycle and gauge transformation data 
relying on props. \ref{theor:cycle2}, \ref{theor:gauge1}, then
the \eqref{twogau10,11} define an action of $2$--gauge transformations on differential $1$--gauge transformations
for any assigned connection doublet.

\begin{defi} \label{def:gaugau}
Let $(a,B)$ be a $(G,H)$--connection doublet and $(\varkappa,\varGamma)$ be a differential 
$(G,H)$--$1$--gauge transformation. For any $(G,H)$--$1$--gauge transformation $\tilde A$, one sets 
\hphantom{xxxxxxxxxxxxxxxx}
\begin{subequations}
\label{twogau13,14}
\begin{align}
&{}^{\tilde A}\varkappa_{|a,B}=t(\tilde A)\varkappa,
\vphantom{\Big]}
\label{twogau13}
\\
&{}^{\tilde A}\varGamma_{|a,B}=\tilde A\varGamma \tilde A^{-1}-d\tilde A\tilde A^{-1}-Q({}^{\varkappa,\varGamma}a,\tilde A).
\vphantom{\Big]}
\label{twogau14}
\end{align}
\end{subequations}
\end{defi}

By prop. \ref{theor:cycle2} \pagebreak and def. \ref{def:gauconn}, the action of the integral $(G,H)$--$1$--gauge transformation 
on the $(G,H)$--cocycles translates into an action of the differential $(G,H)$--$1$--gauge transformations 
corresponding to the integral ones 
onto the $(G,H)$--connection doublets corresponding to the cocycles, as given 
by eqs. \eqref{gauge28,29}. 
$2$--gauge equivalent differential $1$--gauge transformations yield the same gauge transformed 
connection doublet.
 
\begin{prop} \label{prop:gaugau}
Let $(a,B)$ be a $(G,H)$--connection doublet, $(\varkappa,\varGamma)$ be a differential $(G,H)$--$1$--gauge
transformation and $A$ be  $(G,H)$--$2$--gauge transformation. Then, one has \hphantom{xxxxxxxxxxxxxxxxxxxxxxxx}
\begin{subequations}
\label{twogau15,16}
\begin{align}
&{}^{{}^{\tilde A}\varkappa_{|a,B},{}^{\tilde A}\varGamma_{|a,B}}a={}^{\varkappa,\varGamma}a,
\vphantom{\Big]}
\label{twogau15}
\\
&{}^{{}^{\tilde A}\varkappa_{|a,B},{}^{\tilde A}\varGamma_{|a,B}}B={}^{\varkappa,\varGamma}B.
\vphantom{\Big]}
\label{twogau16}
\end{align}
\end{subequations}
\end{prop}

\noindent{\it Proof}. Let $(f,g,W)$ be a cocycle, $(\kappa,\varPsi,\varPhi)$ be a $(f,g,W)$--$1$--gauge transformation
and $A$ be a $2$--gauge transformation. By prop. \ref{prop:twogau0}, 
$({}^A\kappa,{}^A\varPsi,{}^A\varPhi)$ is also a $(f,g,W)$--$1$--gauge transformation. 
By \eqref{gauge26,27}, \eqref{gauge28,29} combined, we have 
$(a_{{}^{\kappa,\varPsi,\varPhi}f,{}^{\kappa,\varPsi,\varPhi}g,{}^{\kappa,\varPsi,\varPhi}W},
B_{{}^{\kappa,\varPsi,\varPhi}f,{}^{\kappa,\varPsi,\varPhi}g,{}^{\kappa,\varPsi,\varPhi}W})$ 
$=({}^{\varkappa_{\kappa,\varPsi,\varPhi},\varGamma_{\kappa,\varPsi,\varPhi}}a_{f,g,W},
{}^{\varkappa_{\kappa,\varPsi,\varPhi},\varGamma_{\kappa,\varPsi,\varPhi}}B_{f,g,W})$
and similarly with $(\kappa,\varPsi,\varPhi)$  replaced by $({}^A\kappa,{}^A\varPsi,{}^A\varPhi)$. 
By \eqref{twogau13,14}, \eqref{twogau15,16}, we have further 
$(\varkappa_{{}^A\kappa,{}^A\varPsi,{}^A\varPhi},\varGamma_{{}^A\kappa,{}^A\varPsi,{}^A\varPhi})=
({}^{\tilde A}\varkappa_{\kappa,\varPsi,\varPhi|a_{f,g,W}B_{f,g,W}},{}^{\tilde A}\varGamma_{\kappa,\varPsi,\varPhi|a_{f,g,W}B_{f,g,W}})$. By prop.
\ref{prop:twogau1}, we have then that
\begin{align}
&({}^{\varkappa_{\kappa,\varPsi,\varPhi},\varGamma_{\kappa,\varPsi,\varPhi}}a_{f,g,W},
{}^{\varkappa_{\kappa,\varPsi,\varPhi},\varGamma_{\kappa,\varPsi,\varPhi}}B_{f,g,W})
\vphantom{\Big]}
\label{twogaue1}
\\
&\hspace{1cm}
=({}^{{}^{\tilde A}\varkappa_{\kappa,\varPsi,\varPhi|a_{f,g,W}B_{f,g,W}},{}^{\tilde A}\varGamma_{\kappa,\varPsi,\varPhi|a_{f,g,W}B_{f,g,W}}}a_{f,g,W},
\vphantom{\Big]}
\nonumber
\\
&\hspace{4cm}
{}^{{}^{\tilde A}\varkappa_{\kappa,\varPsi,\varPhi|a_{f,g,W}B_{f,g,W}},{}^{\tilde A}\varGamma_{\kappa,\varPsi,\varPhi|a_{f,g,W}B_{f,g,W}}}B_{f,g,W}).
\vphantom{\Big]}
\nonumber
\end{align}
By props. \ref{theor:cycle2}, \ref{theor:gauge1},
$(f,g,W)$ and $(\kappa,\varPsi,\varPhi)$ being arbitrary, \eqref{twogau15}, \eqref{twogau16} hold true. 
\hfill $\Box$

\vfil\eject

\section{\normalsize \textcolor{blue}{Higher parallel transport theory}}\label{sec:hiholo}

\hspace{.5cm} In this section, we rederive the higher parallel transport theory
worked out in refs. \cite{Schrei:2009,Schrei:2011,Schrei:2008} and 
\cite{Martins:2007,Martins:2008,Martins:2009} relying on the theory of
Lie crossed module cocycles and their gauge transformation 
developed in sect. \ref{sec:highcocy}. We review first 
the theory of the path and fundamental $2$--groupoids of a manifold 
to recall the reader the basic properties of these which are most
relevant in the following. Next, we show how the $1$-- and
$2$--parallel transport induced by a connection doublet 
can be defined in terms of an associated cocycle. 
Then, we exhibit how $1$--gauge transformation of the 
connection doublet affects the associated parallel transport
by inducing an integral $1$--gauge transformation of the underlying cocycle. The role of 
$2$--gauge transformation is also highlighted. 
The $2$--categorical interpretation of parallel transport and $1$-- and
$2$--gauge transformation thereof is recovered. 
We also touch the issue
of smoothness of the parallel transport. Finally we make explicit 
the equivalence  of our approach to the earlier ones 
recalled above. Again, to help intuition, we present our construction stressing its being an extension of the
ordinary parallel transport theory.

%\vfil\eject

\subsection{\normalsize \textcolor{blue}{Path and fundamental $2$--groupoid}}\label{sec:path}

\hspace{.5cm}  In this subsection, we review the basic notions 
of smooth thin homotopy and homotopy aiming to the definition of 
the path $2$--groupoid of a manifold, one of the essential elements
of higher parallel transport theory.
As this material is not original, we provide no proof of the basic results.

We begin by considering the ordinary path and fundamental groupoids of a manifold $M$. 
Roughly, these are  
groupoids having points and curves 
joining pairs of points as its $0$-- and $1$--cells. We make this more precise next.

\begin{defi}
Let $p_0,p_1$ be points. A curve $\gamma:p_0\rightarrow p_1$ with sitting instants is a
mapping $\gamma\in\Map(\mathbb{R},M)$ such that \pagebreak 
\begin{subequations}
\begin{align}
&\gamma(x)=p_0\qquad \text{for $x<\epsilon$},
\vphantom{\Big]}
\label{path1}
\\
&\gamma(x)=p_1\qquad \text{for $x>1-\epsilon$}
\vphantom{\Big]}
\label{path2}
\end{align}
\end{subequations}
for some $\epsilon>0$ with $\epsilon<1/2$ depending on $\gamma$. 
All curves will have sitting instants unless otherwise stated.
We denote the set of all curves of $M$ by $\Pi_1M$. 
\end{defi}

\begin{defi}
Let $p$ be a point. The unit curve $\iota_p:p\rightarrow p$ of $p$ is defined by
\begin{equation}
\iota_p(x)=p.
\label{path3}
\end{equation}
Let $p_0,p_1$ be points and $\gamma:p_0\rightarrow p_1$ be a curve. The inverse
curve of $\gamma$ is the curve $\gamma^{-1_\circ}:p_1\rightarrow p_0$ defined by \hphantom{xxxxxxxxx}
\begin{equation}
\gamma^{-1_\circ}(x)=\gamma(1-x).
\label{path4}
\end{equation}
Let $p_0,p_1,p_2$ be points and $\gamma_1:p_0\rightarrow p_1$, $\gamma_2:p_1\rightarrow p_2$ 
be curves. The composition of $\gamma_1$, $\gamma_2$ is the curve 
$\gamma_2\circ\gamma_1:p_0\rightarrow p_2$ defined by 
\begin{subequations}
\label{path5,6}
\begin{align}
&\gamma_2\circ\gamma_1(x)=\gamma_1(2x) \qquad \text{for $x\leq 1/2$},
\vphantom{\Big]}
\label{path5}
\\
&\gamma_2\circ\gamma_1(x)=\gamma_2(2x-1) \qquad \text{for $x\geq 1/2$}.
\vphantom{\Big]}
\label{path6}
\end{align}
\end{subequations}
\end{defi}

\noindent 
The above are the type of operations which would be required for $(M,\Pi_1M)$ to be a groupoid, 
but $(M,\Pi_1M)$ is not, as is well--known, as invertibility and associativity do not hold. 
%as the composition of curves is not associative. In a smooth set--up, 
To construct a groupoid out of $(M,\Pi_1M)$, one has to quotient out 
by the relation of either thin homotopy or homotopy. 

\begin{defi}
Let $p_0,p_1$ be points and $\gamma_0,\gamma_1:p_0\rightarrow p_1$ be curves. 
A thin homotopy of $\gamma_0$, $\gamma_1$ is a mapping $h\in\Map(\mathbb{R}^2,M)$
such that
\begin{subequations}
\begin{align}
&h(x,y)=p_0\qquad \text{for $x<\epsilon$},
\vphantom{\Big]}
\label{path7}
\\
&h(x,y)=p_1\qquad \text{for $x>1-\epsilon$},
\vphantom{\Big]}
\label{path8}
\\
&h(x,y)=\gamma_0(x)\qquad \text{for $y<\epsilon$},
\vphantom{\Big]}
\label{path9}
\\
&h(x,y)=\gamma_1(x)\qquad \text{for $y>1-\epsilon$}
\vphantom{\Big]}
\label{path10}
\end{align}
\end{subequations}
for some $\epsilon>0$ with $\epsilon<1/2$ and that %. The homotopy $h$ is said thin if in addition
\begin{equation}
\rank(dh(x,y))\leq 1.
\label{path11} 
\end{equation}
$\gamma_0$, $\gamma_1$ are thin homotopy equivalent, a property 
denoted as $\gamma_1\sim_1\gamma_0$, % ($\gamma_1\sim_{1t}\gamma_0$),  
if there is thin homotopy $h$ of $\gamma_0$, $\gamma_1$. 
If condition \eqref{path11} 
is not imposed, then $h$ is a homotopy of $\gamma_0$, $\gamma_1$
and $\gamma_0$, $\gamma_1$ are homotopy equivalent, $\gamma_0\sim^0{}_1\gamma_1$. 
\end{defi}

\noindent
$\sim_1$, $\sim^0{}_1$ are both equivalence relations.
We denote by $P_1M$ and $P^0{}_1M$ the set of all thin homotopy and homotopy classes 
of curves of $M$.

\begin{prop}
$(M,P_1M)$ and $(M,P^0{}_1M)$ are both groupoids,
the path group\-oid and the fundamental groupoid of $M$. 
\end{prop}

\noindent 
By modding out thin homotopy equivalence,
the algebraic structure we have defined on $\Pi_1M$ induces one of the same form on $P_1M$
satisfying the axioms of invertibility and associativity, rendering $(M,P_1M)$
a true groupoid. Similarly, by modding out homotopy 
equivalence, %%\pagebreak 
$(M,P^0{}_1M)$ also turns out to be a groupoid.
Diagrammatically, the content of these groupoids can be represented as 
\begin{equation}
\xymatrix{{\text{\footnotesize $p_1$}}&{\text{\footnotesize $p_0$}}\ar[l]_\gamma}\!.
\label{}
\end{equation}
where $\gamma$ is understood as a (thin) homotopy class of curves.

Let $M$ be a manifold. 
The path and fundamental $2$--groupoids of $M$ are $2$--groupoids roughly having points, curves 
joining pairs of points and surfaces joining pairs of curves with common endpoints 
as its $0$--, $1$-- and $2$--cells. They are the simplest higher extensions of
path and fundamental groupoids. %We shall define them next.
% definition will be spelled out in detail next.
%is now in order. 

\begin{defi}
For points $p_0,p_1$, a curve $\gamma:p_0\rightarrow p_1$ is defined 
as before. The set of all curves is denoted again by $\Pi_1M$. 

Let $p_0,p_1$ be points and $\gamma_0,\gamma_1:p_0\rightarrow p_1$ be curves. 
A surface $\varSigma:\gamma_0\Rightarrow\gamma_1$ with sitting instants is a map $\varSigma\in\Map(\mathbb{R}^2,M)$
such that  \vskip1mm\pagebreak %%
\begin{subequations}
\begin{align}
&\varSigma(x,y)=p_0\qquad \text{for $x<\epsilon$},
\vphantom{\Big]}
\label{path12}
\\
&\varSigma(x,y)=p_1\qquad \text{for $x>1-\epsilon$},
\vphantom{\Big]}
\label{path13}
\\
&\varSigma(x,y)=\gamma_0(x)\qquad \text{for $y<\epsilon$},
\vphantom{\Big]}
\label{path14}
\\
&\varSigma(x,y)=\gamma_1(x)\qquad \text{for $y>1-\epsilon$}
\vphantom{\Big]}
\label{path15}
\end{align}
\end{subequations}
for some $\epsilon>0$ with $\epsilon<1/2$ depending on $\gamma_0$, $\gamma_1$, $\varSigma$. 
All surfaces will be assumed to have sitting instants unless otherwise stated.
The set of all surfaces is denoted by $\Pi_2M$. 
\end{defi}

\begin{defi} 
For a point $p$, the unit curve $\iota_p:p\rightarrow p$ of $p$ is defined as before.
For points $p_0,p_1$ and a curve $\gamma:p_0\rightarrow p_1$, the inverse
curve $\gamma^{-1_\circ}$ is also defined as before.
For points $p_0,p_1,p_2$ and curves $\gamma_1:p_0\rightarrow p_1$, $\gamma_2:p_1\rightarrow p_2$, 
the composed curve $\gamma_2\circ\gamma_1:p_0\rightarrow p_2$ is again defined as before. 

Let $p_0,p_1$ be points and $\gamma:p_0\rightarrow p_1$ be a curve. The unit surface 
$I_\gamma:\gamma\Rightarrow\gamma$ of $\gamma$ is the surface defined by
\begin{equation}
I_\gamma(x,y)=\gamma(x).
\label{path16}
\end{equation}
Let $p_0,p_1$ be points and $\gamma_0,\gamma_1:p_0\rightarrow p_1$ be curves and 
$\varSigma:\gamma_0\Rightarrow\gamma_1$ be a surface. The vertical inverse of $\varSigma$
is the surface $\varSigma^{-1_\bullet}:\gamma_1\Rightarrow \gamma_0$ %defined by 
\begin{equation}
\varSigma^{-1_\bullet}(x,y)=\varSigma(x,1-y).
\label{path17}
\end{equation}
Let $p_0,p_1$ be points and $\gamma_0,\gamma_1,\gamma_2:p_0\rightarrow p_1$ be curves and 
$\varSigma_1:\gamma_0\Rightarrow\gamma_1$, $\varSigma_2:\gamma_1\Rightarrow\gamma_2$ be 
surfaces. The vertical composition of $\varSigma_1$, $\varSigma_2$
is the surface $\varSigma_2\bullet\varSigma_1:\gamma_0\Rightarrow \gamma_2$ defined by 
\begin{subequations}
\label{path18,19}
\begin{align}
&\varSigma_2\bullet\varSigma_1(x,y)=\varSigma_1(x,2y) \qquad \text{for $y\leq 1/2$},
\vphantom{\Big]}
\label{path18}
\\
&\varSigma_2\bullet\varSigma_1(x,y)=\varSigma_2(x,2y-1) \qquad \text{for $y\geq 1/2$}.
\vphantom{\Big]}
\label{path19}
\end{align}
\end{subequations}
Let $p_0,p_1$ be points and $\gamma_0,\gamma_1:p_0\rightarrow p_1$ be curves and \pagebreak %%
$\varSigma:\gamma_0\Rightarrow\gamma_1$ be a surface. The horizontal inverse of $\varSigma$ 
is the surface $\varSigma^{-1_\circ}:\gamma_0{}^{-1_\circ}\Rightarrow \gamma_1{}^{-1_\circ}$ %defined by
\begin{equation}
\varSigma^{-1_\circ}(x,y)=\varSigma(1-x,y).
\label{path20}
\end{equation}
Let $p_0,p_1,p_2$ be points and $\gamma_0,\gamma_1:p_0\rightarrow p_1$, $\gamma_2,\gamma_3:p_1\rightarrow p_2$ 
be curves and $\varSigma_1:\gamma_0\Rightarrow\gamma_1$, $\varSigma_2:\gamma_2\Rightarrow\gamma_3$ be 
surfaces. The horizontal composition of $\varSigma_1$, $\varSigma_2$
is the surface $\varSigma_2\circ\varSigma_1:\gamma_2\circ\gamma_0\Rightarrow \gamma_3\circ\gamma_1$ defined by
\begin{subequations}
\label{path21,22}
\begin{align}
&\varSigma_2\circ\varSigma_1(x,y)=\varSigma_1(2x,y) \qquad \text{for $x\leq 1/2$},
\vphantom{\Big]}
\label{path21}
\\
&\varSigma_2\circ\varSigma_1(x,y)=\varSigma_2(2x-1,y) \qquad \text{for $x\geq 1/2$}.
\vphantom{\Big]}
\label{path22}
\end{align}
\end{subequations}
\end{defi}

\noindent
The above are the type of operations which would be required for $(M,\Pi_1M$, $\Pi_2M)$ to be a $2$--groupoid, 
but $(M,\Pi_1M,\Pi_2M)$ fails to be one as invertibility and associativity do not hold both for curves and surfaces. 
To construct a $2$--groupoid out of $(M,\Pi_1M,\Pi_2M)$, one has to quotient out by a suitable higher version 
of the relation of either thin homotopy or homotopy. 

\begin{defi}
For points $p_0,p_1$ and curves $\gamma_0,\gamma_1:p_0\rightarrow p_1$ the notions of thin
homotopy $h$ and thin homotopy equivalence of $\gamma_0$, $\gamma_1$ 
are defined exactly as before. 
We denote again by $\sim_1$ thin homotopy equivalence %%\pagebreak 
and by $P_1M$ the set of all 
thin homotopy classes of curves of $M$. 

Let $p_0,p_1$ be points, $\gamma_0,\gamma_1,\gamma_2,\gamma_3:p_0\rightarrow p_1$ be curves
and $\varSigma_0:\gamma_0\Rightarrow\gamma_1$, $\varSigma_1:\gamma_2\Rightarrow\gamma_3$
be surfaces. A thin homotopy of $\varSigma_0$, $\varSigma_1$ is a mapping $H\in\Map(\mathbb{R}^3,M)$
with the property that 
\begin{subequations}
\label{path23,24,25,26,27,28}
\begin{align}
&H(x,y,z)=p_0 \qquad \text{for $x< \epsilon$},
\vphantom{\Big]}
\label{path23}
\\
&H(x,y,z)=p_1 \qquad \text{for $x>1-\epsilon$},
\vphantom{\Big]}
\label{path24}
\\
&H(x,y,z)=H(x,0,z) \qquad \text{for $y<\epsilon$},
\vphantom{\Big]}
\label{path25}
\\
&H(x,y,z)=H(x,1,z) \qquad \text{for $y>1-\epsilon$},
\vphantom{\Big]}
\label{path26}
\\
&H(x,y,z)=\varSigma_0(x,y) \qquad \text{for $z<\epsilon$},
\vphantom{\Big]}
\label{path27}
%%\\
\end{align}
\begin{align}
&H(x,y,z)=\varSigma_1(x,y) \qquad \text{for $z>1-\epsilon$}
\vphantom{\Big]}
\label{path28}
\end{align}
\end{subequations}
for some $\epsilon>0$ and that 
\begin{subequations}
\label{path29,30}
\begin{align}
&\rank(dH(x,0,z)),~\rank(dH(x,1,z))\leq 1,
\vphantom{\Big]}
\label{path29}
\\
&\rank(dH(x,y,z))\leq 2.
\vphantom{\Big]}
\label{path30} %thin2 
\end{align}
\end{subequations}
$\varSigma_0$, $\varSigma_1$ are thin homotopy equivalent, which fact
we write as $\varSigma_1\sim_2\varSigma_0$, 
if there is thin homotopy $H$ of $\varSigma_0$, $\varSigma_1$. 
If condition \eqref{path30} 
is not imposed, then $H$ is a homotopy of $\varSigma_0$, $\varSigma_1$
and $\varSigma_0$, $\varSigma_1$ are homotopy equivalent, $\varSigma_0\sim^0{}_2\varSigma_1$. 
\end{defi}

\noindent
$\sim_2$, $\sim^0{}_2$ are both equivalence relations by 
conditions \eqref{path23}--\eqref{path28}. Condition \eqref{path29} implies that the source and target curves of 
of $\varSigma_0$, $\varSigma_1$ are thin homotopy equivalent, 
$\gamma_0\sim_1\gamma_2$, $\gamma_1\sim_1\gamma_3$.
We denote by $P_2M$ and $P^0{}_2M$ the set of all thin homotopy and homotopy classes 
of surfaces of $M$.

\begin{prop}
$(M,P_1M,P_2M)$ and $(M,P_1M,P^0{}_2M)$ are bot $2$--groupoids,
the path $2$--groupoid and the fundamental $2$--groupoid of $M$, respectively.
\end{prop}

\noindent
By modding out thin homotopy equivalence,
the algebraic structure we have defined on $\Pi_1M$, $\Pi_2M$ induces one of the same 
form on $P_1M$, $P_2M$ satisfying the axioms of invertibility and associativity,
rendering $(M,P_1M,P_2M)$ a true $2$--groupoid.
Similarly, modding out homotopy equivalence, $(M,P_1M,P^0{}_2M)$ also turns out to be a 
$2$--groupoid.
Diagrammatically, the content of these $2$--groupoids can be represented as 
\hphantom{xxxxxxxxxxxxxxx}
\begin{equation}
\xymatrix@C=3pc{
   {\text{\footnotesize $p_1$}}
& {\text{\footnotesize $p_0$}} \ar@/^1pc/[l]^{\gamma_1}="0"
           \ar@/_1pc/[l]_{\gamma_0}="1"
           \ar@{=>}"1"+<0ex,-2.ex>;"0"+<0ex,2.ex>_{\varSigma\,}
}
\label{}
\end{equation}
where $\gamma_0$, $\gamma_1$ is understood as thin homotopy class of curves
and $\varSigma$ as a (thin) homotopy class of surfaces.

Now we are ready to formulate our parallel transport theory. \vskip1mm

\vfil\eject

\subsection{\normalsize \textcolor{blue}{$2$--parallel transport}}\label{sec:twoholo}

\hspace{.5cm}  In this subsection, we shall define and study higher parallel transport.
Our approach is inspired by that of ref. \cite{Schrei:2011}, but relies systematically 
on the cocycle set--up developed in sect. \ref{sec:highcocy}. We assume throughout 
a trivial principal bundle background. 

We begin by reviewing parallel transport in ordinary gauge theory. Let $M$ be a manifold
and $G$ be a Lie group. The basic datum required to define parallel transport is a 
$G$--connection. 

\begin{defi} \label{def:gconn}
A $G$--connection on $M$, or simply a $G$--connection, is a form $\theta\in\Omega^1(M,\mathfrak{g})$. 
We denote the set of $G$--connections by $\Conn(M,G)$. 
\end{defi}

%We are now going to construct the parallel transport along a curve of $M$.

\noindent
If $\gamma$ is a curve and $\theta$ is a $G$--connection on $M$, 
$\gamma^*\theta$ is a $G$--connection in the sense of def. \ref{def:gconn}. By prop. \ref{theor:cycle1},
to $\gamma^*\theta$ there then corresponds a $G$--cocycle $f_{\gamma^*\theta}$.
%According to prop. \ref{theor:cycle1}, there is a canonical one--to--one 
%correspondence between the set $\Conn(\mathbb{R},G)$  
%of $G$--connections on $\mathbb{R}$ and the set $\Cyc(G)$ of $G$--cocycles.

\begin{defi}
Let $\theta$ be a $G$--connection. Let further 
$p_0,p_1$ be points and $\gamma:p_0\rightarrow p_1$ be a curve. 
The parallel transport along $\gamma$ induced by $\theta$ is 
\begin{equation}
F_\theta(\gamma)=f_{\gamma^*\theta}(1,0).
\vphantom{\Big]}
\label{twoholo1}
\end{equation}
%where $f_{\gamma^*a}$ is the $G$--cocycle
%associated with the form $\gamma^*a\in\Omega^1(\mathbb{R},\mathfrak{g})$
%(cf. subsect. \ref{sec:cycle}). 
\end{defi}

%\noindent
Let us fix a $G$--connection $\theta$. 
We have then a mapping $F_\theta:\Pi_1M\rightarrow G$.

\begin{prop} \label{prop:twoholo1}
For any point $p$, one has 
\begin{equation}
F_\theta(\iota_p)=1_G.
\label{twoholo2}
\end{equation}
For any two points $p_0,p_1$ and curve $\gamma:p_0\rightarrow p_1$, one has 
\begin{equation}
F_\theta(\gamma^{-1_\circ})=F_\theta(\gamma)^{-1}.
\label{twoholo3}
\end{equation}
For any three points $p_0,p_1,p_2$ and two curves $\gamma_1:p_0\rightarrow p_1$, $\gamma_2:p_1\rightarrow p_2$, 
\begin{equation}
F_\theta(\gamma_2\circ\gamma_1)=F_\theta(\gamma_2)F_\theta(\gamma_1).
\label{twoholo4}
\end{equation} 
\end{prop}

\noindent{\it Proof}. If $f$ is a $G$--cocycle and $\phi:\mathbb{R}\rightarrow\mathbb{R}$ is a map, then 
the mapping $\phi^*f:\mathbb{R}^2\rightarrow G$ defined by the expression 
\begin{equation}
\phi^*f(x',x)=f(\phi(x'),\phi(x))
\label{twoholoa1}
\end{equation}
satisfies \eqref{cycle2,3} and, so, 
is also a $G$--cocycle, the pull--back $\phi^*f$ of $f$ by $\phi$.
The one--to--one correspondence between $G$--connections
$a$ and $G$--cocycles $f\in\Cyc(G)$
established  by prop. \ref{theor:cycle1} 
is natural with respect to pull-back, as $f_{\phi^*a}=\phi^*f_a$ and 
$a_{\phi^*f}=\phi^*a_f$. 

For illustration, we show \eqref{twoholo4}. 
Define $\phi_1,\phi_2:\mathbb{R}\rightarrow\mathbb{R}$ 
by $\phi_1(x)=x/2$ and $\phi_2(x)=x/2+1/2$. It follows from \eqref{path5,6} that 
$(\gamma_2\circ\gamma_1)\circ\phi_1(x)=\gamma_1(x)$ for $x\leq 1$ and 
$(\gamma_2\circ\gamma_1)\circ\phi_2(x)=\gamma_2(x)$ for $x\geq 0$. Then, 
\begin{align}
F_\theta(&\gamma_2\circ\gamma_1)=f_{\gamma_2\circ\gamma_1{}^*\theta}(1,0)
\vphantom{\Big]}
\label{twoholoa2}
\\
&=f_{\gamma_2\circ\gamma_1{}^*\theta}(1,1/2)f_{\gamma_2\circ\gamma_1{}^*\theta}(1/2,0)
\vphantom{\Big]}
\nonumber
\\
&=f_{\gamma_2\circ\gamma_1{}^*\theta}(\phi_2(1),\phi_2(0))f_{\gamma_2\circ\gamma_1{}^*\theta}(\phi_1(1),\phi_1(0))
\vphantom{\Big]}
\nonumber
\\
&=\phi_2{}^*f_{\gamma_2\circ\gamma_1{}^*\theta}(1,0)\phi_1{}^*f_{\gamma_2\circ\gamma_1{}^*\theta}(1,0)
\vphantom{\Big]}
\nonumber
\\
&=f_{\phi_2{}^*\gamma_2\circ\gamma_1{}^*\theta}(1,0)f_{\phi_1{}^*\gamma_2\circ\gamma_1{}^*\theta}(1,0)
\vphantom{\Big]}
\nonumber
\\
&=f_{(\gamma_2\circ\gamma_1)\circ\phi_2{}^*\theta}(1,0)f_{(\gamma_2\circ\gamma_1)\circ\phi_1{}^*\theta}(1,0)
\vphantom{\Big]}
\nonumber
\\
&=f_{\gamma_2{}^*\theta}(1,0)f_{\gamma_1{}^*\theta}(1,0)=F_\theta(\gamma_2)F_\theta(\gamma_1).
\vphantom{\Big]}
\nonumber
\end{align}
\eqref{twoholo2}, \eqref{twoholo3} are proven by similar techniques. \hfill $\Box$ 

$F_\theta$ has the fundamental property of homotopy invariance as stated by the following proposition.

\begin{prop} \label{theor:twoholo2}
Let $p_0,p_1$ be points and $\gamma_y:p_0\rightarrow p_1$, $y\in\mathbb{R}$, be a smooth $1$--parameter family of curves
such that the mapping $h:\mathbb{R}^2\rightarrow M$ defined by $h(x,y)=\gamma_y(x)$
is a thin homotopy of $\gamma_0$, $\gamma_1$. Then,
%\eqref{twoholo6} 
\begin{equation}
F_\theta(\gamma_1)=F_\theta(\gamma_0). %\vphantom{\ul{\ul{\ul{\ul{x}}}}}
\label{twoholo6}
\end{equation}
\end{prop}

\noindent{\it Proof}. The proof is based on the variational formula 
\begin{align}
f_{\gamma_y{}^*\theta}(x,x_0)^{-1}&\partial_yf_{\gamma_y{}^*\theta}(x,x_0)
\vphantom{\Big]}
\label{twoholo5} %defo1
\\
&=-\int_{x_0}^x d\xi\,f_{\gamma_y{}^*\theta}(\xi,x_0)^{-1}h^*(d\theta+[\theta,\theta]/2)_{yx}(\xi,y)f_{\gamma_y{}^*\theta}(\xi,x_0)
\vphantom{\Big]}
\nonumber
\\
&\hphantom{=}\,-f_{\gamma_y{}^*\theta}(x,x_0)^{-1}h^*\theta_{y}(x,y)
f_{\gamma_y{}^*\theta}(x,x_0)+h^*\theta_{y}(x_0,y),
\vphantom{\Big]}
\nonumber
\end{align}
which is straightforward though lengthy to derive. Since $h$ is a thin homotopy, 
$h^*(d\theta+[\theta,\theta]/2)_{xy}(x,y)=0$, by \eqref{path11}, and $h^*\theta_{y}(1,y)=h^*\theta_{y}(0,y)=0$,
by \eqref{path7}, \eqref{path8}. Hence, by\eqref{twoholo1}, in virtue of \eqref{twoholo5},
\begin{equation}
F_\theta(\gamma_y)^{-1}\partial_yF_\theta(\gamma_y)=f_{\gamma_y{}^*\theta}(1,0)^{-1}\partial_yf_{\gamma_y{}^*\theta}(1,0)
=0,
\label{twoholo5/1}
\end{equation}
from which \eqref{twoholo6} follows. 
\hfill $\Box$ 

\noindent
%In this way, 
The map $F_\theta:\Pi_1M\rightarrow G$ factors so through one $\bar F_\theta:P_1M\rightarrow G$ 
from the path groupoid $1$--cell set $P_1M$ into $G$, giving  a categorical map
$\bar F_\theta:(M,P_1M)\rightarrow BG$
\begin{equation}
\xymatrix{{\text{\footnotesize $p_1$}}&{\text{\footnotesize $p_0$}}\ar[l]_\gamma}
\quad
\xymatrix{\ar@{|->}[r]&}
\quad
\xymatrix@C=3.5pc{{\text{\footnotesize $*$}}&{\text{\footnotesize $*$}}\ar[l]_{\bar F_\theta(\gamma)}}\!,
\label{diagholo1}
\end{equation}
from the path groupoid $(M,P_1M)$ into the 
delooping groupoid $BG$ of the group $G$ %introduced in 
(cf. subsects. \ref{sec:cycle} and \ref{sec:path}).

\begin{prop} \label{prop:twoholo3}
$\bar F_\theta$ is a groupoid functor.
\end{prop}

\noindent{\it Proof}. The statement follows from combining props. \ref{prop:twoholo1}, \ref{theor:twoholo2}. 
Functoriality results from relations \eqref{twoholo2}--\eqref{twoholo4}. \hfill $\Box$ 

\begin{defi} \label{def:flat}
The $G$--connection $\theta$ is said flat if 
\begin{equation}
d\theta+\frac{1}{2}[\theta,\theta]=0. 
\label{twoholo7}
\end{equation}
\end{defi}

\begin{prop} \label{theor:twoholo10} Let $\theta$ be flat.
Let $p_0,p_1$ be points and $\gamma_y:p_0\rightarrow p_1$, $y\in\mathbb{R}$, be a smooth $1$--parameter family of curves
such that \pagebreak the mapping $h:\mathbb{R}^2\rightarrow M$ defined by $h(x,y)=\gamma_y(x)$
is a homotopy of $\gamma_0$, $\gamma_1$. Then, 
%\eqref{twoholo6} 
\begin{equation}
F_\theta(\gamma_1)=F_\theta(\gamma_0).
\label{twoholo6/1}
\end{equation}
\end{prop}

\noindent{\it Proof}. The proof is based on relation %the variational formula 
\eqref{twoholo5} and follows the same
lines as that of prop. \ref{theor:twoholo2} except for the vanishing of the integral term
in the right hand side of \eqref{twoholo5} which is now due to the flatness of $\theta$ 
instead of the thinness of $H$. \hfill $\Box$ 

\noindent
Hence, the map $F_\theta:\Pi_1M\rightarrow G$ factors through one $\bar F^0{}_\theta:P^0{}_1M\rightarrow G$ from the 
fundamental groupoid $1$--cell set $P^0{}_1M$ into $G$ yielding a categorical map
$\bar F_\theta:(M,P^0{}_1M)\rightarrow BG$ of the fundamental groupoid $(M,P^0{}_1M)$
into the delooping groupoid $BG$.% (cf. eq. \eqref{diagholo1}). 

\begin{prop} \label{prop:twoholo5}
When the connection $\theta$ is flat,  
$\bar F^0{}_\theta:(M,P^0{}_1M)\rightarrow BG$ is a groupoid functor.
\end{prop}

\noindent{\it Proof}. The statement follows from combining prop. \ref{prop:twoholo1} and prop. \ref{theor:twoholo10}
with functoriali\-ty resulting again from relations \eqref{twoholo2}--\eqref{twoholo4}. \hfill $\Box$ 

We consider now the higher case. 
Let $M$ be a manifold and $(G,H)$ be a Lie crossed module. 
The basic datum required to define parallel transport is a 
$(G,H)$--connection doublet. 

\begin{defi} \label{def:ghconn}
A $(G,H)$--connection doublet on $M$, or simply a $(G,H)$--con\-nection doublet,
is a pair of forms $(\theta,\varUpsilon)\in\Omega^1(M,\mathfrak{g})\times \Omega^2(M,\mathfrak{h})$ 
satisfying the zero fake curvature condition
\begin{equation}
d\theta+\frac{1}{2}[\theta,\theta]-\dot t(\varUpsilon)=0. 
\label{twoholo8}
\end{equation}
We denote the set of $(G,H)$--connection doublets by $\Conn(M,G,H)$. 
\end{defi}

\noindent
If $\varSigma$ is a surface and $(\theta,\varUpsilon)$ is a $(G,H)$--connection doublet on $M$, then 
$(\varSigma^*\theta$, $\varSigma^*\varUpsilon)$ is a $(G,H)$--connection in the sense 
of def. \ref{def:ghconn}.  By prop. \ref{theor:cycle2}, %\pagebreak 
with $(\varSigma^*\theta,\varSigma^*\varUpsilon)$ there is then associated a $(G,H)$--cocycle 
$(f_{\varSigma^*\theta,\varSigma^*\varUpsilon|0},g_{\varSigma^*\theta,\varSigma^*\varUpsilon|0},
W_{\varSigma^*\theta,\varSigma^*\varUpsilon|0})$.

\begin{defi} 
Let $(\theta,\varUpsilon)$ be a $(G,H)$--connection. Let further 
$p_0,p_1$ be points, $\gamma_0,\gamma_1:p_0\rightarrow p_1$ be curves and
$\varSigma:\gamma_0\Rightarrow \gamma_1$ be a surface. 
The $1$--parallel transport along $\gamma_0,\gamma_1$ and $2$--parallel transport along
$\varSigma$ induced by  $(\theta,\varUpsilon)$ are 
\begin{subequations}
\label{twoholo9,10,11}
\begin{align}
&F_{\theta,\varUpsilon}(\gamma_0)=f_{\varSigma^*\theta,\varSigma^*\varUpsilon|0}(1,0),
\vphantom{\Big]}
\label{twoholo9}
\\
&F_{\theta,\varUpsilon}(\gamma_1)=f_{\varSigma^*\theta,\varSigma^*\varUpsilon|1}(1,0),
\vphantom{\Big]}
\label{twoholo10}
\\
&F_{\theta,\varUpsilon}(\varSigma)=W_{\varSigma^*\theta,\varSigma^*\varUpsilon}(0,1;1,0).
\vphantom{\Big]}
\label{twoholo11}
\end{align}
\end{subequations}
\end{defi}

From the target matching condition 
$(f_{\varSigma^*\theta,\varSigma^*\varUpsilon|0},g_{\varSigma^*\theta,\varSigma^*\varUpsilon|0},
W_{\varSigma^*\theta,\varSigma^*\varUpsilon|0})$ obeys (cf. eq. \eqref{cycle15}), 
one has the following result.

\begin{prop} Let $p_0,p_1$ be points, $\gamma_0,\gamma_1:p_0\rightarrow p_1$ be curves and
$\varSigma:\gamma_0\Rightarrow \gamma_1$ be a surface. Then, one has 
\begin{equation}
F_{\theta,\varUpsilon}(\gamma_1)
=t(F_{\theta,\varUpsilon}(\varSigma))F_{\theta,\varUpsilon}(\gamma_0).
\label{twoholo14}
\end{equation}
\end{prop}

\noindent{\it Proof}. To begin with, we observe that there is $\epsilon>0$ with $\epsilon<1/2$ such that 
\begin{equation}
g_{\varSigma^*\theta,\varSigma^*\varUpsilon|x}(y',y)=1_G
\label{twoholob1}
\end{equation}
for $x<\epsilon$ or $x>1-\epsilon$ and arbitrary $y,y'$. 
This follows from the fact that, by prop. \ref{theor:cycle2}, 
 $g_{\varSigma^*\theta,\varSigma^*\varUpsilon|x}(y',y)$ is the solution 
of the differential problem \eqref{cycle37}, \eqref{cycle40} with $a_y(x,y)$ replaced by 
$\varSigma^*\theta_y(x,y)$ and that $\varSigma^*\theta_y(x,y)=0$ identically for the values of $x$ indicated
on account of \eqref{path7}, \eqref{path8}. 

By \eqref{twoholo9}--\eqref{twoholo11},  using the properties \eqref{cycle15},
\eqref{cycle17} and taking \eqref{twoholob1} into account, we find
\begin{align}
t(F_{\theta,\varUpsilon}(\varSigma))&=t(W_{\varSigma^*\theta,\varSigma^*\varUpsilon}(0,1;1,0))
\vphantom{\Big]}
\label{twoholob2}
\\
&=g_{\varSigma^*\theta,\varSigma^*\varUpsilon|1}(1,0)^{-1}f_{\varSigma^*\theta,\varSigma^*\varUpsilon|1}(0,1)^{-1}
g_{\varSigma^*\theta,\varSigma^*\varUpsilon|0}(1,0)f_{\varSigma^*\theta,\varSigma^*\varUpsilon|0}(0,1)
\vphantom{\Big]}
\nonumber
\\
&=f_{\varSigma^*\theta,\varSigma^*\varUpsilon|1}(1,0)f_{\varSigma^*\theta,\varSigma^*\varUpsilon|0}(1,0)^{-1}
\vphantom{\Big]}
\nonumber
\\
&=F_{\theta,\varUpsilon}(\gamma_1)F_{\theta,\varUpsilon}(\gamma_0)^{-1},
\vphantom{\Big]}
\nonumber
\end{align}
which leads immediately to \eqref{twoholo14}. \hfill $\Box$

\noindent
Physical intuition suggests that it should be possible to express the 
$1$--parallel transport $F_{\theta,\varUpsilon}(\gamma)$ along a curve $\gamma$ 
independently from any other curve $\gamma'$ with the same endpoints and surface
$\varSigma$ connecting $\gamma$ to $\gamma'$. This is indeed the case, as we shall show next. 

\begin{lemma} \label{lemma:para}
Let $p_0,p_1$ be points and $\gamma:p_0\rightarrow p_1$ be a curve. Then,
$f_{I_\gamma{}^*\theta,I_\gamma{}^*\varUpsilon|y}$, where $I_\gamma:\gamma\Rightarrow\gamma$ is the unit surface of $\gamma$ 
(cf. eq. \eqref{path16}), % and $y\in\mathbb{R}$, 
is independent from the value of $y$. 
\end{lemma}

\noindent{\it Proof}. By prop. \ref{theor:cycle2},
$f_{I_\gamma{}^*\theta,I_\gamma{}^*\varUpsilon|y}(x,x_0)$ is the solution 
of the differential problem \eqref{cycle36}, \eqref{cycle39} with 
$a_x(x,y)=I_\gamma{}^*\theta_x(x,y)$. Since $I_\gamma{}^*\theta_x(x,y)=\gamma^*a_x(x)$ is independent from $y$, so is 
$f_{I_\gamma{}^*\theta,I_\gamma{}^*\varUpsilon|y}(x,x_0)$. \hfill $\Box$ 

\begin{defi} \label{def:alt1tr} If $p_0,p_1$ are points and $\gamma:p_0\rightarrow p_1$ is a curve, one sets 
\begin{equation}
F_{\theta,\varUpsilon}(\gamma)=f_{I_\gamma{}^*\theta,I_\gamma{}^*\varUpsilon|y}(1,0).
\label{twoholo13}
\end{equation}
\end{defi}

%\noindent
%\eqref{twoholo13} gives the same result as \eqref{twoholo9}, \eqref{twoholo10}. 

\begin{prop} \label{prop:paraequiv}
If $p_0,p_1$ are points, $\gamma_0,\gamma_1:p_0\rightarrow p_1$ are curves and
$\varSigma:\gamma_0\Rightarrow \gamma_1$ is a surface, then the value of $F_{\theta,\varUpsilon}(\gamma_i)$
computed using \eqref{twoholo9}, \eqref{twoholo10} equals that obtained using \eqref{twoholo13}. 
%\begin{equation}
%F_{\theta,\varUpsilon}(\gamma_i)=F_{\theta,\varUpsilon|y}(\gamma_i).
%\label{twoholo13/1}
%\end{equation}
\end{prop}

\noindent{\it Proof}. By prop. \ref{theor:cycle2},
$f_{\varSigma^*\theta,\varSigma^*\varUpsilon|y}(x,x_0)$ is the solution 
of the differential problem \eqref{cycle36}, \eqref{cycle39} with 
$a_x(x,y)=\varSigma^*\theta_x(x,y)$. 
Likewise, $f_{I_{\gamma_i}{}^*\theta,I_{\gamma_i}{}^*\varUpsilon|y}(x,x_0)$ solves
the differential problem \eqref{cycle36}, \eqref{cycle39} 
with $a_x(x,y)=I_{\gamma_i}{}^*\theta_x(x,y)$. Since $I_{\gamma_i}{}^*\theta_x(x,y)=\varSigma^*\theta_x(x,i)$ 
for $i=1,2$ and any $y$, we have 
$f_{I_{\gamma_i}{}^*\theta,I_{\gamma_i}{}^*\varUpsilon|y}(x,x_0)=f_{\varSigma^*\theta,\varSigma^*\varUpsilon|i}(x,x_0)$.
Hence, \eqref{twoholo9}, \eqref{twoholo10} and \eqref{twoholo13} furnish the same value of $F_{\theta,\varUpsilon|y}(\gamma_i)$. 
\hfill $\Box$ %\varSigma

%\noindent
Let us fix a $(G,H)$--connection doublet $(\theta,\varUpsilon)$. 
We have then two mappings $F_{\theta,\varUpsilon}:\Pi_1M\rightarrow G$
and $F_{\theta,\varUpsilon}:\Pi_2M\rightarrow H$.

\begin{prop} \label{prop:twoholo6} For any point $p$, one has 
\begin{equation}
F_{\theta,\varUpsilon}(\iota_p)=1_G.
\label{twoholo15}
\end{equation}
For any two points $p_0,p_1$ and curve $\gamma:p_0\rightarrow p_1$, one has 
\begin{equation}
F_{\theta,\varUpsilon}(\gamma^{-1_\circ})=F_{\theta,\varUpsilon}(\gamma)^{-1}.
\label{twoholo16}
\end{equation}
For any three $p_0,p_1,p_2$ and two curves $\gamma_1:p_0\rightarrow p_1$, $\gamma_2:p_1\rightarrow p_2$, 
\begin{equation}
F_{\theta,\varUpsilon}(\gamma_2\circ\gamma_1)=F_{\theta,\varUpsilon}(\gamma_2)F_{\theta,\varUpsilon}(\gamma_1).
\label{twoholo17}
\end{equation}
For any two points $p_0,p_1$ and curve $\gamma:p_0\rightarrow p_1$, 
\begin{equation}
F_{\theta,\varUpsilon}(I_\gamma)=1_H.
\label{twoholo18}
\end{equation}
If $p_0,p_1$ are points, $\gamma_0,\gamma_1:p_0\rightarrow p_1$ are curves and
$\varSigma:\gamma_0\Rightarrow\gamma_1$ is a surface, then
\begin{equation}
F_{\theta,\varUpsilon}(\varSigma^{-1\bullet})=F_{\theta,\varUpsilon}(\varSigma)^{-1}.
\label{twoholo19}
\end{equation}
If $p_0,p_1$ are points, $\gamma_0,\gamma_1,\gamma_2:p_0\rightarrow p_1$ are curves and 
$\varSigma_1:\gamma_0\Rightarrow\gamma_1$, $\varSigma_2:\gamma_1\Rightarrow\gamma_2$ are surfaces, then 
\hphantom{xxxxxxxxxxx}
\begin{equation}
F_{\theta,\varUpsilon}(\varSigma_2\bullet\varSigma_1)
=F_{\theta,\varUpsilon}(\varSigma_2)F_{\theta,\varUpsilon}(\varSigma_1).
\label{twoholo20}
\end{equation}
If $p_0,p_1$ are points, $\gamma_0,\gamma_1:p_0\rightarrow p_1$ are curves 
and $\varSigma:\gamma_0\Rightarrow\gamma_1$ is a surface, then
\begin{equation}
F_{\theta,\varUpsilon}(\varSigma^{-1\circ})
=m(F_{\theta,\varUpsilon}(\gamma_0)^{-1})(F_{\theta,\varUpsilon}(\varSigma)^{-1}).
\label{twoholo21}
\end{equation}
If $p_0,p_1,p_2$ are points, $\gamma_0,\gamma_1:p_0\rightarrow p_1$, $\gamma_2,\gamma_3:p_1\rightarrow p_2$ 
are curves and $\varSigma_1:\gamma_0\Rightarrow\gamma_1$, $\varSigma_2:\gamma_2\Rightarrow\gamma_3$
are surfaces, then 
\begin{equation}
F_{\theta,\varUpsilon}(\varSigma_2\circ\varSigma_1)
=F_{\theta,\varUpsilon}(\varSigma_2)m(F_{\theta,\varUpsilon}(\gamma_2))(F_{\theta,\varUpsilon}(\varSigma_1)).
\label{twoholo22}
\end{equation}
\end{prop}

\noindent{\it Proof}. \pagebreak 
For any map $\phi:\mathbb{R}\rightarrow\mathbb{R}$, we define two maps
$l_\phi:\mathbb{R}^2\rightarrow\mathbb{R}^2$,
$r_\phi:\mathbb{R}^2\rightarrow\mathbb{R}^2$ by setting 
$l_\phi(x,y)=(\phi(x),y)$, $r_\phi(x,y)=(x,\phi(y))$. 
If $(f,g,W)$ is a $(G,H)$--cocycle, the maps
$l_\phi{}^*f:\mathbb{R}^2\times \mathbb{R}\rightarrow G$, 
$l_\phi{}^*g:\mathbb{R}\times \mathbb{R}^2\rightarrow G$, 
$l_\phi{}^*W:\mathbb{R}^2\times \mathbb{R}^2\rightarrow H$ given by
\begin{subequations}
\label{twoholoc1,2,3}
\begin{align}
&l_\phi{}^*f(x',x;y)=f(\phi(x'),\phi(x);y),
\vphantom{\Big]}
\label{twoholoc1}
\\
&l_\phi{}^*g(x;y',y)=g(\phi(x);y',y),
\vphantom{\Big]}
\label{twoholoc2}
\\
&l_\phi{}^*W(x',x;y',y)=W(\phi(x'),\phi(x);y',y)
\vphantom{\Big]}
\label{twoholoc3}
\end{align}
\end{subequations}
and those $r_\phi{}^*f:\mathbb{R}^2\times \mathbb{R}\rightarrow G$, 
$r_\phi{}^*g:\mathbb{R}\times \mathbb{R}^2\rightarrow G$, 
$r_\phi{}^*W:\mathbb{R}^2\times \mathbb{R}^2\rightarrow H$ by
\begin{subequations}
\label{twoholoc4,5,6}
\begin{align}
&r_\phi{}^*f(x',x;y)=f(x',x;\phi(y)),  \hspace{1.55cm}
\vphantom{\Big]}
\label{twoholoc4}
\\
&r_\phi{}^*g(x;y',y)=g(x;\phi(y'),\phi(y)),
\vphantom{\Big]}
\label{twoholoc5}
\\
&r_\phi{}^*W(x',x;y',y)=W(x',x;\phi(y'),\phi(y))
\vphantom{\Big]}
\label{twoholoc6}
\end{align}
\end{subequations} %
satisfy \eqref{cycle15} and \eqref{cycle11,12,13,14}
and, consequently, constitute two $(G,H)$--cocycles, the left and right pull-back
$(l_\phi{}^*f,l_\phi{}^*g,l_\phi{}^*W)$, $(r_\phi{}^*f,r_\phi{}^*g,r_\phi{}^*W)$
of $(f,g,W)$ by $\phi$.

The one--to--one correspondence between $(G,H)$--connections 
$(a,B)$ and $(G$, $H)$--cocycles $(f,g,W)$ stated  by prop. \ref{theor:cycle2} 
is natural with respect to left/right pull-back, as one has 
$(f_{l_\phi{}^*a,l_\phi{}^*B},g_{l_\phi{}^*a,l_\phi{}^*B},W_{l_\phi{}^*a,l_\phi{}^*B})$ $=$
$(l_\phi{}^*f_{a,B},l_\phi{}^*g_{a,B},l_\phi{}^*W_{a,B})$ and 
$(a_{l_\phi{}^*f,l_\phi{}^*g,l_\phi{}^*W},B_{l_\phi{}^*f,l_\phi{}^*g,l_\phi{}^*W})
=(l_\phi{}^*a_{f,g,W},\!l_\phi{}^*B_{f,g,W})$ for left pull-back and \linebreak 
$(f_{r_\phi{}^*a,r_\phi{}^*B}$, $g_{r_\phi{}^*a,r_\phi{}^*B}$,  $W_{r_\phi{}^*a,r_\phi{}^*B})
=(r_\phi{}^*f_{a,B},r_\phi{}^*g_{a,B}$, %\linebreak 
$r_\phi{}^*W_{a,B})$ and 
$(a_{r_\phi{}^*f,r_\phi{}^*g,r_\phi{}^*W}$, $B_{r_\phi{}^*f,r_\phi{}^*g,r_\phi{}^*W})
=(r_\phi{}^*a_{f,g,W},r_\phi{}^*B_{f,g,W})$ for right pull-back.

As an illustration, we prove \eqref{twoholo20}.
Define $\phi_1,\phi_2:\mathbb{R}\rightarrow\mathbb{R}$ 
by $\phi_1(x)=x/2$ and $\phi_2(x)=x/2+1/2$. It follows from \eqref{path21,22} that 
$(I_{\gamma_2}\circ I_{\gamma_1})\circ l_{\phi_1}(x,y)=I_{\gamma_1}(x,y)$ for $x\leq 1$ and 
$(I_{\gamma_2}\circ I_{\gamma_1})\circ l_{\phi_2}(x,y)=I_{\gamma_2}(x,y)$ for $x\geq 0$. Then, 
by \eqref{cycle13} and \eqref{cycle17}, we have %one has %wwwwwwww
\begin{align}
F_{\theta,\varUpsilon}&(\varSigma_2\circ\varSigma_1)
=W_{\varSigma_2\circ\varSigma_1{}^*\theta,\varSigma_2\circ\varSigma_1{}^*\varUpsilon|1,0}(0,1)
\vphantom{\Big]}
\label{twoholoc7}
\\
&=W_{\varSigma_2\circ\varSigma_1{}^*\theta,\varSigma_2\circ\varSigma_1{}^*\varUpsilon|1,0}(1/2,1) \hspace{7.8cm}
\vphantom{\Big]}
\nonumber
%\\
\end{align}
\begin{align}
&\hspace{2.15cm}\times 
m(f_{\varSigma_2\circ\varSigma_1{}^*\theta,\varSigma_2\circ\varSigma_1{}^*\varUpsilon|0}(1/2,1)^{-1})
(W_{\varSigma_2\circ\varSigma_1{}^*\theta,\varSigma_2\circ\varSigma_1{}^*\varUpsilon|1,0}(0,1/2))
\vphantom{\Big]}
\nonumber
\\
&=W_{\varSigma_2\circ\varSigma_1{}^*\theta,\varSigma_2\circ\varSigma_1{}^*\varUpsilon|1,0}(\phi_2(0),\phi_2(1))
\vphantom{\Big]}
\nonumber
\\
&\hspace{.05cm}\times
m(f_{\varSigma_2\circ\varSigma_1{}^*\theta,\varSigma_2\circ\varSigma_1{}^*\varUpsilon|0}(\phi_2(0),\phi_2(1))^{-1})
(W_{\varSigma_2\circ\varSigma_1{}^*\theta,\varSigma_2\circ\varSigma_1{}^*\varUpsilon|1,0}(\phi_1(0),\phi_1(1)))
\vphantom{\Big]}
\nonumber
\\
&=l_{\phi_2}{}^*W_{\varSigma_2\circ\varSigma_1{}^*\theta,\varSigma_2\circ\varSigma_1{}^*\varUpsilon|1,0}(0,1)
\vphantom{\Big]}
\nonumber
\\
&\hspace{1.75cm}\times
m(l_{\phi_2}{}^*f_{\varSigma_2\circ\varSigma_1{}^*\theta,\varSigma_2\circ\varSigma_1{}^*\varUpsilon|0}(0,1)^{-1})
(l_{\phi_1}{}^*W_{\varSigma_2\circ\varSigma_1{}^*\theta,\varSigma_2\circ\varSigma_1{}^*\varUpsilon|1,0}(0,1))
\vphantom{\Big]}
\nonumber
\\
&=W_{l_{\phi_2}{}^*\varSigma_2\circ\varSigma_1{}^*\theta,l_{\phi_2}{}^*\varSigma_2\circ\varSigma_1{}^*\varUpsilon|1,0}(0,1)
\vphantom{\Big]}
\nonumber
\\
&\hspace{.75cm}\times
m(f_{l_{\phi_2}{}^*\varSigma_2\circ\varSigma_1{}^*\theta,l_{\phi_2}{}^*\varSigma_2\circ\varSigma_1{}^*\varUpsilon|0}(0,1)^{-1})
(W_{l_{\phi_1}{}^*\varSigma_2\circ\varSigma_1{}^*\theta,l_{\phi_1}{}^*\varSigma_2\circ\varSigma_1{}^*\varUpsilon|1,0}(0,1))
\vphantom{\Big]}
\nonumber
\\
&=W_{(\varSigma_2\circ\varSigma_1)\circ l_{\phi_2}{}^*\theta,(\varSigma_2\circ\varSigma_1)\circ l_{\phi_2}{}^*\varUpsilon|1,0}(0,1)
%\,W_{(\varSigma_2\circ\varSigma_1)\circ l_{\phi_1}{}^*\theta,(\varSigma_2\circ\varSigma_1)\circ l_{\phi_1}{}^*\varUpsilon|1,0}(0,1)
\vphantom{\Big]}
\nonumber
\\
&%=W_{(\varSigma_2\circ\varSigma_1)\circ l_{\phi_2}{}^*\theta,(\varSigma_2\circ\varSigma_1)\circ l_{\phi_2}{}^*\varUpsilon|1,0}(0,1)
\hspace{-.15cm}\times
m(f_{(\varSigma_2\circ\varSigma_1)\circ l_{\phi_2}{}^*\theta,(\varSigma_2\circ\varSigma_1)\circ l_{\phi_2}{}^*\varUpsilon|0}(0,1)^{-1})
(W_{(\varSigma_2\circ\varSigma_1)\circ l_{\phi_1}{}^*\theta,(\varSigma_2\circ\varSigma_1)\circ l_{\phi_1}{}^*\varUpsilon|1,0}(0,1))
\vphantom{\Big]}
\nonumber
\\
&=W_{\varSigma_2{}^*\theta,\varSigma_2{}^*\varUpsilon|1,0}(0,1)
m(f_{\varSigma_2{}^*\theta,\varSigma_2{}^*\varUpsilon|0}(0,1)^{-1})
(W_{\varSigma_1{}^*\theta,\varSigma_1{}^*\varUpsilon|1,0}(0,1))
\vphantom{\Big]}
\nonumber
\\
&\hspace{6cm}
=F_{\theta,\varUpsilon}(\varSigma_2)m(F_{\theta,\varUpsilon}(\gamma_2))(F_{\theta,\varUpsilon}(\varSigma_1))
\vphantom{\Big]}
\nonumber
\end{align}
\eqref{twoholo20} is proven by a similar procedure involving this time
right pull-back. The other relations are shown by using similar techniques. \hfill $\Box$

Analogously to the ordinary case, $F_{\theta,\varUpsilon}$ is thin homotopy invariant as established by the following theorem.
%has the fundamental properties of thin homotopy invariance. %, though the proof is much more involved. 

\begin{prop} \label{theor:twoholo3}
Let $p_0,p_1$ be points and $\gamma_{0z},\gamma_{1z}:p_0\rightarrow p_1$ and 
$\varSigma_z:\gamma_{0z}\Rightarrow\gamma_{1z}$, $z\in\mathbb{R}$ be $1$--parameter families of curves 
and surfaces such that the mapping $H:\mathbb{R}^3\rightarrow M$ defined by $H(x,y,z)=\varSigma_z(x,y)$
is a thin homotopy of $\varSigma_0$, $\varSigma_1$. Then, one has the identities 
\begin{subequations}
\label{twoholo28,29,30}
\begin{align}
&F_{\theta,\varUpsilon}(\gamma_{01})=F_{\theta,\varUpsilon}(\gamma_{00}),
\vphantom{\Big]}
\label{twoholo28}
\\
&F_{\theta,\varUpsilon}(\gamma_{11})=F_{\theta,\varUpsilon}(\gamma_{10}),
\vphantom{\Big]}
\label{twoholo29}
\\
&F_{\theta,\varUpsilon}(\varSigma_1)=F_{\theta,\varUpsilon}(\varSigma_0).
\vphantom{\Big]}
\label{twoholo30}
\end{align}
\end{subequations}
\end{prop}

\noindent{\it Proof}. The proof is based on the variational formulae \pagebreak 
\begin{align}
&f_{\varSigma_z{}^*\theta,\varSigma_z{}^*\varUpsilon|y}(x,x_0)^{-1}
\partial_zf_{\varSigma_z{}^*\theta,\varSigma_z{}^*\varUpsilon|y}(x,x_0)%\hspace{5cm}
\vphantom{\Big]}
\label{twoholo23} %defo2
\\
&=-\int_{x_0}^x d\xi\,f_{\varSigma_z{}^*\theta,\varSigma_z{}^*\varUpsilon|y}(\xi,x_0)^{-1}
\dot t(H^*\varUpsilon_{zx}(\xi,y,z))
f_{\varSigma_z{}^*\theta,\varSigma_z{}^*\varUpsilon|y}(\xi,x_0)
\vphantom{\Big]}
\nonumber
\\
&\hphantom{=}\,-f_{\varSigma_z{}^*\theta,\varSigma_z{}^*\varUpsilon|y}(x,x_0)^{-1}H^*\theta_{z}(x,y,z)
f_{\varSigma_z{}^*\theta,\varSigma_z{}^*\varUpsilon|y}(x,x_0)+H^*\theta_{z}(x_0,y,z),
\vphantom{\Big]}
\nonumber
\\
&g_{\varSigma_z{}^*\theta,\varSigma_z{}^*\varUpsilon|x}(y,y_0)^{-1}
\partial_zg_{\varSigma_z{}^*\theta,\varSigma_z{}^*\varUpsilon|x}(y,y_0)
\vphantom{\Big]}
\label{twoholo24} %defo2
\\
&=-\int_{y_0}^y d\eta\,g_{\varSigma_z{}^*\theta,\varSigma_z{}^*\varUpsilon|x}(\eta,y_0)^{-1}
\dot t(H^*\varUpsilon_{zy}(x,\eta,z))
g_{\varSigma_z{}^*\theta,\varSigma_z{}^*\varUpsilon|x}(\eta,y_0)
\vphantom{\Big]}
\nonumber
\\
&\hphantom{=}\,-g_{\varSigma_z{}^*\theta,\varSigma_z{}^*\varUpsilon|x}(y,y_0)^{-1}H^*\theta_{z}(x,y,z)
g_{\varSigma_z{}^*\theta,\varSigma_z{}^*\varUpsilon|x}(y,y_0)+H^*\theta_{z}(x,y_0,z),
\vphantom{\Big]}
\nonumber
\\
&W_{\varSigma_z{}^*\theta,\varSigma_z{}^*\varUpsilon}(x,x_0;y,y_0)^{-1}
\partial_zW_{\varSigma_z{}^*\theta,\varSigma_z{}^*\varUpsilon}(x,x_0;y,y_0)
\vphantom{\Big]}
\label{twoholo25} %defo2
\\
&=-\int_{x_0}^xd\xi \int_{y_0}^yd\eta\, W_{\varSigma_z{}^*\theta,\varSigma_z{}^*\varUpsilon}(x,x_0;\eta,y_0)^{-1}
\dot m(g_{\varSigma_z{}^*\theta,\varSigma_z{}^*\varUpsilon|x_0}(\eta,y_0)^{-1}
\vphantom{\Big]}
\nonumber
\\
&\hspace{.5cm}\times f_{\varSigma_z{}^*\theta,\varSigma_z{}^*\varUpsilon|\eta}(\xi,x_0)^{-1})
(H^*(d\varUpsilon+[\theta,\varUpsilon])_{xyz}(\xi,\eta,z))W_{\varSigma_z{}^*\theta,\varSigma_z{}^*\varUpsilon}(x,x_0;\eta,y_0)
\vphantom{\Big]}
\nonumber
\\
&\hphantom{=}\,-\int_{x_0}^xd\xi\Big[W_{\varSigma_z{}^*\theta,\varSigma_z{}^*\varUpsilon}(x,x_0;y,y_0)^{-1}
\dot m(g_{\varSigma_z{}^*\theta,\varSigma_z{}^*\varUpsilon|x_0}(y,y_0)^{-1}
\vphantom{\Big]}
\nonumber
\\
&\hspace{1cm}\times f_{\varSigma_z{}^*\theta,\varSigma_z{}^*\varUpsilon|y}(\xi,x_0)^{-1})
(H^*\varUpsilon_{xz}(\xi,y,z))W_{\varSigma_z{}^*\theta,\varSigma_z{}^*\varUpsilon}(x,x_0;y,y_0)
\vphantom{\Big]}
\nonumber
\\
&\hspace{1cm}-\dot m(f_{\varSigma_z{}^*\theta,\varSigma_z{}^*\varUpsilon|y_0}(\xi,x_0)^{-1})
(H^*\varUpsilon_{xz}(\xi,y_0,z))\Big]
\vphantom{\Big]}
\nonumber
\\
&\hphantom{=}\,+\int_{y_0}^yd\eta\Big[W_{\varSigma_z{}^*\theta,\varSigma_z{}^*\varUpsilon}(x,x_0;y,y_0)^{-1}
\vphantom{\Big]}
\nonumber
\\
&\hspace{1cm}
\times\dot m(g_{\varSigma_z{}^*\theta,\varSigma_z{}^*\varUpsilon|x_0}(\eta,y_0)^{-1})(H^*\varUpsilon_{zy}(x_0,\eta,z))
W_{\varSigma_z{}^*\theta,\varSigma_z{}^*\varUpsilon}(x,x_0;y,y_0)
\vphantom{\Big]}
\nonumber
\\
&\hspace{1cm}-\dot m(f_{\varSigma_z{}^*\theta,\varSigma_z{}^*\varUpsilon|y_0}(x,x_0)^{-1}
g_{\varSigma_z{}^*\theta,\varSigma_z{}^*\varUpsilon|x}(\eta,y_0)^{-1})
(H^*\varUpsilon_{zy}(x,\eta,z))\Big]
\vphantom{\Big]}
\nonumber
\\
&\hspace{1cm}+Q(H^*\theta_z(x_0,y_0,z),W_{\varSigma_z{}^*\theta,\varSigma_z{}^*\varUpsilon}(x,x_0;y,y_0)^{-1}),
\vphantom{\Big]}
\nonumber
\end{align}
which are straightforward albeit very lengthy to obtain. Since $H$ is a thin homotopy, 
$H^*\varUpsilon_{zx}(x,i,z)=0$ for $i=0,1$, by \eqref{path29},
$H^*(d\varUpsilon+[\theta,\varUpsilon])_{xyz}(x,y,z)=0$, by \eqref{path30}, 
and $H^*\theta_z(i,j,z)=0$ and $H^*\varUpsilon_{yx}(i,y,z)=0$ for $i,j=0,1$,
by \eqref{path23}, \eqref{path24}. Therefore, by \eqref{twoholo9}--\eqref{twoholo11},
in virtue of \eqref{twoholo23},  \eqref{twoholo25}, we have \hphantom{xxxxxxxxxxxxxxxxxxxxx}
\begin{subequations}
\begin{align}
&F_{\theta,\varUpsilon}(\gamma_{0z})^{-1}\partial_zF_{\theta,\varUpsilon}(\gamma_{0z})=0,
%\vphantom{\ul{\ul{\ul{\ul{\ul{\ul{g}}}}}}}
\vphantom{\Big]}
\label{}
%\\
\end{align}
\begin{align}
&F_{\theta,\varUpsilon}(\gamma_{1z})^{-1}\partial_zF_{\theta,\varUpsilon}(\gamma_{1z})=0,
\vphantom{\Big]}
\label{}
\\
&F_{\theta,\varUpsilon}(\varSigma_z)^{-1}\partial_zF_{\theta,\varUpsilon}(\varSigma_z)=0,
\vphantom{\Big]}
\label{}
\end{align}
\end{subequations}
from which \eqref{twoholo28}--\eqref{twoholo30} follow. \hfill $\Box$

\noindent

The thin homotopy invariance of $1$--parallel transport holds also 
if the latter is defined autonomously according to def. \ref{def:alt1tr}. 

\begin{prop} \label{theor:twoholo3/1}
Let $p_0,p_1$ be points and $\gamma_y:p_0\rightarrow p_1$, $y\in\mathbb{R}$, be a smooth $1$--parameter family of curves
such that the mapping $h:\mathbb{R}^2\rightarrow M$ defined by $h(x,y)=\gamma_y(x)$
is a thin homotopy of $\gamma_0$, $\gamma_1$. Then,
%\eqref{twoholo6} 
\begin{equation}
F_{\theta,\varUpsilon}(\gamma_1)=F_{\theta,\varUpsilon}(\gamma_0). 
\label{twoholo6/2}
\end{equation}
\end{prop}

\noindent{\it Proof}. Under the assumptions made, the $1$--parameter family of surfaces 
$I_{\gamma_z}:\gamma_z\Rightarrow\gamma_z$ is such that $H(x,y,z)=I_{\gamma_z}(x,y)=\gamma_z(x)$
is a thin homotopy of $I_{\gamma_0}$, $I_{\gamma_1}$. The statement then follows from prop. 
\ref{theor:twoholo3} with $\gamma_{0z}=\gamma_{1z}=\gamma_z$ and $\varSigma_z=I_{\gamma_z}$. 
 \hfill $\Box$

\noindent
The maps $\bar F_{\theta,\varUpsilon}:\Pi_1M\rightarrow G$, $\bar F_{\theta,\varUpsilon}:\Pi_2M\rightarrow H$ factor therefore through
others $\bar F_{\theta,\varUpsilon}:P_1M\rightarrow G$, $\bar F_{\theta,\varUpsilon}:P_2M\rightarrow H$ from the path groupoid $1$-- and $2$--cell 
sets $P_1M$, $P_2M$ into $G$, $H$, respectively, and, so, it induces a categorical map
$\bar F_{\theta,\varUpsilon}:(M,P_1M,P_2M)\rightarrow B_0(G,H)$
\begin{equation}
\xymatrix@C=3pc@R=3.3pc{
   {\text{\footnotesize $p_0$}}
& {\text{\footnotesize $p_1$}} \ar@/^1pc/[l]^{\gamma_1}="0"
           \ar@/_1pc/[l]_{\gamma_0}="1"
           \ar@{=>}"1"+<0ex,-2.ex>;"0"+<0ex,2.ex>_{\varSigma\,}
}
\quad
\xymatrix{\ar@{|->}[r]&}
\quad
\xymatrix@C=9pc@R=3.3pc{
   {\text{\footnotesize $*$}}
& {\text{\footnotesize $*$}} \ar@/^1pc/[l]^{\bar F_{\theta,\Upsilon}(\gamma_1)}="0"
           \ar@/_1pc/[l]_{\bar F_{\theta,\Upsilon}(\gamma_0)}="1"
           \ar@{=>}"1"+<0ex,-2.ex>;"0"+<0ex,2.ex>_{\bar F_{\theta,\Upsilon}(\varSigma)\,}
}
\label{diagholo2}
\end{equation}
of the path $2$--groupoid $(M,P_1M,,P_2M)$ into the 
delooping $2$--groupoid $B_0(G,H)$ of the Lie crossed module $(G,H)$. 
(cf. subsects. \ref{sec:cycle} and \ref{sec:path}).

\begin{prop} \label{prop:twoholo7}
$\bar F_{\theta,\varUpsilon}$ is a $2$--groupoid $2$--functor. 
\end{prop}

\noindent{\it Proof}. The statement follows from combining props. \ref{prop:twoholo6}, \ref{theor:twoholo3}. 
Functoriality results from relations \eqref{twoholo15}--\eqref{twoholo22}. \hfill $\Box$ 

\begin{defi}  \label{def:hflat}
The $(G,H)$--connection $(\theta,\varUpsilon)$ is said flat if 
\begin{equation}
d\varUpsilon+[\theta,\varUpsilon]=0
\label{twoholo27}
\end{equation}
\end{defi}

\begin{prop} \label{theor:twoholo11} Let $(\theta,\varUpsilon)$ be flat.
Let $p_0,p_1$ be points and $\gamma_{0z},\gamma_{1z}:p_0\rightarrow p_1$ and 
$\varSigma_z:\gamma_{0z}\Rightarrow\gamma_{1z}$, $z\in\mathbb{R}$ be $1$--parameter families of curves 
and surfaces such that the mapping $H:\mathbb{R}^3\rightarrow M$ defined by $H(x,y,z)=\varSigma_z(x,y)$
is a homotopy of $\varSigma_0$, $\varSigma_1$. Then, one has the identities 
\begin{subequations}
\label{twoholo28,29,30/1}
\begin{align}
&F_{\theta,\varUpsilon}(\gamma_{01})=F_{\theta,\varUpsilon}(\gamma_{00}),
\vphantom{\Big]}
\label{twoholo28/1}
\\
%\end{align}
%\begin{align}
&F_{\theta,\varUpsilon}(\gamma_{11})=F_{\theta,\varUpsilon}(\gamma_{10}),
\vphantom{\Big]}
\label{twoholo29/1}
\\
&F_{\theta,\varUpsilon}(\varSigma_1)=F_{\theta,\varUpsilon}(\varSigma_0).
\vphantom{\Big]}
\label{twoholo30/1}
\end{align}
\end{subequations}
\end{prop}

\noindent{\it Proof}. The proof is based on the variational formulae \eqref{twoholo23}, \eqref{twoholo25}  and follows the same
lines as that of prop. \ref{theor:twoholo3} except for the vanishing of the double integral term
in the right hand side of \eqref{twoholo25} which is now due to the flatness of $(\theta,\varUpsilon)$ 
instead of the thinness of $H$. \hfill $\Box$ 

\noindent
Prop. \ref{theor:twoholo3/1} of course keeps holding unchanged. 

\noindent
In this way, the maps $\bar F_{\theta,\varUpsilon}:\Pi_1M\rightarrow G$, $\bar F_{\theta,\varUpsilon}:\Pi_2M\rightarrow H$ 
factor through others $\bar F^0{}_{\theta,\varUpsilon}:P_1M\rightarrow G$, $\bar F^0{}_{\theta,\varUpsilon}:P^0{}_2M\rightarrow H$ 
from the fundamental groupoid $1$-- and $2$--cell 
sets $P_1M$, $P^0{}_2M$ into $G$, $H$, respectively, yielding so a  a categorical map
$\bar F^0{}_{\theta,\varUpsilon}:(M,P_1M,P^0{}_2M)\rightarrow B_0(G,H)$
of the fundamental $2$--groupoid $(M,P_1M,,P^0{}_2M)$ into the 
delooping $2$--groupoid $B_0(G,H)$. 

\begin{prop} \label{prop:twoholo8}
When the connection doublet $(\theta,\varUpsilon)$ is flat,  
$\bar F^0{}_{\theta,\varUpsilon}:(M,P_1M$, $P^0{}_2M)\rightarrow B_0(G,H)$ is a 
$2$--groupoid $2$--functor. \end{prop}

\noindent{\it Proof}. The statement follows from combining prop. \ref{prop:twoholo6} and prop. \ref{theor:twoholo11}
with functoriali\-ty resulting again from relations \eqref{twoholo15}--\eqref{twoholo22}. \hfill $\Box$ 

We now turn to the analysis of $1$--gauge transformation of parallel
transport. \vskip1mm

\vfil\eject

\subsection{\normalsize \textcolor{blue}{$2$--parallel transport and $1$--gauge transformation}}
\label{sec:gauholo}

\hspace{.5cm} 
In this subsection, we shall analyze $1$--gauge transformation of higher parallel
transport relying on the cocycle $1$--gauge transformation set--up 
of sect. \ref{sec:hiholo}.

We begin by reviewing gauge transformation in ordinary gauge theory. 
Let $M$ be a manifold and $G$ be a Lie group. 

\begin{defi} A $G$--gauge transformation is a map $g\in\Map(M,G)$. 
We denote by $\Gau(M,G)$ the set of all gauge transformations.
\end{defi}

$G$--gauge transformations act on $G$--connections (cf. def. \ref{def:gconn}).

\begin{defi}
Let $a$ be a $G$--connection and $g$ be a $G$--gauge transformation.
The gauge transformed $G$--connection ${}^g\theta$ is 
\begin{equation}
{}^g\theta=\Ad g (a)-dgg^{-1}.
\vphantom{\Big]}
\label{gauholo1}
\end{equation}
\end{defi}

\begin{prop} If $\theta$ is a flat $G$--connection, then, for any $G$--gauge transformation $g$,
${}^g\theta$ is also a flat $G$--connection (cf. def. \ref{def:flat}).
\end{prop}

\noindent{\it Proof}. Indeed, using \eqref{gauholo1}, one computes
\begin{equation}
d{}^g\theta+\frac{1}{2}[{}^g\theta,{}^g\theta]=\Ad g\Big(d\theta+\frac{1}{2}[\theta,\theta]\Big)=0,
\vphantom{\Big]}
\label{gauholo1/0}
\end{equation}
which shows the flatness of ${}^g\theta$. \hfill $\Box$

The following theorem is a classic result.

\begin{prop} \label{theor:pargau1} Let $\theta$ be a $G$--connection and 
$g$ be a $G$--gauge transformation. 
Let further $p_0$, $p_1$ be points and $\gamma:p_0\rightarrow p_1$ be a curve. 
Then, the parallel transports $F_\theta(\gamma)$ and $F_{{}^g\theta}(\gamma)$ along $\gamma$ 
are related as 
\begin{equation}
F_{{}^g\theta}(\gamma)=g(p_1)F_\theta(\gamma)g(p_0)^{-1}.
\vphantom{\Big]}
\label{gauholo2}
\end{equation}
\end{prop}

\noindent
{\it Proof}. According to prop. \ref{theor:cycle1}, there exists a \pagebreak 
one--to--one correspondence between $\mathfrak{g}$--valued $1$--forms $a$ on $\mathbb{R}$ and 
$G$--cocycles $f$. By \eqref{gauge30}, \eqref{gauge31}, the action of a gauge transformation 
$\varkappa$ on a cocycle $f$ is such that $a_{{}^\varkappa f}={}^\varkappa a_f$. Then,
%Taking these properties into account, we have 
%Using the one--to--one correspondence between $\mathfrak{g}$--valued $1$--forms on $\mathbb{R}$ 
%and $G$--cocycles established by prop. \ref{theor:cycle1} and \eqref{gauge30}, \eqref{gauge31}, we have
\begin{equation}
{}^\varkappa a=a_{{}^\varkappa f}\big|_{f=f_a}.
\label{gauholo0}
\end{equation}
From this relation, it follows so that 
\begin{equation}
f_{{}^\varkappa a}%=f_{{}^\varkappa a_{\bar f}}\big|_{\bar f=f_a}=f_{a_{{\bar f}}}\big|_{\bar f={}^\varkappa f_a}
={}^\varkappa f_a.
\label{gauholoa1}
\end{equation}
%for any $\mathfrak{g}$--valued $1$--form $a$ and cocycle gauge transformation $\varkappa$.
Setting $a=\gamma^*\theta$ and $\varkappa=\gamma^*g$ in the above relation, we obtain
\begin{equation}
f_{{}^{\gamma^*g} \gamma^*\theta}={}^{\gamma^*g} f_{\gamma^*\theta}.
\label{gauholoa2}
\end{equation}
From here, noting that $\gamma^*{}^g\theta={}^{\gamma^*g}\gamma^*\theta$, we find 
\begin{align}
&F_{{}^g\theta}(\gamma)=f_{\gamma^*{}^g\theta}(1,0)=f_{{}^{\gamma^*g} \gamma^*\theta}(1,0)
={}^{\gamma^*g} f_{\gamma^*\theta}(1,0)
\vphantom{\Big]}
\label{gauholoa3}
\\
&\hspace{2.5cm}=\gamma^*g(1)f_{\gamma^*\theta}(1,0)\gamma^*g(0)^{-1}
=g(p_1)F_\theta(\gamma)g(p_0)^{-1}
\vphantom{\Big]}
\nonumber
\end{align}
as was to be shown. \hfill $\Box$

Recall that, for a $G$--connection $\theta$, the mapping $F_\theta:\Pi_1M\rightarrow G$
induces a groupoid functor $\bar F_\theta:(M,P_1M)\rightarrow BG$ of the path groupoid $(M,P_1M)$ of $M$ in
the delooping $BG$ of $G$ in virtue of its thin homotopy invariance
(cf. prop. \ref{prop:twoholo3}). Likewise, when the $G$--connection $\theta$ is flat,
by its homotopy invariance, $F_\theta$ induces a groupoid functor $\bar F^0{}_\theta:(M,P^0{}_1M)\rightarrow BG$ 
of the fundamental groupoid  $(M,P^0{}_1M)$ of $M$ into $BG$ (cf. prop. \ref{prop:twoholo5}). 

\begin{prop}
For any $G$--connection $\theta$, 
a $G$--gauge transformation $g$ encodes a natural transformation $\bar F_\theta\Rightarrow \bar F_{{}^g\theta}$
of functors. 
If $\theta$ is flat, then $g$ yields a natural transformation $\bar F^0{}_\theta\Rightarrow \bar F^0{}_{{}^g\theta}$.
\end{prop}

\noindent{\it Proof}. By \eqref{gauholo2}, the diagram 
\begin{equation}
\xymatrix@C=3pc{
{\text{\footnotesize $*$}} \ar[d]_{g(p_1)}  %\ar@{}[dr]^(.25){}="a"^(.75){}="b" \ar@{=>} "a";"b"_X 
& {\text{\footnotesize $*$}}\ar[l]_{F_\theta(\gamma)} \ar[d]^{g(p_0)}
\\                 
{\text{\footnotesize $*$}} & {\text{\footnotesize $*$}}\ar[l]^{F_{{}^g\theta}(\gamma)}            
}
\label{gaunat1}
\end{equation}
commutes, 
identifying $g$ as a natural transformation $\bar F_\theta\Rightarrow \bar F_{{}^g\theta}$
or $\bar F^0{}_\theta\Rightarrow \bar F^0{}_{{}^g\theta}$. \hfill $\Box$ %tttt

We now shift to higher gauge theory, introduce the notion of $1$--gauge transformation
and study %in detail 
its action on connection doublets and $2$--parallel transport.
 
Let $M$ be a manifold and $(G,H)$ be a Lie crossed module. 

\begin{defi} A differential $(G,H)$--$1$--gauge transformation is a pair
of a map $g\in\Map(M,G)$ and a $1$--form $J\in\Omega^1(M,\mathfrak{h})$. 
We denote by $\Gau_1(M,G,H)$ the set of all differential $1$--gauge transformations. 
\end{defi}

Differential $(G,H)$--$1$--gauge transformations  act on $(G,H)$--connections doublets 
(cf. def. \ref{def:ghconn}).

\begin{defi}
Let $(\theta,\varUpsilon)$ be a $(G,H)$--connection doublet and $(g,J)$  be a 
$(G$, $H)$--$1$--gauge transformation.
The gauge transformed $(G,H)$--connection doublet $({}^{g,J}\theta,{}^{g,J}\varUpsilon)$ is 
\begin{subequations}
\label{gauholo3,4}
\begin{align}
&{}^{g,J}\theta=\Ad g(\theta)-dgg^{-1}-\dot t(J),
\vphantom{\Big]}
\label{gauholo3}
\\
&{}^{g,J}\varUpsilon=\dot m(g)(\varUpsilon)-dJ
-\frac{1}{2}[J,J]-\widehat{m}(\Ad g(\theta) -dgg^{-1}-\dot t(J),J).
\vphantom{\Big]}
\label{gauholo4}
\end{align}
\end{subequations}
\end{defi}
It can be checked that this gauge transformation is compatible with the zero fake
curvature condition \eqref{twoholo8}. 

\begin{prop} If $(\theta,\varUpsilon)$ is a flat $(G,H)$--connection doublet, then, for any $(G,H)$--$1$--gauge transformation 
$(g,J)$, $({}^{g,J}\theta,{}^{g,J}\varUpsilon)$ is also a flat $(G,H)$--connection doublet
(cf. def. \ref{def:hflat}).
\end{prop}

\noindent{\it Proof}. Indeed, using \eqref{gauholo3,4}, taking \eqref{twoholo8} into account, one finds
\begin{equation}
d{}^{g,J}\varUpsilon+[{}^{g,J}\theta,{}^{g,J}\varUpsilon]
=\dot m(g)(d\varUpsilon+[\theta,\varUpsilon])=0,
\vphantom{\Big]}
\label{gauholoh1/0}
\end{equation}
which shows the flatness of $({}^{g,J}\theta,{}^{g,J}J)$. \hfill $\Box$

Recall that, by prop. \ref{theor:cycle2}, \pagebreak with a $(G,H)$--connection doublet $(a,B)$ in the sense 
of def. \ref{def:r2ghconn} there is associated a $(G,H)$--cocycle $(f_{a,B},g_{a,B},W_{a,B})$. Further, 
by prop. \ref{theor:gauge1}, with a differential $(G,H)$--$1$--gauge transformation 
$(\varkappa,\varGamma)$ in the sense of def. \ref{def:r2dghgau}, there is associated an
$(f_{a,B},g_{a,B},W_{a,B})$--$1$--gauge transformation $(\kappa_{\varkappa,\varGamma;a,B},\varPsi_{\varkappa,\varGamma;a,B},
\varPhi_{\varkappa,\varGamma;a,B})$. This depends not only on the differential transformation $(\varkappa,\varGamma)$ but also
on the connection doublet $(a,B)$, when this latter is allowed to vary. 
The following basic result extends prop. \ref{theor:pargau1} to higher gauge theory in a non trivial manner.

\begin{prop} \label{theor:pargau2} Let $(\theta,\varUpsilon)$ be a $(G,H)$--connection doublet and
$(g,J)$ be a $(G$, $H)$--$1$--gauge transformation. 
Let further $p_0,p_1$ be points, $\gamma_0,\gamma_1:p_0\rightarrow p_1$ be curves and
$\varSigma:\gamma_0\Rightarrow \gamma_1$ be a surface. Then, we have
\begin{subequations}
\label{gauholo5,6,7}
\begin{align}
&F_{{}^{g,J}\theta,{}^{g,J}\varUpsilon}(\gamma_0)=g(p_1)
t(G_{g,J;\theta,\Upsilon}(\gamma_0))F_{\theta,\varUpsilon}(\gamma_0)g(p_0)^{-1},
\vphantom{\Big]}
\label{gauholo5}
\\
&F_{{}^{g,J}\theta,{}^{g,J}\varUpsilon}(\gamma_1)=g(p_1)
t(G_{g,J;\theta,\Upsilon}(\gamma_1))F_{\theta,\varUpsilon}(\gamma_1)g(p_0)^{-1},
\vphantom{\Big]}
\label{gauholo6}
\\
&F_{{}^{g,J}\theta,{}^{g,J}\varUpsilon}(\varSigma)
=m(g(p_1))\big(G_{g,J;\theta,\Upsilon}(\gamma_1)
F_{\theta,\varUpsilon}(\varSigma)G_{g,J;\theta,\Upsilon}(\gamma_0)^{-1}\big),
\vphantom{\Big]}
\label{gauholo7}
\end{align}
%=m(g(p_1))\big(m(F_{\theta,\varUpsilon}(\gamma_1))(G_{g,J}(\gamma_1))^{-1}
%\vphantom{\Big]}
%\label{gauholo7}
%\\
%&\hspace{4.8cm}
%\times F_{\theta,\varUpsilon}(\varSigma)
%m(F_{\theta,\varUpsilon}(\gamma_0))(G_{g,J}(\gamma_0))\big).
%\vphantom{\Big]}
%\nonumber
%\end{align}
\end{subequations}
where $G_{g,J;\theta,\Upsilon}(\gamma_0)$, $G_{g,J;\theta,\Upsilon}(\gamma_1)$ are given by 
\begin{subequations}
\label{gauholo8,9}
\begin{align}
&G_{g,J;\theta,\Upsilon}(\gamma_0)=\varPsi_{\varSigma^*g,\varSigma^*J;\varSigma^*\theta,\varSigma^*\varUpsilon|0}(0,1),
\vphantom{\Big]}
\label{gauholo8}
\\
&G_{g,J;\theta,\Upsilon}(\gamma_1)=\varPsi_{\varSigma^*g,\varSigma^*J;\varSigma^*\theta,\varSigma^*\varUpsilon|1}(0,1).
\vphantom{\Big]}
\label{gauholo9}
\end{align}
\end{subequations}
\end{prop}

\noindent{\it Proof}. By \eqref{gauge26,27}, \eqref{gauge28,29}, the one--to--one correspondence 
between $(G,H)$--cocycles $(f,g,W)$ and $(G,H)$ connections $(a,B)$ (in the sense of def. \ref{def:ghconn})
on one hand and integral $(f,g,W)$--$1$--gauge transformations $(\kappa,\varPsi,\varPhi)$ and differential 
$(G,H)$--$1$--gauge transformations (in the sense of def. \ref{def:r2dghgau}) on the other is such that
$a_{{}^{\kappa,\varPsi,\varPhi}f,{}^{\kappa,\varPsi,\varPhi}g,{}^{\kappa,\varPsi,\varPhi}W}=
{}^{\varkappa_{\kappa,\varPsi,\varPhi},\varGamma_{\kappa,\varPsi,\varPhi}}a_{f,g,W}$,
$B_{{}^{\kappa,\varPsi,\varPhi}f,{}^{\kappa,\varPsi,\varPhi}g,{}^{\kappa,\varPsi,\varPhi}W}=
{}^{\varkappa_{\kappa,\varPsi,\varPhi},\varGamma_{\kappa,\varPsi,\varPhi}}B_{f,g,W}$.
Using these results, it is readily checked that 
\begin{subequations}
\begin{align}
&{}^{\varkappa,\varGamma} a=a_{{}^{\kappa,\varPsi,\varPhi}f,{}^{\kappa,\varPsi,\varPhi}g,{}^{\kappa,\varPsi,\varPhi}W}
\vphantom{\Big]}
\label{}
\\
&\hspace{2.75cm}
\big|_{\kappa=\kappa_{\varkappa,\varGamma;a,B},\varPsi=\varPsi_{\varkappa,\varGamma;a,B},\varPhi=\varPhi_{\varkappa,\varGamma;a,B};f=f_{a,B},g=g_{a,B},W=W_{a,B}},
\vphantom{\Big]}
\nonumber
%\\
\end{align}
\begin{align}
&{}^{\varkappa,\varGamma} B=B_{{}^{\kappa,\varPsi,\varPhi}f,{}^{\kappa,\varPsi,\varPhi}g,{}^{\kappa,\varPsi,\varPhi}W}
\vphantom{\Big]}
\label{}
\\
&\hspace{2.75cm}
\big|_{\kappa=\kappa_{\varkappa,\varGamma;a,B},\varPsi=\varPsi_{\varkappa,\varGamma;a,B},\varPhi=\varPhi_{\varkappa,\varGamma;a,B};f=f_{a,B},g=g_{a,B},W=W_{a,B}}.
\vphantom{\Big]}
\nonumber
\end{align}
\end{subequations}
From these relation, it follows immediately that 
\begin{subequations}
\label{gauholob1,2,3}
\begin{align}
&f_{{}^{\varkappa,\varGamma}a,{}^{\varkappa,\varGamma}B}
={}^{\kappa_{\varkappa,\varGamma;a,B},\varPsi_{\varkappa,\varGamma;a,B},\varPhi_{\varkappa,\varGamma;a,B}}f_{a,B},
\vphantom{\Big]}
\label{gauholob1}
\\
&g_{{}^{\varkappa,\varGamma}a,{}^{\varkappa,\varGamma}B}
={}^{\kappa_{\varkappa,\varGamma;a,B},\varPsi_{\varkappa,\varGamma;a,B},\varPhi_{\varkappa,\varGamma;a,B}}g_{a,B},
\vphantom{\Big]}
\label{gauholob2}
\\
&W_{{}^{\varkappa,\varGamma}a,{}^{\varkappa,\varGamma}B}
={}^{\kappa_{\varkappa,\varGamma;a,B},\varPsi_{\varkappa,\varGamma;a,B},\varPhi_{\varkappa,\varGamma;a,B}}W_{a,B}.
\vphantom{\Big]}
\label{gauholob3}
\end{align}
\end{subequations} 
Setting $a=\varSigma^*\theta$, $B=\varSigma^*\varUpsilon$ and 
$\varkappa=\varSigma^*g$, $\varGamma=\varSigma^*J$ in 
the \eqref{gauholob1,2,3}, we obtain 
\begin{subequations}
\label{gauholob5,6,7}
\begin{align}
&f_{{}^{\varSigma^*g,\varSigma^*J}\varSigma^*\theta,{}^{\varSigma^*g,\varSigma^*J}\varSigma^*\varUpsilon}
\vphantom{\Big]}
\label{gauholob5}
\\
&\hspace{3cm}
={}^{\kappa_{\varSigma^*g,\varSigma^*J;\varSigma^*\theta,\varSigma^*\varUpsilon},
\varPsi_{\varSigma^*g,\varSigma^*J;\varSigma^*\theta,\varSigma^*\varUpsilon},
\varPhi_{\varSigma^*g,\varSigma^*J;\varSigma^*\theta,\varSigma^*\varUpsilon}}f_{\varSigma^*\theta,\varSigma^*\varUpsilon},
\vphantom{\Big]}
\nonumber
\\
&g_{{}^{\varSigma^*g,\varSigma^*J}\varSigma^*\theta,{}^{\varSigma^*g,\varSigma^*J}\varSigma^*\varUpsilon}
\vphantom{\Big]}
\label{gauholob6}
\\
&\hspace{3cm}
={}^{\kappa_{\varSigma^*g,\varSigma^*J;\varSigma^*\theta,\varSigma^*\varUpsilon},
\varPsi_{\varSigma^*g,\varSigma^*J;\varSigma^*\theta,\varSigma^*\varUpsilon},
\varPhi_{\varSigma^*g,\varSigma^*J;\varSigma^*\theta,\varSigma^*\varUpsilon}}g_{\varSigma^*\theta,\varSigma^*\varUpsilon},
\vphantom{\Big]}
\nonumber
\\
&W_{{}^{\varSigma^*g,\varSigma^*J}\varSigma^*\theta,{}^{\varSigma^*g,\varSigma^*J}\varSigma^*\varUpsilon}
\vphantom{\Big]}
\label{gauholob7}
\\
&\hspace{3cm}
={}^{\kappa_{\varSigma^*g,\varSigma^*J;\varSigma^*\theta,\varSigma^*\varUpsilon},
\varPsi_{\varSigma^*g,\varSigma^*J;\varSigma^*\theta,\varSigma^*\varUpsilon},
\varPhi_{\varSigma^*g,\varSigma^*J;\varSigma^*\theta,\varSigma^*\varUpsilon}}W_{\varSigma^*\theta,\varSigma^*\varUpsilon}.
\vphantom{\Big]}
\nonumber
\end{align}
\end{subequations} 

We can now complete the proof of the theorem.  We show relation \eqref{gauholo7} only,
the proof of \eqref{gauholo5}, \eqref{gauholo6} being analogous.
We showed earlier that $g_{\varSigma^*\theta,\varSigma^*\varUpsilon|x}(y',y)=1_G$ 
for $x<\epsilon$ or $x>1-\epsilon$ and arbitrary $y,y'$ (cf. eq. \eqref{twoholob1}).
Similarly, we can show that $\varPhi_{\varSigma^*g,\varSigma^*J;\varSigma^*\theta,\varSigma^*\varUpsilon|x}(y',y)=1_H$
for the same range of $x$ and $y,y'$ values, by considering
%a similar reasoning based on the consideration of 
the differential problem \eqref{gauge23}, \eqref{gauge25} with $\varkappa$, $\varGamma$ replaced by 
$\varSigma^*g$, $\varSigma^*J$ and observing 
%the observation that 
that $\varSigma^*\varGamma_y(x,y)=0$ identically for the values of $x$ indicated
on account of \eqref{path7}, \eqref{path8}. Then, from \eqref{gauholob7}, using \eqref{gauge13} and 
noting that by \eqref{gauholo3,4} %\linebreak 
$\varSigma^*{}^{g,J}\theta=
{}^{\varSigma^*g,\varSigma^*J}\varSigma^*\theta$, 
$\varSigma^*{}^{g,J}\varUpsilon={}^{\varSigma^*g,\varSigma^*J}\varSigma^*\varUpsilon$, 
we find 
\begin{align}
F_{{}^{g,J}\theta,{}^{g,J}\varUpsilon}(\varSigma)
&=W_{\varSigma^*{}^{g,J}\theta,\varSigma^*{}^{g,J}\varUpsilon}(0,1;1,0)\hspace{5.3cm}
\vphantom{\Big]}
\label{gauholob8}
%\\
\end{align}
\begin{align}
&=W_{{}^{\varSigma^*g,\varSigma^*J}\varSigma^*\theta,{}^{\varSigma^*g,\varSigma^*J}\varSigma^*\varUpsilon}(0,1;1,0)
\vphantom{\Big]}
\nonumber
\\
\hspace{1.7cm}
&={}^{\kappa_{\varSigma^*g,\varSigma^*J;\varSigma^*\theta,\varSigma^*\varUpsilon},
\varPsi_{\varSigma^*g,\varSigma^*J;\varSigma^*\theta,\varSigma^*\varUpsilon},
\varPhi_{\varSigma^*g,\varSigma^*J;\varSigma^*\theta,\varSigma^*\varUpsilon}}W_{\varSigma^*\theta,\varSigma^*\varUpsilon}(0,1;1,0)
\vphantom{\Big]}
\nonumber
\\
&=m(\kappa_{\varSigma^*g,\varSigma^*J;\varSigma^*\theta,\varSigma^*\varUpsilon}(1;0))\big(
\varPhi_{\varSigma^*g,\varSigma^*J;\varSigma^*\theta,\varSigma^*\varUpsilon|1}(1,0)
\vphantom{\Big]}
\nonumber
\\
&\hspace{.75cm}\times m(g_{\varSigma^*\theta,\varSigma^*\varUpsilon|1}(1,0)^{-1})
(\varPsi_{\varSigma^*g,\varSigma^*J;\varSigma^*\theta,\varSigma^*\varUpsilon|1}(0,1))
\vphantom{\Big]}
\nonumber
\\
&\hspace{.75cm}\times W_{\varSigma^*\theta,\varSigma^*\varUpsilon}(0,1;1,0)
\vphantom{\Big]}
\nonumber
\\
&\hspace{.75cm}\times 
m(f_{\varSigma^*\theta,\varSigma^*\varUpsilon|0}(0,1)^{-1})
(\varPhi_{\varSigma^*g,\varSigma^*J;\varSigma^*\theta,\varSigma^*\varUpsilon|0}(1,0)^{-1})
\vphantom{\Big]}
\nonumber
\\
&\hspace{.75cm}\times 
\varPsi_{\varSigma^*g,\varSigma^*J;\varSigma^*\theta,\varSigma^*\varUpsilon|0}(0,1)^{-1}\big)
\vphantom{\Big]}
\nonumber
\\
&=m(g(p_1))\big(G_{g,J;\theta,\Upsilon}(\gamma_1)
F_{\theta,\varUpsilon}(\varSigma)G_{g,J;\theta,\Upsilon}(\gamma_0)^{-1}\big),
\vphantom{\Big]}
\nonumber
\end{align}  
showing \eqref{gauholo7}. \hfill $\Box$

In prop. \ref{theor:pargau2}, a new object appears, 
$G_{g,J;\theta,\Upsilon}(\gamma)$. As it turns out,
it has a number of relevant properties which are the topic of the rest of 
this subsection. 

%Analogously to $1$--parallel transport, 
$G_{g,J;\theta,\Upsilon}(\gamma)$
can be defined for any curve $\gamma$ 
independently from any other curve $\gamma'$ with the same endpoints and surface
$\varSigma$ connecting $\gamma$ to $\gamma'$.

\begin{lemma} \label{lemma:gau}
Suppose that $p_0,p_1$ are points and $\gamma:p_0\rightarrow p_1$ is a curve. Then,
$\varPsi_{I_\gamma{}^*g,I_\gamma{}^*J;I_\gamma{}^*\theta,I_\gamma{}^*\varUpsilon|y}$, 
where $I_\gamma:\gamma\Rightarrow\gamma$ is the unit surface of $\gamma$ 
(cf. eq. \eqref{path16}), is independent from $y$. 
\end{lemma}

\noindent{\it Proof}. 
By prop. \ref{theor:gauge1}, $\varPsi_{I_\gamma{}^*g,I_\gamma{}^*J;I_\gamma{}^*\theta,I_\gamma{}^*\varUpsilon|y}$
is the solution of the differential problem \eqref{gauge22}, \eqref{gauge24} with 
$f(x,x_0;y)=f_{I_\gamma{}^*\theta,I_\gamma{}^*\varUpsilon}(x,x_0;y)$, $\varkappa(x,y)=I_\gamma{}^*g(x,y)$ and 
$\varGamma_x(x,y)=I_\gamma{}^*J_x(x,y)$.
Now, by lemma \ref{lemma:para}, $f_{I_\gamma{}^*\theta,I_\gamma{}^*\varUpsilon}(x,x_0;y)$ is independent from $y$. 
Further, $I_\gamma{}^*g(x,y)=\gamma^*g(x)$, $I_\gamma{}^*J_x(x,y)=\gamma^*J_x(x)$ are also independent from $y$.
So, $\varPsi_{I_\gamma{}^*g,I_\gamma{}^*J;I_\gamma{}^*\theta,I_\gamma{}^*\varUpsilon|y}$ is $y$ independent. \hfill $\Box$

\begin{defi} If $p_0,p_1$ are points and $\gamma:p_0\rightarrow p_1$ is a curve, one sets 
\begin{equation}
G_{g,J;\theta,\Upsilon}(\gamma)
=\varPsi_{I_\gamma{}^*g,I_\gamma{}^*J;I_\gamma{}^*\theta,I_\gamma{}^*\varUpsilon|y}(0,1).
%\vphantom{\ul{\ul{\ul{\ul{\ul{g}}}}}}
\label{gauholo13}
\end{equation}
%where $I_\gamma:\gamma\Rightarrow\gamma$ is the unit surface of $\gamma$ (cf. eq. \eqref{path16}) and $y\in\mathbb{R}$.
\end{defi}

\noindent
\eqref{gauholo13} gives the same result as \eqref{gauholo8}, \eqref{gauholo9}.

\begin{prop} \label{prop:altgaudef}
If $p_0,p_1$ are points, $\gamma_0,\gamma_1:p_0\rightarrow p_1$ are curves and
$\varSigma:\gamma_0\Rightarrow \gamma_1$ is a surface, then 
then the value of $G_{g,J;\theta,\Upsilon}(\gamma_i)$
computed using  \eqref{gauholo8}, \eqref{gauholo9} equals that obtained using \eqref{gauholo13}.
%\begin{equation}
%G_{g,J;\theta,\Upsilon}(\gamma_i)=G_{g,J;\theta,\Upsilon|y}(\gamma_i).
%\label{gauholo13/1}
%\end{equation}
\end{prop}

\noindent{\it Proof}. By prop. \ref{theor:cycle2}, 
$\varPsi_{\varSigma^*g,\varSigma^*J;\varSigma^*\theta,\varSigma^*\varUpsilon|y}(x',x)$ is the solution 
of the differential problem \eqref{gauge22}, \eqref{gauge24} with $f(x,x_0;y)
=f_{\varSigma^*\theta,\varSigma^*\varUpsilon|y}(x,x_0)$,
$\varkappa(x,y)=\varSigma^*g(x,y)$,
$\varGamma_x(x,y)=\varSigma^*J_x(x,y)$. 
Likewise, $\varPsi_{I_{\gamma_i}{}^*g,I_{\gamma_i}{}^*J;I_{\gamma_i}{}^*\theta,I_{\gamma_i}{}^*\varUpsilon|y}(x',x)$ solves
the differential problem \eqref{gauge22}, \eqref{gauge24} with $f(x,x_0;y)
=f_{I_{\gamma_i}{}^*\theta,I_{\gamma_i}{}^*\varUpsilon|y}(x,x_0)$,
$\varkappa(x,y)=I_{\gamma_i}{}^*g(x,y)$,
$\varGamma_x(x,y)=I_{\gamma_i}{}^*J_x(x,y)$. %\varSigma^*
Now, we have $f_{I_{\gamma_i}{}^*\theta,I_{\gamma_i}{}^*\varUpsilon|y}(x,x_0)$ 
$=f_{\varSigma^*\theta,\varSigma^*\varUpsilon|i}(x,x_0)$ (see the proof of prop. \ref{prop:paraequiv})
and also $I_{\gamma_i}{}^*g(x,y)=\varSigma^*g(x,i)$,
$I_{\gamma_i}{}^*J_x(x,y)=\varSigma^*J_x(x,i)$ 
for $i=1,2$ and any $y$. So, 
$\varPsi_{I_{\gamma_i}{}^*g,I_{\gamma_i}{}^*J;I_{\gamma_i}{}^*\theta,I_{\gamma_i}{}^*\varUpsilon|y}(x,x_0)$
$=\varPsi_{\varSigma^*g,\varSigma^*J;\varSigma^*\theta,\varSigma^*\varUpsilon|i}(x,x_0)$.
From this relation, recalling \eqref{gauholo8}, \eqref{gauholo9} and \eqref{gauholo13}, 
the statement follows. \hfill $\Box$ 

%\noindent
Let us fix a $(G,H)$--connection doublet $(\theta,\varUpsilon)$
and a $(G,H)$--$1$--gauge transformation $(g,J)$. 
We have then a mapping $G_{g;J;\theta,\varUpsilon}:\Pi_1M\rightarrow H$.

\begin{prop} \label{prop:gauholo8/1}
For any two points $p_0,p_1$ and curve $\gamma:p_0\rightarrow p_1$, one has 
\begin{equation}
F_{{}^{g,J}\theta,{}^{g,J}\varUpsilon}(\gamma)=g(p_1)
t(G_{g,J;\theta,\Upsilon}(\gamma))F_{\theta,\varUpsilon}(\gamma)g(p_0)^{-1},
\label{gauholo16/1}
\end{equation} 
\end{prop}

\noindent{\it Proof}.  This follows from \eqref{gauholo5}, \eqref{gauholo5}, 
setting $\varSigma=I_\gamma$ and using \eqref{twoholo13} and 
\eqref{gauholo13}. \hfill $\Box$

\begin{prop} \label{prop:gauholo8} For any point $p$, one has 
\begin{equation}
G_{g,J;\theta,\varUpsilon}(\iota_p)=1_H.
\label{gauholo15}
\end{equation}
For any two points $p_0,p_1$ and curve $\gamma:p_0\rightarrow p_1$, one has 
\begin{equation}
G_{g,J;\theta,\varUpsilon}(\gamma^{-1_\circ})=m(F_{\theta,\varUpsilon}(\gamma)^{-1})(G_{g,J;\theta,\varUpsilon}(\gamma)^{-1}).
\vphantom{\ul{\ul{\ul{\ul{\ul{x}}}}}}
\label{gauholo16}
\end{equation}
For any three $p_0,p_1,p_2$ and two curves $\gamma_1:p_0\rightarrow p_1$, $\gamma_2:p_1\rightarrow p_2$, 
\begin{equation}
G_{g,J;\theta,\varUpsilon}(\gamma_2\circ\gamma_1)=G_{g,J;\theta,\varUpsilon}(\gamma_2)
m(F_{\theta,\varUpsilon}(\gamma_2))(G_{g,J;\theta,\varUpsilon}(\gamma_1)).
\label{gauholo17}
\end{equation}
\end{prop}

\noindent{\it Proof}.  The proof is analogous to that of prop. \ref{prop:twoholo6},
relying on the pull--back action of the map 
$l_\phi:\mathbb{R}^2\rightarrow\mathbb{R}^2$, $l_\phi(x,y)=(\phi(x),y)$, 
%$r_\phi:\mathbb{R}^2\rightarrow\mathbb{R}^2$, 
induced by a function $\phi:\mathbb{R}\rightarrow\mathbb{R}$. 
The left pull--back $(l_\phi{}^*f,l_\phi{}^*g,l_\phi{}^*W)$ 
of a $(G,H)$--cocycle $(f,g,W)$ is the $(G,H)$--cocycle defined 
by eqs. \eqref{twoholoc1,2,3}. The left pull--back 
$(l_\phi{}^*\kappa,l_\phi{}^*\varPsi,l_\phi{}^*\varPhi)$
of an $(f,g,W)$--$1$--gauge transformation $(\kappa,\varPsi,\varPhi)$ 
is the $(l_\phi{}^*f,l_\phi{}^*g,l_\phi{}^*W)$--gauge transformation given by 
\hphantom{xxxxxxxxxxxxx}
\begin{subequations}
\label{gauholoc1,2,3}
\begin{align}
&l_\phi{}^*\kappa(x;y)=\kappa(\phi(x);y),
\vphantom{\Big]}
\label{gauholoc1}
\\
&l_\phi{}^*\varPsi(x',x;y)=\varPsi(\phi(x'),\phi(x);y), 
\vphantom{\Big]}
\label{gauholoc2}
\\
&l_\phi{}^*\varPhi(x;y',y)=\varPhi(\phi(x);y',y). 
\vphantom{\Big]}
\label{gauholoc3}
\end{align}
\end{subequations}
The verification of the validity of the cocycle relations %\eqref{cycle15}, \eqref{cycle11,12,13,14}
\eqref{gauge5,6} is straightforward. 

The one--to--one correspondence between form pairs % \pagebreak 
$(a,B)\in\Omega^1(\mathbb{R}^2,\mathfrak{g})\times \Omega^2(\mathbb{R}^2,\mathfrak{h})$ 
and $(G,H)$--cocycles $(f,g,W)\in\Cyc(G,H)$ established  by prop. \ref{theor:cycle2} 
is natural with respect to left pull-back.
Likewise, the one--to--one correspondence between pairs 
$(\varkappa,\varGamma)\in\Map(M,G)\times \Omega^1(\mathbb{R}^2,\mathfrak{h})$
and $(f_{a,B},g_{a,B},W_{a,B})$--gauge transformation is natural, meaning that the relations 
$(\kappa_{l_\phi{}^*\varkappa,l_\phi{}^*\varGamma;l_\phi{}^*a,l_\phi{}^*B}$,
$\varPsi_{l_\phi{}^*\varkappa,l_\phi{}^*\varGamma;l_\phi{}^*a,l_\phi{}^*B},
\varPhi_{l_\phi{}^*\varkappa,l_\phi{}^*\varGamma;l_\phi{}^*a,l_\phi{}^*B})$ $=
(l_\phi{}^*\kappa_{\varkappa,\varGamma;a,B},l_\phi{}^*\varPsi_{\varkappa,\varGamma;a,B},
l_\phi{}^*\varPhi_{\varkappa,\varGamma;a,B})$ as well as
$(\varkappa_{l_\phi{}^*\kappa,l_\phi{}^*\varPsi,l_\phi{}^*\varPhi}$, $\varGamma_{l_\phi{}^*\kappa,l_\phi{}^*\varPsi,l_\phi{}^*\varPhi})
=$ $(l_\phi{}^*\varkappa_{\kappa,\varPsi,\varPhi},l_\phi{}^*\varGamma_{\kappa,\varPsi,\varPhi})$
hold.

Given these results, the proof of relations \eqref{gauholo15}, \eqref{gauholo16}, \eqref{gauholo17}
is totally analogous to that of \eqref{twoholo18}, \eqref{twoholo21}, \eqref{twoholo22}.
For instance, the verification of \eqref{gauholo17} proceeds along the same lines as that 
of \eqref{twoholo22} as indicated in \eqref{twoholoc7}: replace 
$\varSigma_i$ by $I_{\gamma_i}$ and $W_{\varSigma_i{}^*\theta,\varSigma_i{}^*\varUpsilon}$ by 
$\varPsi_{I_{\gamma_i}{}^*g,I_{\gamma_i}{}^*J;I_{\gamma_i}{}^*\theta,I_{\gamma_i}{}^*\varUpsilon}$
and use \eqref{gauge5}. % systematically. 
\hfill $\Box$

%\vspace{1mm} 
Naturally, thin homotopy invariance holds for gauge transformation along a curve. 

\begin{prop} \label{theor:gauholo3/1}
Let $p_0,p_1$ be points and $\gamma_y:p_0\rightarrow p_1$, $y\in\mathbb{R}$, be a smooth $1$--parameter family of curves
such that the mapping $h:\mathbb{R}^2\rightarrow M$ defined by $h(x,y)=\gamma_y(x)$
is a thin homotopy of $\gamma_0$, $\gamma_1$. Then,
%\eqref{twoholo6} 
\begin{equation}
G_{g,J;\theta,\varUpsilon}(\gamma_1)=G_{g,J;\theta,\varUpsilon}(\gamma_0). 
\label{gauholo6/2}
\end{equation}
\end{prop}

\noindent{\it Proof}. 
The proof is based on the variational formula 
\begin{align}
&\partial_z\varPsi_{I_{\gamma_z}{}^*g,I_{\gamma_z}{}^*J;I_{\gamma_z}{}^*\theta,I_{\gamma_z}{}^*\varUpsilon|y}(x,x_0)
\varPsi_{I_{\gamma_z}{}^*g,I_{\gamma_z}{}^*J;I_{\gamma_z}{}^*\theta,I_{\gamma_z}{}^*\varUpsilon|y}(x,x_0)^{-1}
\vphantom{\Big]}
\label{gauholok1}
\\
&=-\int_{x_0}^xd\xi\,\varPsi_{I_{\gamma_z}{}^*g,I_{\gamma_z}{}^*J;I_{\gamma_z}{}^*\theta,I_{\gamma_z}{}^*\varUpsilon|y}(\xi,x_0)
\dot m(f_{I_{\gamma_z}{}^*\theta,I_{\gamma_z}{}^*\varUpsilon|y}(\xi,x_0)^{-1})  
\vphantom{\Big]}
\nonumber
\\
&\hphantom{=\,}\bigg(H^*(\dot m(g^{-1})(dJ+[J,J]/2
+\widehat{m}(\Ad g(\theta)-dgg^{-1}-\dot t(J),J)))_{zx}(\xi,y,z)
%\vphantom{\ul{\ul{\ul{\ul{\ul{\ul{g}}}}}}}
\vphantom{\Big]}
\nonumber
\\
&\hphantom{=\,\bigg(}+\widehat{m}\Big(\int_{x_0}^\xi d\xi_0\,f_{I_{\gamma_z}{}^*\theta,I_{\gamma_z}{}^*\varUpsilon|y}(\xi,\xi_0)
\dot t(H^*\varUpsilon_{zx}(\xi_0,y,z))f_{I_{\gamma_z}{}^*\theta,I_{\gamma_z}{}^*\varUpsilon|y}(\xi,\xi_0)^{-1}
\vphantom{\Big]}
\nonumber
\\
&\hphantom{=\,\bigg(+\widehat{m}}-f_{I_{\gamma_z}{}^*\theta,I_{\gamma_z}{}^*\varUpsilon|y}(\xi,x_0)
H^*\theta_z(x_0,y,z))f_{I_{\gamma_z}{}^*\theta,I_{\gamma_z}{}^*\varUpsilon|y}(\xi,x_0)^{-1},
\vphantom{\Big]}
\nonumber
\\
&\hphantom{=\,\bigg(+\widehat{m}\Big(-f_y}H^*(\dot m(g^{-1})(J))_x(\xi,y,z)\Big)\bigg)
\varPsi_{I_{\gamma_z}{}^*g,I_{\gamma_z}{}^*J;I_{\gamma_z}{}^*\theta,I_{\gamma_z}{}^*\varUpsilon|y}(\xi,x_0)^{-1}
\vphantom{\Big]}
\nonumber
\\
&\hphantom{=\,}-\varPsi_{I_{\gamma_z}{}^*g,I_{\gamma_z}{}^*J;I_{\gamma_z}{}^*\theta,I_{\gamma_z}{}^*\varUpsilon|y}(x,x_0)
\dot m(f_{I_{\gamma_z}{}^*\theta,I_{\gamma_z}{}^*\varUpsilon|y}(x,x_0)^{-1})
\vphantom{\Big]}
\nonumber
\\
&\hspace{1.5cm}(H^*(\dot m(g^{-1})(J))_z(x,y,z))
\varPsi_{I_{\gamma_z}{}^*g,I_{\gamma_z}{}^*J;I_{\gamma_z}{}^*\theta,I_{\gamma_z}{}^*\varUpsilon|y}(x,x_0)^{-1}
\vphantom{\Big]}
\nonumber
\\
&\hphantom{=\,}+H^*(\dot m(g^{-1})(J))_z(x_0,y,z),
\vphantom{\Big]}
\nonumber
\end{align}
where $H:\mathbb{R}^3\rightarrow M$ is the mapping defined by 
$H(x,y,z)=I_{\gamma_z}(x,y)=\gamma_z(x)$. 
Under the assumptions made, the $1$--parameter family of surfaces 
$I_{\gamma_z}:\gamma_z\Rightarrow\gamma_z$ is such that $H$ 
is a thin homotopy of $I_{\gamma_0}$, $I_{\gamma_1}$ with the property that
$\rank(dH(x,y,z))\leq 1$. So, 
$H^*(dJ+[J,J]/2
+\widehat{m}(\Ad g(\theta)-dgg^{-1}-\dot t(J),J))=0$ and 
$H^*\varUpsilon_{zx}(x,y,z)=0$. Further, $H^*\theta_z(i,y,z)=0$
and $H^*J_z(i,y,z)=0$ for $i=0,1$, 
by \eqref{path23}, \eqref{path24}. 
So, by \eqref{gauholo13} and \eqref{gauholok1}, we have
\begin{equation}
\partial_zG_{g,J;\theta,\varUpsilon}(\gamma_z)G_{g,J;\theta,\varUpsilon}(\gamma_z)^{-1}=0
\label{gauholok2}
\end{equation}
from which \eqref{gauholo6/2} follows immediately.  \hfill $\Box$

Recall that, for a $(G,H)$--connection doublet $(\theta,\varUpsilon)$, the mappings 
$\bar F_{\theta,\varUpsilon}:\Pi_1M\rightarrow G$, 
$\bar F_{\theta,\varUpsilon}:\Pi_2M\rightarrow H$ induce a $2$--groupoid functor
$\bar F_{\theta,\varUpsilon}:(M,P_1M,P_2M)$ $\rightarrow B_0(G,H)$ 
of the path $2$--groupoid $(M,P_1M,P_2M)$ of $M$ into the 
delooping $2$--groupoid $B_0(G,H)$ of the Lie crossed module $(G,H)$
by their thin homotopy invariance (cf. prop. \ref{prop:twoholo7}). 
Furthermore, when the $(G,H)$--connection doublet $(\theta,\varUpsilon)$ is flat, the 
$\bar F_{\theta,\varUpsilon}$ induce a $2$--groupoid functor
$\bar F^0{}_{\theta,\varUpsilon}:(M,P_1M,P^0{}_2M)\rightarrow B_0(G,H)$ 
of the fundamental $2$--groupoid $(M,P_1M,P^0{}_2M)$ of $M$ into $B_0(G,H)$
by their homotopy invariance (cf. prop. \ref{prop:twoholo8}).
By what found above, the map $G_{g,J;\theta,\varUpsilon}:\Pi_1M\rightarrow H$ factors through one 
$\bar G_{g,J;\theta,\varUpsilon}:P_1M\rightarrow H$ from the path groupoid $1$--cell 
set $P_1M$ into $H$.

\begin{prop} \label{prop:pseudo}
For any $(G,H)$--connection doublet $(\theta,\varUpsilon)$,
a $(G,H)$--$1$--gauge transformation $(g,J)$ encodes a pseudonatural transformation 
$\bar G_{g,J;\theta,\varUpsilon}:\bar F_{\theta,\varUpsilon}\Rightarrow \bar F_{{}^{g,J}\theta,{}^{g,J}\varUpsilon}$
of $2$--functors. 
If $(\theta,\varUpsilon)$ is flat, then $(g,J)$ yields a pseudonatural transformation 
$\bar G^0{}_{g,J;\theta,\varUpsilon}:\bar F^0{}_{\theta,\varUpsilon}\Rightarrow \bar F^0{}_{{}^{g,J}\theta,{}^{g,J}\varUpsilon}$.
\end{prop}

\noindent{\it Proof}. By \eqref{gauholo16/1}, for any curve $\gamma:p_0\rightarrow p_1$
we have a $2$--cell of $B_0(G,H)$
\begin{equation}
\xymatrix@C=5pc@R=2.pc{
{\text{\footnotesize $*$}} \ar[d]_{g(p_1)}  
\ar@{}[dr]^(.13){}="a"^(.88){}="b" \ar@{=>} "a";"b"|-{\tilde G_{g,J;\theta,\varUpsilon}(\gamma)}
& {\text{\footnotesize $*$}}\ar[l]_{F_{\theta,\varUpsilon}(\gamma)} \ar[d]^{g(p_0)}
\\                 
{\text{\footnotesize $*$}} & {\text{\footnotesize $*$}}\ar[l]^{F_{{}^{g,J}\theta,{}^{g,J}\varUpsilon}(\gamma)}            
}
\label{gauholol1}
\end{equation}
where $\tilde G_{g,J;\theta,\varUpsilon}(\gamma)$ is given by 
\begin{equation}
\tilde G_{g,J;\theta,\varUpsilon}=m(g(p_1))(G_{g,J;\theta,\varUpsilon})
\label{gauholol2}
\end{equation}
The $2$--cells \eqref{gauholol1} define a pseudonatural transformation 
$\bar F_{\theta,\varUpsilon}\Rightarrow \bar F_{{}^{g,J}\theta,{}^{g,J}\varUpsilon}$ if 
\vspace{1truemm}
\begin{equation}
\xymatrix@C=5.5pc@R=2.pc{
{\text{\footnotesize $*$}} \ar[d]_{g(p_2)} \ar@{}[dr]^(.13){}="a"^(.88){}="b" \ar@{=>} "a";"b"|-{\tilde G_{g,J;\theta,\varUpsilon}(\gamma_2)}  
& {\text{\footnotesize $*$}} \ar[d]|-{g(p_1)\vphantom{\ul{\dot f}}}
\ar[l]_{F_{\theta,\varUpsilon}(\gamma_2)} \ar@{}[dr]^(.13){}="a"^(.88){}="b" \ar@{=>} "a";"b"|-{\tilde G_{g,J;\theta,\varUpsilon}(\gamma_1)} 
& {\text{\footnotesize $*$}}\ar[l]_{F_{\theta,\varUpsilon}(\gamma_1)} \ar[d]^{g(p_0)}
\\                 
{\text{\footnotesize $*$}} & {\text{\footnotesize $*$}} \ar[l]^{F_{{}^{g,J}\theta,{}^{g,J}\varUpsilon}(\gamma_2)}
& {\text{\footnotesize $*$}}\ar[l]^{F_{{}^{g,J}\theta,{}^{g,J}\varUpsilon}(\gamma_1)}   
}
\,
\xymatrix{~\ar@{}[d]|-{\text{\normalsize $=\vphantom{\dot h}$}}
\\
~}
\,
\xymatrix@C=6.pc@R=2.pc{
{\text{\footnotesize $*$}} \ar[d]_{g(p_2)} 
\ar@{}[dr]^(.13){}="a"^(.88){}="b" \ar@{=>} "a";"b"|-{\tilde G_{g,J;\theta,\varUpsilon}(\gamma_2\circ\gamma_1)}
& {\text{\footnotesize $*$}} \ar[l]_{F_{\theta,\varUpsilon}(\gamma_2\circ \gamma_1)} \ar[d]^{g(p_0)}
\\                 
{\text{\footnotesize $*$}} & {\text{\footnotesize $*$}}\ar[l]^{F_{{}^{g,J}\theta,{}^{g,J}\varUpsilon}(\gamma_2\circ\gamma_1)}          
}
\label{gauholol3}
\end{equation} 
\vskip 0truemm \eject\noindent
for any pair of curves $\gamma_1:p_0\rightarrow p_1$, $\gamma_2:p_1\rightarrow p_2$ and
\begin{equation}
\xymatrix@C=7.pc@R=2.pc{
{\text{\footnotesize $*$}} \ar[d]_{g(p_1)}  \ar@{}[dr]^(.13){}="a"^(.88){}="b" \ar@{=>} "a";"b"|-{\tilde G_{g,J;\theta,\varUpsilon}(\gamma_0)}
& {\text{\footnotesize $*$}}\ar[l]_{F_{\theta,\varUpsilon}(\gamma_0)} \ar[d]^{g(p_0)}
\\                 
{\text{\footnotesize $*$}} & {\text{\footnotesize $*$}}\ar[l]|-{\,\,F_{{}^{g,J}\theta,{}^{g,J}\varUpsilon}(\gamma_0)\,\,}="0"  
\ar@/^2.5pc/[l]^{F_{{}^{g,J}\theta,{}^{g,J}\varUpsilon}(\gamma_1)}="1" \ar@{=>}"0"
+<0ex,-1.75ex>;"1"+<0ex,2.25ex>|-{F_{{}^{g,J}\theta,{}^{g,J}\varUpsilon}(\varSigma)}     
}
\hspace{1mm}
\xymatrix{~\ar@{}[d]|-{\text{\normalsize $=\vphantom{\dot h}$}}
\\
~}
\hspace{1mm}
\xymatrix@C=7.pc@R=2.pc{
{\text{\footnotesize $*$}} \ar[d]_{g(p_1)}  \ar@{}[dr]^(.13){}="a"^(.88){}="b" \ar@{=>} "a";"b"|-{\tilde G_{g,J;\theta,\varUpsilon}(\gamma_1)} 
& {\text{\footnotesize $*$}}\ar[l]|-{\,\,F_{\theta,\varUpsilon}(\gamma_1)\,\,}="0"  
\ar@/_2.5pc/[l]_{F_{\theta,\varUpsilon}(\gamma_0)}="1" \ar@{=>}"1"+<0ex,-2ex>;"0"+<0ex,1.75ex>|-{F_{\theta,\varUpsilon}(\varSigma)}
\ar[d]^{g(p_0)}
\\                 
{\text{\footnotesize $*$}} & {\text{\footnotesize $*$}}\ar[l]^{F_{{}^{g,J}\theta,{}^{g,J}\varUpsilon}(\gamma_1)}         
}
\label{gauholol4}
\end{equation}
for any surface $\varSigma:\gamma_0\rightarrow\gamma_1$ hold, where the diagrams are composed by the usual
pasting algorithm. These conditions are in fact satisfied. \eqref{gauholol3} holds as a consequence of 
\eqref{gauholo17}. \eqref{gauholol4} follows from relation \eqref{gauholo7}. The first part of the 
proposition follows. The proof of the second half is essentially identical.  
\hfill  $\Box$

%\vfil\eject

\subsection{\normalsize \textcolor{blue}{$2$--parallel transport and $2$--gauge transformation}}
\label{sec:twogauholo}

\hspace{.5cm} In this subsection, we shall study $2$--gauge
transformation in higher parallel transport theory. This has no
analogue in ordinary gauge theory. 

Let $M$ be a manifold and $(G,H)$ be a Lie crossed module.

\begin{defi}
A $(G,H)$--$2$--gauge transformation is a mapping $\tilde\varOmega\in \Map(M$, $H)$.
We denote by $\Gau_2(M,G,H)$ the set of all $2$--gauge transformations. 
\end{defi}

$(G,H)$--$2$--gauge transformations act on $(G,H)$--$1$--gauge transformations, the action depending on 
an assigned $(G,H)$--connection doublet.

\begin{defi} Let $(\theta,\varUpsilon)$ be a $(G,H)$--connection doublet, 
$(g,J)$ be a $(G,H)$--$1$--gauge transformation
and $\tilde\varOmega$ a $(G,H)$--$2$--gauge transformation. The $2$--gauge transformed
$1$--gauge transformation $({}^{\tilde \varOmega} g_{|\theta,\varUpsilon},{}^{\tilde \varOmega}J_{|\theta,\varUpsilon})$ is 
\begin{subequations}
\label{twogauholo1,2}
\begin{align}
&{}^{\tilde \varOmega}g_{|\theta,\varUpsilon}=t(\tilde \varOmega )g,
\vphantom{\Big]}
\label{twogauholo1}
\\
&{}^{\tilde \varOmega}J_{|\theta,\varUpsilon}=\tilde \varOmega J \tilde \varOmega ^{-1}-d\tilde \varOmega \tilde \varOmega ^{-1}-
Q({}^{g,J}\theta,\tilde \varOmega ).
\vphantom{\Big]}
\label{twogauholo2}
\end{align}
\end{subequations}
where ${}^{g,J}\theta$ is given by \eqref{gauholo3} and $\tilde \varOmega$ is defined by 
\begin{equation}
\tilde \varOmega =m(g)(\varOmega). 
\label{twogauholo3}
\end{equation}
\end{defi}

$2$--gauge equivalent $1$--gauge transformations yield the the same gauge transformed 
connection doublet. 

\begin{prop}
Let $(\theta,\varUpsilon)$ be a $(G,H)$--connection doublet, $(g,J)$ be a $(G,H)$--$1$--gauge
transformation and $\varOmega$ be a $(G,H)$--$2$--gauge transformation. Then, 
\begin{subequations}
\label{twogauholo4,5}
\begin{align}
&{}^{{}^{\tilde \varOmega}g_{|\theta,\varUpsilon},{}^{\tilde \varOmega}J_{|\theta,\varUpsilon}}\theta={}^{g,J}\theta,
\vphantom{\Big]}
\label{twogauholo4}
\\
&{}^{{}^{\tilde \varOmega}g_{|\theta,\varUpsilon},{}^{\tilde \varOmega}J_{|\theta,\varUpsilon}}\varUpsilon={}^{g,J}\varUpsilon.
\vphantom{\Big]}
\label{twogauholo5}
\end{align}
\end{subequations}
\end{prop}

\noindent{\it Proof}. This is straightforwardly verified evaluating \eqref{gauholo3}, 
\eqref{gauholo4} for the $1$--gauge transformation $({}^{\tilde \varOmega}g_{|\theta,\varUpsilon},{}^{\tilde \varOmega}J_{|\theta,\varUpsilon})$
and using the zero fake curvature condition \eqref{twoholo8}.
\hfill $\Box$

The action of $2$--gauge transformations on $1$--gauge transformations
translates into one on the map
$G_{g,J;\theta,\varUpsilon}:\Pi_1M\rightarrow H$.

\begin{prop}
Let $(\theta,\varUpsilon)$ be a $(G,H)$--connection doublet, $(g,J)$ be a $(G,H)$--$1$--gauge
transformation and $\varOmega$ be a $(G,H)$--$2$--gauge transformation. Then, 
for any curve $\gamma:p_0\rightarrow p_1$, one has 
\begin{equation}
G_{{}^{\tilde \varOmega} g_{|\theta,\varUpsilon},{}^{\tilde \varOmega}J_{|\theta,\varUpsilon};\theta,\varUpsilon}(\gamma)
=\varOmega(p_1)^{-1}G_{g,J;\theta,\varUpsilon}(\gamma)m(F_{\theta,\varUpsilon}(\gamma))(\varOmega(p_0))
\label{twogauholo6}
\end{equation}
where $\varOmega$ is related to $\tilde \varOmega$ by \hphantom{xxxxxxxxxxxxxxxx}
\begin{equation}
\varOmega=m(g^{-1})(\tilde\varOmega). 
\label{twogauholo7}
\end{equation}
\end{prop}

\noindent{\it Proof}. In the course of the proof of prop. \ref{prop:gaugau}, it was found that 
$(\varkappa_{{}^A\kappa,{}^A\varPsi,{}^A\varPhi}$, $\varGamma_{{}^A\kappa,{}^A\varPsi,{}^A\varPhi})
=({}^{\tilde A}\varkappa_{\kappa,\varPsi,\varPhi|a_{f,g,W}B_{f,g,W}},{}^{\tilde A}\varGamma_{\kappa,\varPsi,\varPhi|a_{f,g,W}B_{f,g,W}})$
for any $(G,H)$--cocycle $(f$, $g,W)$, $(f,g,W)$--$1$--gauge transformation $(\kappa,\varPsi,\varPhi)$ 
and $(G,H)$--$2$--gauge transformation $A$, where $A$ and $\tilde A$ are related by \eqref{twogau12}. 
Setting $(f,g,W)=(f_{a,B},g_{a,B}$, $W_{a,B})$ and 
$(\kappa,\varPsi,\varPhi)=(\kappa_{\varkappa,\varGamma,a,B},\varPsi_{\varkappa,\varGamma,a,B},\varPhi_{\varkappa,\varGamma,a,B})$
in this relation, 
where $(a,B)$ and $(\varkappa,\varGamma)$ are a $(G,H)$--connection doublet and a differential $(G,H)$--$1$--gauge
transformation in the sense of defs. \ref{def:r2dghgau} and \ref{def:r2dghgau}, respectively, we find that
\begin{align}
&(\kappa_{{}^{\tilde A}\varkappa_{|a,B},{}^{\tilde A}\varGamma_{|a,B};a,B},\varPsi_{{}^{\tilde A}\varkappa_{|a,B},{}^{\tilde A}\varGamma_{|a,B};a,B},
\varPhi_{{}^{\tilde A}\varkappa_{|a,B},{}^{\tilde A}\varGamma_{|a,B};a,B})
\vphantom{\Big]}
\label{twogauholox1}
\\
&\hspace{5cm}=
({}^A\kappa_{\varkappa,\varGamma;a,B},{}^A\varPsi_{\varkappa,\varGamma;a,B},
{}^A\varPhi_{\varkappa,\varGamma;a,B})
\vphantom{\Big]}
\nonumber
\end{align}
Using the mid component of \eqref{twogauholox1} and the cocycle relation \eqref{twogau2}
and the definitions \eqref{twoholo13} and \eqref{gauholo8,9}, we find
\begin{align}
G_{{}^{\tilde \varOmega} g_{|\theta,\varUpsilon},{}^{\tilde \varOmega}J_{|\theta,\varUpsilon};\theta,\varUpsilon}(\gamma)
&=\varPsi_{I_\gamma{}^*{}^{\tilde \varOmega} g_{|\theta,\varUpsilon},I_\gamma{}^*{}^{\tilde \varOmega}J_{|\theta,\varUpsilon};I_\gamma{}^*\theta,I_\gamma{}^*\varUpsilon|y}(0,1)
\vphantom{\Big]}
\label{twogauholox2}
\\
&=\varPsi_{{}^{I_\gamma{}^*\tilde \varOmega} I_\gamma{}^*g_{|I_\gamma{}^*\theta,I_\gamma{}^*\varUpsilon},
{}^{I_\gamma{}^*\tilde \varOmega}I_\gamma{}^*J_{|I_\gamma{}^*\theta,I_\gamma{}^*\varUpsilon};I_\gamma{}^*\theta,I_\gamma{}^*\varUpsilon|y}(0,1)
\vphantom{\Big]}
\nonumber
\\
&={}^{I_\gamma{}^*\varOmega}\varPsi_{I_\gamma{}^*g,I_\gamma{}^*J;I_\gamma{}^*\theta,I_\gamma{}^*\varUpsilon|y}(0,1)
\vphantom{\Big]}
\nonumber
\\
&=I_\gamma{}^*\varOmega_{|y}(1)^{-1}\varPsi_{I_\gamma{}^*g,I_\gamma{}^*J;I_\gamma{}^*\theta,I_\gamma{}^*\varUpsilon|y}(0,1)
\vphantom{\Big]}
\nonumber
\\
&\hspace{1.5cm}\times m(f_{I_\gamma{}^*\theta,I_\gamma{}^*\varUpsilon|y}(0,1)^{-1})(I_\gamma{}^*\varOmega_{|y}(0))
\vphantom{\Big]}
\nonumber
\\
&=\varOmega(p_1)^{-1}G_{g,J;\theta,\varUpsilon}(\gamma)m(F_{\theta,\varUpsilon}(\gamma))(\varOmega(p_0)).
\vphantom{\Big]}
\nonumber
\end{align}
\eqref{twogauholo6} is so proven  \hfill $\Box$

Recall that, for a $(G,H)$--connection doublet $(\theta,\varUpsilon)$
and a $(G,H)$--$1$--gauge transformation $(g,J)$, the map 
$G_{g,J;\theta,\varUpsilon}:\Pi_1M\rightarrow H$ furnishes the data of a pseudonatural transformation 
$\bar G_{g,J;\theta,\varUpsilon}:\bar F_{\theta,\varUpsilon}\Rightarrow \bar F_{{}^{g,J}\theta,{}^{g,J}\varUpsilon}$ of
the parallel transport $2$-functor $\bar F_{\theta,\varUpsilon}$ of $(\theta,\varUpsilon)$
to that $\bar F_{{}^{g,J}\theta,{}^{g,J}\varUpsilon}$ of $({}^{g,J}\theta,{}^{g,J}\varUpsilon)$
and likewise one $\bar G^0{}_{g,J;\theta,\varUpsilon}$ $:\bar F^0{}_{\theta,\varUpsilon}\Rightarrow \bar F^0{}_{{}^{g,J}\theta,{}^{g,J}\varUpsilon}$
when $(\theta,\varUpsilon)$ is flat (cf. prop. \ref{prop:pseudo}).

\begin{prop} \label{prop:modi}
For every $(G,H)$--connection doublet $(\theta,\varUpsilon)$
and $(G,H)$--$1$-- gauge transformation $(g,J)$, a
$(G,H)$--$2$--gauge transformation $\tilde \varOmega$
encodes a modification $\bar H_{g,J;\theta,\varUpsilon;\tilde\varOmega}:\bar G_{g,J;\theta,\varUpsilon}\Rrightarrow
\bar G_{{}^{\tilde \varOmega} g_{|\theta,\varUpsilon},{}^{\tilde \varOmega}J_{|\theta,\varUpsilon};\theta,\varUpsilon}$ of pseudonatural transformations. 
If $(\theta,\varUpsilon)$ is flat, then $\tilde \varOmega$ yields a pseudonatural transformation modification 
$\bar H^0{}_{g,J;\theta,\varUpsilon;\tilde\varOmega}:\bar G^0{}_{g,J;\theta,\varUpsilon}\Rrightarrow
\bar G^0{}_{{}^{\tilde \varOmega} g_{|\theta,\varUpsilon},{}^{\tilde \varOmega}J_{|\theta,\varUpsilon};\theta,\varUpsilon}$. 
\end{prop}

\noindent{\it Proof}. By \eqref{twogauholo1}, for any point $p$ we
have a $2$--cell of $B_0(G,H)$, 
\begin{equation}
\xymatrix@C=3.5pc@R=2.pc{
   {\text{\footnotesize $*$}}
& {\text{\footnotesize $*$}} \ar@/^1.2pc/[l]^{{}^{\tilde \varOmega} g_{|\theta,\varUpsilon}(p)}="0"
           \ar@/_1.2pc/[l]_{g(p)}="1"
           \ar@{=>}"1"+<0ex,-2.ex>;"0"+<0ex,2.ex>|-{\tilde\varOmega(p)}
}
\label{twogauholoz1}
\end{equation}
$\tilde\varOmega$ defines a modification 
$\bar H_{g,J;\theta,\varUpsilon;\tilde\varOmega}:\bar G_{g,J;\theta,\varUpsilon}\Rrightarrow
G_{{}^{\tilde \varOmega} g_{|\theta,\varUpsilon},{}^{\tilde \varOmega}J_{|\theta,\varUpsilon};\theta,\varUpsilon}$ if 
\begin{align}
&\xymatrix@R=2.5pc@C=8.5pc{
{\text{\footnotesize $*$}} \ar[d]_{g(p_1)}  
\ar@{}[dr]^(.13){}="a"^(.88){}="b" \ar@{=>} "a";"b"|-{\!\!\tilde G_{g,J;\theta,\varUpsilon}(\gamma)}
& {\text{\footnotesize $*$}}\ar[l]_{F_{\theta,\varUpsilon}(\gamma)} \ar[d]|-{g(p_0)\vphantom{fg}}="0"  \ar@/^5.25pc/[d]^{{}^{\tilde \varOmega} g_{|\theta,\varUpsilon}(p_0)}="1"  
\ar@{=>}"0"+<3ex,0ex>;"1"+<-5.3ex,0ex>|-{\tilde \varOmega(p_0)}
\\                 
{\text{\footnotesize $*$}} & {\text{\footnotesize $*$}}\ar[l]^{F_{{}^{g,J}\theta,{}^{g,J}\varUpsilon}(\gamma)}             
}
%\hspace{4cm}
\label{twogauholoz2}
\\
%\hspace{2cm}
\xymatrix{&
\\
~\ar@{}[r]|-{\text{\normalsize $=\vphantom{\ul{\ul{\ul{\ul{\ul{g}}}}}}$}}
&
~}
&\hspace{-4mm}
\xymatrix@R=2.5pc@C=8.5pc{
{\text{\footnotesize $*$}} \ar[d]|-{{}^{\tilde \varOmega} g_{|\theta,\varUpsilon}(p_1)\vphantom{fg}}="0"  \ar@/_5.25pc/[d]_{g(p_1)}="1" 
\ar@{=>}"1"+<3.5ex,0ex>;"0"+<-4.5ex,0ex>|-{\tilde \varOmega(p_1)}
 \ar@{}[dr]^(.13){}="a"^(.88){}="b" \ar@{=>} "a";"b"
|-{\,\,\,\,\,\,\,\,\tilde G_{{}^{\tilde \varOmega} g_{|\theta,\varUpsilon},{}^{\tilde \varOmega}J_{|\theta,\varUpsilon};\theta,\varUpsilon}(\gamma)}
& {\text{\footnotesize $*$}}\ar[l]_{F_{\theta,\varUpsilon}(\gamma)} \ar[d]^{{}^{\tilde \varOmega} g_{|\theta,\varUpsilon}(p_0)}
\\                 
{\text{\footnotesize $*$}} & {\text{\footnotesize $*$}}\ar[l]^{F_{{}^{g,J}\theta,{}^{g,J}\varUpsilon}(\gamma)}           
}
\nonumber
\end{align}
for every curve $\gamma:p_0\rightarrow p_1$, where $\tilde G_{g,J;\theta,\varUpsilon}$ is given in 
\eqref{gauholol2} and similarly $\tilde G_{{}^{\tilde \varOmega} g_{|\theta,\varUpsilon},{}^{\tilde \varOmega}J_{|\theta,\varUpsilon};\theta,\varUpsilon}$
and the diagrams are composed by the usual pasting algorithm. 
This conditions is indeed fulfilled as a consequence of \eqref{twogauholo6}. The first part of the proposition follows. 
The proof of the second half is essentially identical. \hfill $\Box$

%\vfil\eject

\subsection{\normalsize \textcolor{blue}{Smoothness properties of parallel transport}}\label{sec:smooth}

\hspace{.5cm} In this subsection, we shall examine the smoothness properties of the 
parallel transport functors constructed in the preceding sections.

Let $M$ be a manifold and $G$ be a Lie group.

\begin{prop} \label{prop:gsmooth} Let $\theta$ be a $G$--connection. Then, the parallel transport functor
$\bar F_\theta:(M,P_1M)\rightarrow BG$ is smooth in the diffeological sense: if $\gamma_\alpha$
is a family of curves depending smoothly on a set of parameters $\alpha$ varying in a bounded closed domain
$A$ of $\mathbb{R}^d$ for some $d$, then the mapping $\alpha\in A\rightarrow F_\theta(\gamma_\alpha)\in G$ 
is smooth. When the connection $\theta$ is flat, the same property holds for the parallel transport functor
$\bar F^0{}_\theta:(M,P^0{}_1M)\rightarrow BG$. 
\end{prop}

\noindent{\it Proof}. Let $a_\alpha$ be a $G$--connection in the sense of def.
\ref{def:r2gconn} depending smoothly on a set of parameters $\alpha$ varying in a bounded closed domain
$A$ of $\mathbb{R}^d$ for some $d$. Then the $G$--cocycle $f_{a_\alpha}$ given by \eqref{cycle8}
solving the differential problem \eqref{cycle9}, \eqref{cycle10} with $a$ replaced by $a_\alpha$
depends smoothly on $\alpha$ meaning that the mapping $\alpha\in A\rightarrow f_{a_\alpha}(x',x)\in G$
is smooth for any fixed $x,x'\in\mathbb{R}$. 

Let now $\theta$ be a $G$--connection and $\gamma_\alpha$ be a family of curves depending smoothly on $\alpha\in A$.
Then, $\gamma_\alpha{}^*\theta$ is a  $G$--connection in the sense of def. \ref{def:r2gconn} depending smoothly on 
$\alpha$. By \eqref{twoholo1}, then, $\alpha\rightarrow F_\theta(\gamma_\alpha)=f_{\gamma_\alpha{}^*\theta}(1,0)$ is smooth.
The statement follows. The flat case is treated similarly.
\hfill $\Box$

The above results extend straightforwardly to higher parallel transport.
Let $M$ be a manifold and $(G,H)$ be a Lie crossed module.

\begin{prop} \label{prop:ghsmooth} Let $(\theta,\varUpsilon)$ be a $(G,H)$--connection doublet. Then, the parallel transport $2$--functor
$\bar F_{\theta,\varUpsilon}:(M,P_1M,P_2M)\rightarrow B_0(G,H)$ is smooth in the diffeological sense: if 
$\varSigma_\alpha:\gamma_{0\alpha}\Rightarrow \gamma_{1\alpha}$ is a family of surfaces depending smoothly 
on a set of parameters $\alpha$ varying in a bounded closed domain $A$ of $\mathbb{R}^d$ for some $d$, 
then the mappings $\alpha\in A\rightarrow F_{\theta,\varUpsilon}(\gamma_{0\alpha})\in G$,
$\alpha\in A\rightarrow F_{\theta,\varUpsilon}(\gamma_{1\alpha})\in G$,
$\alpha\in A\rightarrow F_{\theta,\varUpsilon}(\varSigma_\alpha)\in H$ are smooth. 
When the connection doublet $(\theta,\varUpsilon)$ is flat, the same property holds 
for the parallel transport functor $\bar F^0{}_{\theta,\varUpsilon}:(M,P_1M,P^0{}_2M)\rightarrow B_0(G,H)$. 
\end{prop}

\noindent{\it Proof}. Let $(a_\alpha,B_\alpha)$ be a $(G,H)$--connection doublet in the sense of def.
\ref{def:r2ghconn} depending smoothly on a set of parameters $\alpha$ varying in a bounded closed domain
$A$ of $\mathbb{R}^d$ for some $d$. Then, the $(G,H)$--cocycle $(f_{a_\alpha,B_\alpha},g_{a_\alpha,B_\alpha},W_{a_\alpha,B_\alpha})$ 
given by \eqref{cycle33,34,35}
solving the differential problem \eqref{cycle36,37,38}, \eqref{cycle39,40,41} with $a$, $B$ replaced by $a_\alpha$, $B_\alpha$
depends smoothly on $\alpha$ meaning that the mapping $\alpha\in A\rightarrow f_{a_\alpha,B_\alpha}(x',x;y)$ $\in G$,
$\alpha\in A\rightarrow g_{a_\alpha,B_\alpha}(x;y',y)\in G$, $\alpha\in A\rightarrow W_{a_\alpha,B_\alpha}(x',x;y',y)\in H$
are all smooth for any fixed $x,x',y,y'\in\mathbb{R}$. 

Let now $(\theta,\varUpsilon)$ be a $(G,H)$--connection doublet and $\varSigma_\alpha:\gamma_{0\alpha}\Rightarrow \gamma_{1\alpha}$ 
be a family of surfaces depending smoothly on $\alpha\in A$. 
Then, $(\varSigma_\alpha{}^*\theta,\varSigma_\alpha{}^*\varUpsilon)$ is a  $(G,H)$--connection doublet 
in the sense of def. \ref{def:r2ghconn} depending smoothly on 
$\alpha$. By \eqref{twoholo9,10,11}, then, $\alpha\rightarrow F_{\theta,\varUpsilon}(\gamma_{0\alpha})
=f_{\varSigma_\alpha{}^*\theta,\varSigma_\alpha{}^*\varUpsilon|0}(1,0)$, $\alpha\rightarrow F_{\theta,\varUpsilon}(\gamma_{1\alpha})
=f_{\varSigma_\alpha{}^*\theta,\varSigma_\alpha{}^*\varUpsilon|1}(1,0)$  and $\alpha\rightarrow F_{\theta,\varUpsilon}(\varSigma_\alpha)
=W_{\varSigma_\alpha{}^*\theta,\varSigma_\alpha{}^*\varUpsilon}(0,1;1,0)$ are smooth.
The statement follows. The flat case is treated similarly. \hfill $\Box$

The above proposition has a counterpart at the level of $1$--gauge transformations. 

\begin{prop} \label{prop:gaughsmooth} Let $(\theta,\varUpsilon)$ be a $(G,H)$--connection doublet and $(g,J)$ a $(G,H)$--$1$--gauge
transformation. Then, the gauge pseudonatural transformation $\bar G_{g,J;\theta,\varUpsilon}:\bar F_{\theta,\varUpsilon}
\Rightarrow \bar F_{{}^{g,J}\theta,{}^{g,J}\varUpsilon}$ is smooth in the diffeological sense: if 
$\gamma_\alpha$ is a family of curves depending smoothly 
on a set of parameters $\alpha$ varying in a bounded closed domain $A$ of $\mathbb{R}^d$ for some $d$, 
then the mapping $\alpha\in A\rightarrow G_{g,J;\theta,\varUpsilon}(\gamma_{0\alpha})\in H$ is  smooth. 
When the connection doublet $(\theta,\varUpsilon)$ is flat, the same property holds 
for the gauge pseudonatural transformation $\bar G^0{}_{g,J;\theta,\varUpsilon}:\bar F^0{}_{\theta,\varUpsilon}
\Rightarrow \bar F^0{}_{{}^{g,J}\theta,{}^{g,J}\varUpsilon}$.
\end{prop}

\noindent{\it Proof}. The statement is proven by a reasoning analogous to that showing
prop. \ref{prop:ghsmooth} relying on the smoothness properties of the solution of 
the differential problem \eqref{gauge22}, \eqref{gauge24} and using \eqref{gauholo13}. 
\hfill $\Box$

%\vfil\eject

\subsection{\normalsize \textcolor{blue}{Relation to other formulations}}\label{sec:other}

\hspace{.5cm} In this subsect, we shall analyze the relation between our formulation of higher 
parallel transport and other formulations appeared in the
literature. This is an important point. 

Let $M$ be a manifold and $(G,H)$ be a Lie crossed module. 
According to Schreiber and Waldorf \cite{Schrei:2009,Schrei:2011,Schrei:2008}, higher parallel transport 
is constructed as follows. 

\begin{defi} Let  $(\theta,\varUpsilon)$ be 
a $(G,H)$--connection. For a curve $\gamma$, the $1$--pa\-rallel transport along $\gamma$ is given by \hphantom{xxxxxxxxxx}
\begin{equation}
F_{SW\theta,\varUpsilon}(\gamma)=f_{SW\theta,\varUpsilon;\gamma}(1),
\label{other1}
\end{equation}
where $f_{SW\theta,\varUpsilon;\gamma}(x)$ is the solution of the differential problem 
\begin{align}
&d_xu(x)u(x)^{-1}=-\gamma^*\theta_x(x),
\vphantom{\Big]}
\label{other2}
\\
&u(0)=1_G
\vphantom{\Big]}
\label{other3}
\end{align}
with $u:\mathbb{R}\rightarrow G$ a smooth mapping.
For a surface $\varSigma$, the $2$--parallel transport along $\varSigma$ is given by \hphantom{xxxxxxxxxxx}
\begin{equation}
F_{SW\theta,\varUpsilon}(\varSigma)=W_{SW\theta,\varUpsilon;\varSigma}(1),
\label{other4}
\end{equation}
where $W_{SW\theta,\varUpsilon;\varSigma}(y)$ is the solution of the differential problem 
\begin{align}
&\partial_yE(y)E(y){}^{-1}
=\int_0^1d\xi\,\dot m(F_{SW\theta,\varUpsilon}(\gamma_{\varSigma \xi,y}))\varSigma^*\varUpsilon_{xy}(\xi,y),
\vphantom{\Big]}
\label{other5}
\\
&E(0)=1_H
\vphantom{\Big]}
\label{other6}
\end{align}
with $E:\mathbb{R}\rightarrow H$ a smooth mapping.
Here, $\gamma_{\varSigma \xi,y}:\varSigma(\xi,y)\rightarrow
\varSigma(1,y)$ is the curve defined by the expression 
\begin{equation}
\gamma_{\varSigma \xi,y}(x)=\varSigma(\xi+(1-\xi)\varphi(x),y),
\label{other7}
\end{equation}
where $\varphi:\mathbb{R}\rightarrow\mathbb{R}$ is a smooth function such that $\varphi(x)=0$ for $x<\epsilon$ and
$\varphi(x)=1$ for $x>1-\epsilon$ for some small $\epsilon>0$. 
\end{defi}

The function $\varphi$ is introduced to ensure that $\gamma_{\varSigma  \xi,y}$ 
has sitting instants. Its choice is immaterial, as a change of 
it amounts to a thin homotopy that leaves $F_{SW\theta,\varUpsilon}(\gamma_{\varSigma \xi,y})$ invariant. 
The following proposition holds. 

\begin{prop} For any curve $\gamma$,
\begin{equation}
F_{SW\theta,\varUpsilon}(\gamma)=F_{\theta,\varUpsilon}(\gamma).
\label{other8}
\end{equation}
Similarly, for any surface $\varSigma$,
\begin{equation}
F_{SW\theta,\varUpsilon}(\varSigma)=F_{\theta,\varUpsilon}(\varSigma).
\label{other9}
\end{equation}
\end{prop}

\noindent{\it Proof}. We show first \eqref{other8}. As $\gamma^*\theta_x(x)=I_\gamma{}^*\theta_x(x,y)$ for any $y$, 
the differential problem \eqref{other2}, \eqref{other3} is identical to that  \eqref{cycle36}, \eqref{cycle39} with 
$a_x(x,y)=I_\gamma{}^*\theta_x(x,y)$  %$(a,B)=(I_\gamma{}^*\theta,I_\gamma{}^*\varUpsilon)$ 
and $x_0=0$, which is solved precisely by $f_{I_\gamma{}^*\theta,I_\gamma{}^*\varUpsilon|y}(x,0)$. So, 
\begin{equation}
f_{SW\theta,\varUpsilon;\gamma}(x)=f_{I_\gamma{}^*\theta,I_\gamma{}^*\varUpsilon|y}(x,0).
\label{othera1}
\end{equation}
\eqref{other8} then follows from \eqref{other1} and \eqref{twoholo13}. 

The proof of \eqref{other9} requires more work but follows a similar route.
We begin with noticing that  
$f_{SW\theta,\varUpsilon;\gamma_{\varSigma \xi,y}}(x)$ is the solution of the differential problem 
\eqref{other2}, \eqref{other3} with $\gamma=\gamma_{\varSigma \xi,y}$. Since
\begin{equation}
\gamma_{\varSigma \xi,y}{}^*\theta_x(x)=(1-\xi)d_x\varphi(x)\varSigma^*\theta_x(\xi+(1-\xi)\varphi(x),y)
\label{othera2}
\end{equation}
by \eqref{other7}, the differential problem can thus more explicitly be stated as 
\begin{align}
&d_xu(x)u(x)^{-1}=-(1-\xi)d_x\varphi(x)\varSigma^*\theta_x(\xi+(1-\xi)\varphi(x),y),
\vphantom{\Big]}
\label{othera3}
\\
&u(0)=1_G.
\vphantom{\Big]}
\label{othera4}
\end{align}
Comparing this with the differential problem 
\eqref{cycle36}, \eqref{cycle39} with 
$a_x(x,y)=\varSigma^*\theta_x(x,y)$ and $x_0=\xi$, solved by
$f_{\varSigma^*\theta,\varSigma^*\varUpsilon|y}(x,\xi)$, we find that 
\begin{equation}
f_{SW\theta,\varUpsilon;\gamma_{\varSigma \xi,y}}(x)=f_{\varSigma^*\theta,\varSigma^*\varUpsilon|y}(\xi+(1-\xi)\varphi(x),\xi).
\vphantom{\ul{\ul{\ul{\ul{\ul{g}}}}}}
\label{othera5}
\end{equation}
From \eqref{other1} with $\gamma=\gamma_{\varSigma \xi,y}$, it follows that 
\begin{equation}
F_{SW\theta,\varUpsilon}(\gamma_{\varSigma \xi,y})%=f_{SW\theta,\varUpsilon;\gamma_{\varSigma \xi,y}}(1)
=f_{\varSigma^*\theta,\varSigma^*\varUpsilon|y}(\xi,1)^{-1}.
\label{othera6}
\end{equation}
Recalling \eqref{twoholob1}, we also have that 
\begin{equation}
1_G=g_{\varSigma^*\theta,\varSigma^*\varUpsilon|1}(y,0)^{-1}.
\label{othera7}
\end{equation}
Taking \eqref{othera6}, \eqref{othera7} into account, we can recast the differential problem 
\eqref{other5}, \eqref{other6} in the form 
\begin{align}
&\partial_yE(y)E(y){}^{-1}
\vphantom{\Big]}
\label{othera8}
\\
&\hspace{2cm}
=\int_0^1d\xi\,\dot m(g_{\varSigma^*\theta,\varSigma^*\varUpsilon|1}(y,0)^{-1}
f_{\varSigma^*\theta,\varSigma^*\varUpsilon|y}(\xi,1)^{-1})\varSigma^*\varUpsilon_{xy}(\xi,y),
\vphantom{\Big]}
\nonumber
\\
&E(0)=1_H.
\vphantom{\Big]}
\label{othera9}
\end{align}
This is equivalent to the first form of the differential problem 
\eqref{cycle38}, \eqref{cycle41} with 
$v_{|x_0,y_0}(y)=g_{\varSigma^*\theta,\varSigma^*\varUpsilon|x_0}(y,y_0)$,
$u_{|y,x_0}(x)=f_{\varSigma^*\theta,\varSigma^*\varUpsilon|y}(x,x_0)$
and $B_{xy}(x,y)=\varSigma^*\varUpsilon_{xy}(x,y)$ after integrating with respect to $x$ and 
setting $x=0$, $x_0=1$ and $y_0=0$. From here, it follows that 
\begin{equation}
W_{SW\theta,\varUpsilon;\varSigma}(y)=W_{\varSigma^*\theta,\varSigma^*\varUpsilon|0,1}(y,0).
\label{othera10}
\end{equation}
\eqref{other9} then follows from \eqref{other4} and \eqref{twoholo11}. 
\hfill $\Box$ 

The prescription given by Martins and Picken in
\cite{Martins:2007,Martins:2008}  
for the computation of
higher parallel transport is essentially equivalent to that of
Schreiber and Waldorf and, consequently, to ours.

\vfil\eject

\appendix

\section{\normalsize \textcolor{blue}{Double categories}}\label{sec:dcat}

\hspace{.5cm} In this appendix, we present the basic notions and
results of double category theory, which is required by our cocycle 
based formulation of parallel transport theory. Most of the material
is not original and is included to help the reader. (See for instance \cite{Fiore:2007aa}.)
However, to the
best of our knowledge, the notions of double natural transformation
and modification we present and use in the main body of the paper are 
original. We also define the plane rectangle double groupoid
playing an essential role in our construction and recall the definition
of the edge symmetric double  groupoid of a crossed module
for its relevance. 

%\vfil\eject

\subsection{\normalsize \textcolor{blue}{Double categories}}\label{sec:dcdef}

\hspace{.5cm} Double categories are categories internal to the
category of categories \cite{Ehresmann:1963aa}.
They are however more conveniently defined as follows. 

\begin{defi} 
A double category $D$ consists of the following elements
\begin{enumerate}

\item A set of objects $a,~b,~c,~ \dots$.

\item For each pair of objects $a$, $b$ a set of horizontal arrows and one of vertical arrows,
\vspace{-.3cm}
\begin{equation}
\vbox{
\xymatrix{{\text{\footnotesize $b$}}&{\text{\footnotesize $a$}}\ar[l]_x}
\vspace{-.75cm}}
\qquad
\xymatrix{{\text{\footnotesize $b$}}
\\
{\text{\footnotesize $a$}}\ar[u]_x}
\label{dcdef1}
\end{equation}
\vspace{-.3cm}  

\item For each quadruple of objects $a,~b,~c,~ d$, pair of horizontal arrows
$\xymatrix{{\text{\footnotesize $b$}}&{\text{\footnotesize $a$}}\ar[l]_y}\!,$ 
$\xymatrix{{\text{\footnotesize $d$}}&{\text{\footnotesize $c$}}\ar[l]_u}$
and pair of vertical arrows 
$\xymatrix{{\text{\footnotesize $c$}}&{\text{\footnotesize $a$}}\ar[l]_x}\!,$
$\xymatrix{{\text{\footnotesize $d$}}&{\text{\footnotesize $b$}}\ar[l]_v}$ (here written horizontally for
convenience), 
a set of arrow squares
\begin{equation}
\xymatrix{
{\text{\footnotesize $d$}}  & {\text{\footnotesize $c$}}\ar[l]_u  \ar@{}[dl]^(.25){}="a"^(.75){}="b" \ar@{=>} "a";"b"_X %|-{X}
\\                 
{\text{\footnotesize $b$}} \ar[u]^v & {\text{\footnotesize $a$}}\ar[u]_x\ar[l]^y             
}
\label{dcdef2}
\end{equation}

\end{enumerate}
Objects and horizontal arrows form an ordinary category with composition $\circ_h$ and identity assigning map $\id_h$.
Similarly, objects and vertical arrows form a category with composition $\circ_v$ and identity assigning map 
$\id_v$. Furthermore, arrow squares can be composed both horizontally and vertically compatibly 
with the composition of horizontal and vertical arrows, 
\begin{equation}
\xymatrix{
{\text{\footnotesize $f$}} & {\text{\footnotesize $e$}} \ar[l]_v \ar@{}[dl]^(.25){}="a"^(.75){}="b" \ar@{=>} "a";"b"_Y  
& {\text{\footnotesize $d$}}\ar[l]_u  \ar@{}[dl]^(.25){}="a"^(.75){}="b" \ar@{=>} "a";"b"_X %|-{X}
\\                 
{\text{\footnotesize $c$}} \ar[u]^t  & {\text{\footnotesize $b$}} \ar[l]^y \ar[u]|-{s\vphantom{\ul{\dot f}}} 
& {\text{\footnotesize $a$}} \ar[u]_r\ar[l]^x             
}
\quad
\xymatrix{~\ar@{}[d]|-{\text{\normalsize $=\vphantom{\dot h}$}}
\\
~}
\quad
\xymatrix{
{\text{\footnotesize $f$}}  && {\text{\footnotesize $d$}} \ar[ll]_{v\circ_hu}  
\ar@{}[dll]^(.25){}="a"^(.75){}="b" \ar@{=>} "a";"b"_{Y\circ_hX} %|-{X}
\\                 
{\text{\footnotesize $c$}} \ar[u]^t && {\text{\footnotesize $a$}}\ar[u]_r\ar[ll]^{y\circ_hx}             
}
\label{dcdef3}
\end{equation}
\begin{equation}
\hspace{.85cm}\xymatrix{
{\text{\footnotesize $f$}} & {\text{\footnotesize $e$}} \ar[l]_z 
\ar@{}[dl]^(.25){}="a"^(.75){}="b" \ar@{=>} "a";"b"_Y %|-{X}
\\
{\text{\footnotesize $d$}} \ar[u]^t & {\text{\footnotesize $c$}} \ar[l]|-{\,\,y\,\,} \ar[u]_r 
\ar@{}[dl]^(.25){}="a"^(.75){}="b" \ar@{=>} "a";"b"_X %|-{X}
\\                 
{\text{\footnotesize $b$}} \ar[u]^s & {\text{\footnotesize $a$}} \ar[u]_q\ar[l]^x             
}
\quad
\xymatrix{~\ar@{}[d]|-{\text{\normalsize $=\vphantom{\dot h}$}}
\\
~}
\quad
\xymatrix{
{\text{\footnotesize $f$}}  && {\text{\footnotesize $e$}} \ar[ll]_z  
\ar@{}[dll]^(.25){}="a"^(.75){}="b" \ar@{=>} "a";"b"_{Y\circ_v X} %|-{X}
\\                 
{\text{\footnotesize $b$}} \ar[u]^{t\circ_v s} && {\text{\footnotesize $a$}}\ar[u]_{r\circ_v q}\ar[ll]^x             
}
\nonumber%\label{dcdef4}
\end{equation}
%\vfil\eject\noindent
Compatible horizontal and vertical identity arrow squares are also defined,
\begin{equation}
\xymatrix{
{\text{\footnotesize $b$}}  && {\text{\footnotesize $b$}}\ar[ll]_{\id_{hb}}  
\ar@{}[dll]^(.25){}="a"^(.75){}="b" \ar@{=>} "a";"b"_{\Id_{hx}} %|-{X}
\\                 
{\text{\footnotesize $a$}} \ar[u]^x && {\text{\footnotesize $a$}}\ar[u]_x\ar[ll]^{\id_{ha}}             
}
\quad 
\xymatrix{
{\text{\footnotesize $b$}}  && {\text{\footnotesize $a$}}\ar[ll]_x  
\ar@{}[dll]^(.25){}="a"^(.75){}="b" \ar@{=>} "a";"b"_{\Id_{vx}} %|-{X}
\\                 
{\text{\footnotesize $b$}} \ar[u]^{\id_{vb}} && {\text{\footnotesize $a$}}\ar[u]_{\id_{va}}\ar[ll]^x             
}
\label{dcdef5}
\end{equation}
Vertical arrows and arrow squares connecting them form an ordinary category with composition $\circ_h$
and identity assigning map $\Id_h$. Similarly, horizontal arrows and arrow squares form a category 
with composition $\circ_v$ and identity assigning map $\Id_v$.
Finally the exchange law holds, which means that the result of the composition
of the four arrow squares of the form \hphantom{xxxxxxxxxxx}
\begin{equation}
\xymatrix{
{\text{\footnotesize $i$}} & {\text{\footnotesize $h$}} \ar[l]_v 
\ar@{}[dl]^(.25){}="a"^(.75){}="b" \ar@{=>} "a";"b"_U  
& {\text{\footnotesize $g$}} \ar[l]_u  
\ar@{}[dl]^(.25){}="a"^(.75){}="b" \ar@{=>} "a";"b"_Z %|-{X}
\\
{\text{\footnotesize $f$}} \ar[u]^s 
& {\text{\footnotesize $e$}} \ar[l]|-{\,\,w\,\,} \ar[u]|-{q\vphantom{\ul{\dot q}}}  
\ar@{}[dl]^(.25){}="a"^(.75){}="b" \ar@{=>} "a";"b"_Y  
& {\text{\footnotesize $d$}}\ar[l]|-{\,\,z\,\,} \ar[u]_n  
\ar@{}[dl]^(.25){}="a"^(.75){}="b" \ar@{=>} "a";"b"_X %|-{X}
\\                 
{\text{\footnotesize $c$}} \ar[u]^r  & {\text{\footnotesize $b$}}  \ar[l]^y \ar[u]|-{p\vphantom{\ul{\dot p}}}  
& {\text{\footnotesize $a$}}\ar[u]_m\ar[l]^x             
}
\label{dcdef6}
\end{equation}
does not depend on whether the horizontal composition of the bottom and top pairs of squares 
or the vertical composition of the right and left pairs of squares is performed first.
\end{defi}

The transpose of a double category $D$, which switches the vertical and horizontal arrows, 
is again a double category $TD$. 

\begin{defi} A double groupoid $D$ is a double category in which the horizontal and vertical arrow
categories are groupoid with inverse operations ${}^{-1_h}$, ${}^{-1_v}$, respectively, 
and each arrow square has an horizontal and vertical inverse compatible with the arrow inversions
\begin{equation}
\xymatrix{
{\text{\footnotesize $c$}}  && {\text{\footnotesize $d$}}\ar[ll]_{u^{-1_h}}  
\ar@{}[dll]^(.25){}="a"^(.75){}="b" \ar@{=>} "a";"b"_{X^{-1_h}} %|-{X}
\\                 
{\text{\footnotesize $a$}} \ar[u]^x && {\text{\footnotesize $b$}}\ar[u]_v\ar[ll]^{y^{-1_h}}             
}
\quad
\xymatrix{
{\text{\footnotesize $b$}}  && {\text{\footnotesize $a$}}\ar[ll]_y  
\ar@{}[dll]^(.25){}="a"^(.75){}="b" \ar@{=>} "a";"b"_{X^{-1_v}} %|-{X}
\\                 
{\text{\footnotesize $d$}} \ar[u]^{v^{-1_v}} && {\text{\footnotesize $c$}}\ar[u]_{x^{-1_v}}\ar[ll]^u             
}
\label{dcdef7}
\end{equation}
\end{defi}
%such that the categories $D_h$ and $D_v$ are also groupoids. 

%\vfil\eject

\subsection{\normalsize \textcolor{blue}{Double functors}}\label{sec:dcfnctr}

\hspace{.5cm} Double functors are structure preserving maps of double
categories. 

Let $D$, $E$ be double categories.

\begin{defi}
A double category functor $F:D\rightarrow E$ consists of the following elements
\begin{enumerate}

\item 
A mapping $\xymatrix{{\text{\footnotesize $a$}}\ar@{|->}[r]& {\text{\footnotesize $F(a)$}}}$ %$a\mapsto F(a)$ 
of the set of objects of $D$ into that of $E$. 

\item Mappings \hphantom{xxxxxxxx}
\vspace{-.3cm}
\begin{equation}
\vbox{
\xymatrix{{\text{\footnotesize $b$}}&{\text{\footnotesize $a$}}\ar[l]_x}
\vspace{-.75cm}}
\quad
\vbox{
\xymatrix{
\ar@{|->}[r]&
}
\vspace{-.74cm}}
\quad
\vbox{
\xymatrix{{\text{\footnotesize $F(b)$}}&{\text{\footnotesize $F(a)$}}\ar[l]_{F(x)}}
\vspace{-.75cm}}
\qquad
\xymatrix{{\text{\footnotesize $b$}}
\\
{\text{\footnotesize $a$}}\ar[u]_x}
\quad
\vbox{
\xymatrix{
\ar@{|->}[r]&
}
\vspace{-.74cm}}
\quad
\xymatrix{{\text{\footnotesize $F(b)$}}
\\
{\text{\footnotesize $F(a)$}}\ar[u]_{F(x)}}
\label{dcfnctr1}
\end{equation}
of the sets of horizontal and vertical arrows of $D$ into those of $E$, respectively,
compatible with the mapping of objects.

\item A mapping \hphantom{xxxxxxxx}
\begin{equation}
\xymatrix{
{\text{\footnotesize $d$}}  & {\text{\footnotesize $c$}}\ar[l]_u  
\ar@{}[dl]^(.25){}="a"^(.75){}="b" \ar@{=>} "a";"b"_X %|-{X}
\\                 
{\text{\footnotesize $b$}} \ar[u]^v & {\text{\footnotesize $a$}}\ar[u]_x\ar[l]^y             
}
\quad
\vbox{
\xymatrix{
\ar@{|->}[r]&
}
\vspace{-.74cm}}
\quad
\xymatrix{
{\text{\footnotesize $F(d)$}}  && {\text{\footnotesize $F(c)$}}\ar[ll]_{F(u)}  
\ar@{}[dll]^(.25){}="a"^(.75){}="b" \ar@{=>} "a";"b"_{F(X)} %|-{X}
\\                 
{\text{\footnotesize $F(b)$}} \ar[u]^{F(v)} && {\text{\footnotesize $F(a)$}}\ar[u]_{F(x)}\ar[ll]^{F(y)}             
}
\label{dcfnctr2}
\end{equation}
of the set of arrow squares of $D$ into that of $E$ compatible with the mappings of objects 
and arrows.
\end{enumerate}
These mappings are required to preserve all types of compositions and units.
\end{defi}

Let $D$, $E$ be double groupoids.

\begin{defi}
A double groupoid functor functor $F:D\rightarrow E$  is a double category functor
that preserves all types of inverses. 
\end{defi}

\begin{prop} Small double categories and double functors with the obvious composition 
and identity assigning map constitute a category. Small double groupoids and double functors form 
a full subcategory of it. 
\end{prop}

%\vfil\eject

\subsection{\normalsize \textcolor{blue}{Edge $2$--categories of double categories}}\label{sec:dcedge}

%Let $D$ be a double category. % or a double groupoid. 

\hspace{.5cm} Edge categories are $2$--categories canonically associated with 
double categories playing an important role in many double categorical constructions. 

\begin{prop} With a double category $D$ there are associated two strict $2$--categories $H\!D$ and $V\!D$,
called edge $2$--categories of $D$.

The $2$--category $H\!D$ is defined as follows.

\begin{enumerate}

\item The $0$--cells of $H\!D$ are the objects of $D$.

\item The $1$--cells of $H\!D$ are the horizontal arrows of $D$.
%\begin{equation}
%\xymatrix{{\text{\footnotesize $b$}}&{\text{\footnotesize $a$}}\ar[l]_x}\!.
%\label{dcedge}
%\end{equation}

\item The $2$--cells of $H\!D$ are the arrow squares of $D$ of the
  form \pagebreak 
\begin{equation}
\vbox{
\xymatrix@C=3pc{
   {\text{\footnotesize $b$}}
& {\text{\footnotesize $a$}} \ar@/^1pc/[l]^x="0"
           \ar@/_1pc/[l]_y="1"
           \ar@{=>}"1"+<0ex,-2.ex>;"0"+<0ex,2.ex>_X
}
\vspace{-1.3cm}}
\quad
\xymatrix{~\ar@{}[d]|-{\text{\normalsize $\equiv\vphantom{\dot h}$}}
\\
~}
\quad
\xymatrix{
{\text{\footnotesize $b$}}  & {\text{\footnotesize $a$}}\ar[l]_y  
\ar@{}[dl]^(.25){}="a"^(.75){}="b" \ar@{=>} "a";"b"_X %|-{X}
\\                 
{\text{\footnotesize $b$}} \ar[u]^{\id_{vb}} & {\text{\footnotesize $a$}}\ar[u]_{\id_{va}}\ar[l]^x            
}
\label{dcedge1}
\end{equation}
\end{enumerate}
The composition of two $1$--cells of $H\!D$ is the composition of the corresponding horizontal arrows of $D$. 
The identity $1$--cells of $H\!D$ are the horizontal identity arrows of $D$.
The horizontal composition of two $2$--cells of $H\!D$ is the horizontal composition of the corresponding arrow squares 
of $D$. The vertical composition of two $2$--cells of $H\!D$ is the vertical composition of the corresponding arrow squares 
of $D$. The unit $2$--cells of $H\!D$ are the vertical unit squares of $D$. 

The $2$--category $V\!D$ is defined as follows.
\begin{enumerate}

\item The $0$--cells of $V\!D$ are the objects of $D$.

\item The $1$--cells of $V\!D$ are the vertical arrows of $D$.

\item The $2$--cells of $V\!D$ are the arrow squares of $D$ of the form 
\begin{equation}
\vbox{
\xymatrix@C=3pc{
   {\text{\footnotesize $b$}}
& {\text{\footnotesize $a$}} \ar@/^1pc/[l]^y="0"
           \ar@/_1pc/[l]_x="1"
           \ar@{=>}"1"+<0ex,-2.ex>;"0"+<0ex,2.ex>_X
}
\vspace{-1.3cm}}
%\xymatrix{
%{\text{\footnotesize $b$}}
%\\
%{\text{\footnotesize $a$}}\ar@/^1pc/[u]^y="1"
%          \ar@/_1pc/[u]_x="0"
% \ar@{=>}"0"+<-2.ex,0.ex>;"1"+<2.ex,0.ex>_X
%}
\quad
\xymatrix{~\ar@{}[d]|-{\text{\normalsize $\equiv\vphantom{\dot h}$}}
\\
~}
\quad
\xymatrix{
{\text{\footnotesize $b$}}  & {\text{\footnotesize $b$}}\ar[l]_{\id_{hb}}  
\ar@{}[dl]^(.25){}="a"^(.75){}="b" \ar@{=>} "a";"b"_X %|-{X}
\\                 
{\text{\footnotesize $a$}} \ar[u]^y & {\text{\footnotesize $a$}}\ar[u]_x\ar[l]^{\id_{ha}}             
}
\label{dcedge2}
\end{equation}
\end{enumerate} 
The composition of two $1$--cells of $V\!D$ is the composition of the corresponding vertical arrows of $D$. 
The identity $1$--cells of $V\!D$ are the vertical identity arrows of $D$.
The horizontal composition of two $2$--cells of $V\!D$ is the vertical composition of the corresponding arrow squares 
of $D$. The vertical composition of two $2$--cells of $V\!D$ is the horizontal composition of the corresponding arrow squares 
of $D$. The unit $2$--cells of $V\!D$ are the horizontal unit squares of $D$. 
\end{prop}

We denote by $H\!D_0$ and $V\!D_0$ the ordinary categories underlying $H\!D$ and $V\!D$. 
$H\!D_0$ is the category whose $0$-- and $1$ cells are the objects and horizontal arrows of 
$D$ with the composition $\circ_h$ and identity assigning map $\id_h$ inherited from $D$. 
Similarly, $V\!D_0$ is the category whose $0$-- and $1$ cells are the objects and vertical arrows of 
$D$ with the composition $\circ_v$ and identity assigning map $\id_v$ inherited from $D$. 

\begin{prop}
If $D$ is a double groupoid, then $H\!D$ and $V\!D$ are $2$--groupoids. 

The inverse of a $1$--cell of $H\!D$ is the inverse of the corresponding horizontal arrow of $D$. 
The horizontal inverse of a $2$--cell of $H\!D$ is the horizontal inverse of the corresponding arrow square
of $D$. The vertical inverse of a $2$--cell of $H\!D$ is the vertical inverse of the corresponding arrow square
of $D$.

The inverse of a $1$--cell of $V\!D$ is the inverse of the corresponding vertical arrow of $D$. 
The horizontal inverse of a $2$--cell of $V\!D$ is the vertical inverse of the corresponding arrow square
of $D$. The vertical inverse of a $2$--cell of $V\!D$ is the horizontal inverse of the corresponding arrow square
of $D$.
\end{prop} 

In such a case, $H\!D_0$ and $V\!D_0$ are ordinary groupoids. 

\begin{defi}
A double category $D$ is said edge symmetric if there is an invertible $2$--functor
$S:V\!D\rightarrow H\!D$. Similarly, for a double groupoid $D$. 
\end{defi}

$S$ induces an invertible functor $S_0:V\!D_0\rightarrow H\!D_0$.

\begin{prop}
A double functor $F:D\rightarrow E$ of %%\pagebreak 
two double categories or groupoids $D$, $E$ induces 
strict $2$--functors
$H\!F:H\!D\rightarrow H\!E$, $V\!F:V\!D\rightarrow V\!E$ of the associated horizontal and vertical $2$--categories
or $2$--groupoids $H\!D$, $H\!E$ and  $V\!D$, $V\!E$, respectively.
\end{prop}

The edge $2$--categories of double categories enter in an essential
way in the definition of the notion of folding.

%\vfil\eject

\subsection{\normalsize \textcolor{blue}{Folding of edge symmetric double categories}}\label{sec:dcedsym}

\hspace{.5cm}
Let $D$ be an edge symmetric double category or a double groupoid. Then, as we explained in
subapp. \ref{sec:dcedge}, we have an invertible functor of $V\!D_0$ into $H\!D_0$,\pagebreak 
\vspace{-.35cm}
\begin{equation}
\xymatrix{
{\text{\footnotesize $b$}}
\\
{\text{\footnotesize $a$}}\ar[u]_x
}
\hspace{.4cm}
\vbox{
\xymatrix{\ar@{|->}[r]
&
}
\vspace{-.75cm}}
\hspace{.4cm}
\vbox{
\xymatrix{
\text{\footnotesize $b$}&{\text{\footnotesize $a$}}\ar[l]_{\tilde x}
}
\vspace{-.75cm}}
\label{dcedsym1}
\end{equation}

\begin{defi} A horizontal folding of $D$  consists of a single datum.
\begin{enumerate}

\item A mapping of the set arrow squares of $D$ into that of $2$--cells of $H\!D$ 
\begin{equation}
\xymatrix{
{\text{\footnotesize $d$}}  & {\text{\footnotesize $c$}}\ar[l]_u  \ar@{}[dl]^(.25){}="a"^(.75){}="b" \ar@{=>} "a";"b"_X %|-{X}
\\                 
{\text{\footnotesize $b$}} \ar[u]^v & {\text{\footnotesize $a$}}\ar[u]_x\ar[l]^y             
}
\hspace{.4cm}
\vbox{
\xymatrix{\ar@{|->}[r]
&
}
\vspace{-.75cm}}
\hspace{.4cm}
\xymatrix@C=3pc@R=2.15pc{
{\text{\footnotesize $d$}}  & {\text{\footnotesize $a$}}\ar[l]_{u\circ_h \tilde x}  
\ar@{}[dl]^(.25){}="a"^(.75){}="b" \ar@{=>} "a";"b"_{\tilde X} %|-{X}
\\                 
{\text{\footnotesize $d$}} \ar[u]^{\id_{vd}} & {\text{\footnotesize $a$}}\ar[u]_{\id_{va}}\ar[l]^{\tilde v\circ_h y}            
}
\label{dcedsym2}
\end{equation}
\end{enumerate}
The following axioms
\begin{equation}
\xymatrix{
{\text{\footnotesize $f$}} & {\text{\footnotesize $e$}} \ar[l]_v 
\ar@{}[dl]^(.25){}="a"^(.75){}="b" \ar@{=>} "a";"b"_Y  
& {\text{\footnotesize $d$}}\ar[l]_u  \ar@{}[dl]^(.25){}="a"^(.75){}="b" \ar@{=>} "a";"b"_X %|-{X}
\\                 
{\text{\footnotesize $c$}} \ar[u]^t  & {\text{\footnotesize $b$}} \ar[l]^y \ar[u]|-{s\vphantom{\ul{\dot f}}} 
& {\text{\footnotesize $a$}} \ar[u]_r\ar[l]^x             
}
\quad
\vbox{
\xymatrix{
\ar@{|->}[r]&
}
\vspace{-.74cm}}
\quad
\xymatrix{
{\text{\footnotesize $f$}}  &&& {\text{\footnotesize $a$}}\ar[lll]_{v\circ_h u\circ_h \tilde r}  
\ar@{}[dlll]^(.25){}="a"^(.75){}="b" \ar@{=>} "a";"b"_{\Id_{vv}\circ_h \tilde X\,\,\,\,\,\,\,} %|-{X}
\\
{\text{\footnotesize $f$}} \ar[u]^{\id_{vf}} &&& 
{\text{\footnotesize $a$}}\ar[lll]|-{\,\,v\circ_h \tilde s\circ_h x\,\,} \ar[u]_{\id_{va}} 
\ar@{}[dlll]^(.25){}="a"^(.75){}="b" \ar@{=>} "a";"b"_{\tilde Y\circ_h \Id_{vx}\,\,\,\,\,\,\,} %|-{X}
\\                 
{\text{\footnotesize $f$}}  \ar[u]^{\id_{vf}} &&& {\text{\footnotesize $a$}}\ar[u]_{\id_{va}}\ar[lll]^{\tilde t\circ_h y\circ_h x}             
}
\label{dcedsym3}
\end{equation}
\begin{equation}
\xymatrix{
{\text{\footnotesize $f$}} & {\text{\footnotesize $e$}} \ar[l]_z \ar@{}[dl]^(.25){}="a"^(.75){}="b" \ar@{=>} "a";"b"_Y %|-{X}
\\
{\text{\footnotesize $d$}} \ar[u]^t & {\text{\footnotesize $c$}} \ar[l]|-{\,\,y\,\,} \ar[u]_r 
\ar@{}[dl]^(.25){}="a"^(.75){}="b" \ar@{=>} "a";"b"_X %|-{X}
\\                 
{\text{\footnotesize $b$}} \ar[u]^s & {\text{\footnotesize $a$}} \ar[u]_q\ar[l]^x             
}
\quad
\vbox{
\xymatrix{
\ar@{|->}[r]&
}
\vspace{-1.4cm}}
\quad
\xymatrix{
{\text{\footnotesize $f$}}  &&& {\text{\footnotesize $a$}}\ar[lll]_{z\circ_h \tilde r\circ_h \tilde q}  
\ar@{}[dlll]^(.25){}="a"^(.75){}="b" \ar@{=>} "a";"b"_{\tilde Y\circ_h \Id_{v\tilde q}\,\,\,\,\,\,\,} %|-{X}
\\
{\text{\footnotesize $f$}} \ar[u]^{\id_{vf}} &&& 
{\text{\footnotesize $a$}}\ar[lll]|-{\,\,\tilde t\circ_h y\circ_h \tilde q\,\,} \ar[u]_{\id_{va}} 
\ar@{}[dlll]^(.25){}="a"^(.75){}="b" \ar@{=>} "a";"b"_{\Id_{v\tilde t}\circ_h \tilde X\,\,\,\,\,\,\,} %|-{X}
\\                 
{\text{\footnotesize $f$}}  \ar[u]^{\id_{vf}} &&& {\text{\footnotesize $a$}}\ar[u]_{\id_{va}}\ar[lll]^{\tilde t\circ_h \tilde s\circ_h x} 
}
\label{dcedsym4}
\end{equation}
\begin{equation}
\xymatrix{
{\text{\footnotesize $b$}}  && {\text{\footnotesize $b$}}\ar[ll]_{\id_{hb}}  
\ar@{}[dll]^(.25){}="a"^(.75){}="b" \ar@{=>} "a";"b"_{\Id_{hx}} %|-{X}
\\                 
{\text{\footnotesize $a$}} \ar[u]^x && {\text{\footnotesize $a$}}\ar[u]_x\ar[ll]^{\id_{ha}}             
}
\quad
\vbox{
\xymatrix{
\ar@{|->}[r]&
}
\vspace{-.74cm}}
\quad
\xymatrix{
{\text{\footnotesize $b$}}  && {\text{\footnotesize $a$}}\ar[ll]_{\tilde x}  
\ar@{}[dll]^(.25){}="a"^(.75){}="b" \ar@{=>} "a";"b"_{\Id_{v\tilde x}} %|-{X}
\\                 
{\text{\footnotesize $b$}} \ar[u]^{\id_{vb}} && {\text{\footnotesize $a$}}\ar[u]_{\id_{va}}\ar[ll]^{\tilde x}          
}
\label{dcedsym5}
\end{equation}
must be fulfilled. For a double groupoid $D$ we have further
\begin{equation}
\xymatrix{
{\text{\footnotesize $c$}}  && {\text{\footnotesize $d$}}\ar[ll]_{u^{-1_h}}  
\ar@{}[dll]^(.25){}="a"^(.75){}="b" \ar@{=>} "a";"b"_{X^{-1_h}} %|-{X}
\\                 
{\text{\footnotesize $a$}} \ar[u]^x && {\text{\footnotesize $b$}}\ar[u]_v\ar[ll]^{y^{-1_h}}             
}
\,\,%\quad
\vbox{
\xymatrix{
\ar@{|->}[r]&
}
\vspace{-.73cm}}
\,\,%\quad
\xymatrix{
{\text{\footnotesize $c$}}  &&&&&& {\text{\footnotesize $b$}}\ar[llllll]_{u^{-1_h}\circ_h \tilde v} 
\ar@{}[dllllll]^(.25){}="a"^(.75){}="b" \ar@{=>} "a";"b"_{\Id_{v\tilde x}\circ_h \tilde X^{-1_h}\circ_h \Id_{v\tilde v}\hphantom{xxxxxxxxxx}} %|-{X}
\\                 
{\text{\footnotesize $c$}} \ar[u]^{\id_{vc}} &&&&&& {\text{\footnotesize $b$}}\ar[u]_{\id_{vb}}\ar[llllll]^{\tilde x\circ_h y^{-1_h}}          
}
\label{dcedsym6}
\end{equation}
\begin{equation}
\xymatrix{
{\text{\footnotesize $b$}}  && {\text{\footnotesize $a$}}\ar[ll]_y  \ar@{}[dll]^(.25){}="a"^(.75){}="b" \ar@{=>} "a";"b"_{X^{-1_v}} %|-{X}
\\                 
{\text{\footnotesize $d$}} \ar[u]^{v^{-1_v}} && {\text{\footnotesize $c$}}\ar[u]_{x^{-1_v}}\ar[ll]^u             
}
\,\,%\quad
\vbox{
\xymatrix{
\ar@{|->}[r]&
}
\vspace{-.73cm}}
\,\,%\quad
\xymatrix{
{\text{\footnotesize $b$}}  &&&&&& {\text{\footnotesize $c$}}\ar[llllll]_{y\circ_h \tilde x^{-1_h}} 
\ar@{}[dllllll]^(.25){}="a"^(.75){}="b" \ar@{=>} "a";"b"_{\Id_{v\tilde v^{-1_h}}\circ_h \tilde X^{-1_v}\circ_h \Id_{v\tilde x^{-1_h}}\hphantom{xxxxxxxx}} %|-{X}
\\                 
{\text{\footnotesize $b$}} \ar[u]^{\id_{vb}} &&&&&& {\text{\footnotesize $c$}}\ar[u]_{\id_{vc}}\ar[llllll]^{\tilde v^{-1_h}\circ_h u}          
}
\label{dcedsym7}
\end{equation}
A vertical folding is defined similarly. 
\end{defi}

The foldings considered below will tacitly be assumed to be horizontal.
The analysis can however easily be performed for vertical foldings too. 

%The definitions of double natural transformation and modification given below assume 
%a horizontal folding. The definitions for a vertical folding also exists.

%We shall consider only horizontal foldings 
%aiming to define double natural transformations.

%\vfil\eject

\subsection{\normalsize \textcolor{blue}{Double natural transformations}}\label{sec:dcnatr}

\hspace{.5cm} In double category theory, there is a standard notion of
double natural transformation,
which has two variants.
%Further, a double natural transformation can be either horizontal or vertical. 
This notion
however does not fit our purposes. % for reasons explained in the main body of the paper. 
Here, we present a new one, which is original to the best
of our knowledge. 

Let $D$, $E$ be double categories or groupoids. Further, let $E$ be edge symmetric and 
equipped with a folding (cf. apps. \ref{sec:dcedge}, \ref{sec:dcedsym}). 
Let $F,G:D\rightarrow E$ be two double functors (cf. subapp. \ref{sec:dcfnctr}). 

\begin{defi} A double natural transformation $\rho:F\Rightarrow G$ consists of the following
data. 
\begin{enumerate}

\item A mapping 
of the set of object of $D$ into the set of vertical arrows of $E$,
\vspace{-.2cm}
\begin{equation}
\vbox{
\xymatrix{{\text{\footnotesize $a$}}\hspace{.3cm}\ar@{|->}[r]
&
}
\vspace{-.75cm}}
\xymatrix{
{\text{\footnotesize $G(a)$}}
\\
{\text{\footnotesize $F(a)$}}\ar[u]_{\rho(a)}
}
\label{dcnatr1}
\end{equation}

\item Two compatible functors from the horizontal and vertical arrow categories of $D$ into 
the horizontal truncation category $E_h$ of $E$.
\begin{equation}
\vbox{
\hbox{\xymatrix{
{\text{\footnotesize $b$}}&{\text{\footnotesize $a$}}\ar[l]_x
}
\quad\xymatrix{\ar@{|->}[r]
&
}\quad
}
\vspace{-.75cm}}
\xymatrix{
{\text{\footnotesize $G(b)$}}  & {\text{\footnotesize $G(a)$}}\ar[l]_{G(x)}  
\ar@{}[dl]^(.25){}="a"^(.75){}="b" \ar@{=>} "a";"b"_{\rho(x)} %|-{X}
\\                 
{\text{\footnotesize $F(b)$}} \ar[u]^{\rho(b)} & {\text{\footnotesize $F(a)$}}\ar[u]_{\rho(a)}\ar[l]^{F(x)}             
}
\label{dcnatr2}
\end{equation}
\begin{equation}
\xymatrix{
{\text{\footnotesize $b$}}
\\
{\text{\footnotesize $a$}}\ar[u]_x
}
\quad\vbox{
\xymatrix{\ar@{|->}[r]
&
}
\vspace{-.75cm}}\quad
\xymatrix{
{\text{\footnotesize $G(b)$}}  & {\text{\footnotesize $G(a)$}}\ar[l]_{\widetilde{G(x)}}  
\ar@{}[dl]^(.25){}="a"^(.75){}="b" \ar@{=>} "a";"b"_{{\bar\rho}(x)} %|-{X}
\\                 
{\text{\footnotesize $F(b)$}} \ar[u]^{\rho(b)} & {\text{\footnotesize $F(a)$}}\ar[u]_{\rho(a)}\ar[l]^{\widetilde{F(x)}}             
}
\nonumber%\label{dcnatr3}
\end{equation}
\end{enumerate}
Above $E_h$ is the category whose objects are the vertical arrows of $E$ and whose morphisms
are the arrow squares of $E$ connecting them with the composition $\circ_h$
and identity assigning map $\Id_h$ inherited form $E$. %%\pagebreak 
The data  must fulfill a special naturality condition. For any arrow square \hphantom{xxxxxxxxx}
\begin{equation}
\xymatrix{
{\text{\footnotesize $d$}}  & {\text{\footnotesize $b$}}\ar[l]_y  
\ar@{}[dl]^(.25){}="a"^(.75){}="b" \ar@{=>} "a";"b"_X %|-{X}
\\                 
{\text{\footnotesize $c$}} \ar[u]^u & {\text{\footnotesize $a$}}\ar[u]_x\ar[l]^z             
}
\label{dcnatr4}
\end{equation}
one has %\pagebreak 
\begin{equation}
\xymatrix@C=3pc@R=2.3pc{
{\text{\footnotesize $G(d)$}} & {\text{\footnotesize $G(b)$}} \ar[l]_{G(y)} 
\ar@{}[dl]^(.25){}="a"^(.75){}="b" \ar@{=>} "a";"b"_{\rho(y)}  
& {\text{\footnotesize $G(a)$}} \ar[l]_{\widetilde{G(x)}}  
\ar@{}[dl]^(.25){}="a"^(.75){}="b" \ar@{=>} "a";"b"_{{\bar\rho}( x)} %|-{X}
\\
{\text{\footnotesize $F(d)$}} \ar[u]^{\rho(d)}
& {\text{\footnotesize $F(b)$}} \ar[l]|-{\,\,F(y)\,\,} \ar[u]|-{\rho(b)\vphantom{\ul{\dot f}}}  
& {\text{\footnotesize $F(a)$}}\ar[l]|-{\,\,\widetilde{F(x)}\,\,} \ar[u]_{\rho(a)}  
\ar@{}[dll]^(.25){}="a"^(.75){}="b" \ar@{=>} "a";"b"_{\widetilde{F(X)}} %|-{X}
\\                 
{\text{\footnotesize $F(d)$}} \ar[u]^{\id_{vF(d)}}  && {\text{\footnotesize $F(a)$}}\ar[u]_{\id_{vF(a)}}
\ar[ll]^{\widetilde{F(u)}\circ_h F(z)}            
}
\,\,
\vbox{
\xymatrix{
\\
\ar@{}[u]|-{\text{\normalsize $=$}}
}
\vspace{-2cm}}
\,\,
\xymatrix@C=3pc@R=2.3pc{
{\text{\footnotesize $G(d)$}} &  
& {\text{\footnotesize $G(a)$}} \ar[ll]_{G(y)\circ_h\widetilde{G(x)}}  
\ar@{}[dll]^(.25){}="a"^(.75){}="b" \ar@{=>} "a";"b"_{\widetilde{G(X)}} %|-{X}
\\
{\text{\footnotesize $G(d)$}} \ar[u]^{\id_{vG(d)}} 
& {\text{\footnotesize $G(c)$}} \ar[l]|-{\,\,\widetilde{G(u)}\,\,}   
\ar@{}[dl]^(.25){}="a"^(.75){}="b" \ar@{=>} "a";"b"_{{\bar\rho}( u)}  
& {\text{\footnotesize $G(a)$}}\ar[l]|-{\,\,G(z)\,\,} \ar[u]_{\id_{vG(a)}}   
\ar@{}[dl]^(.25){}="a"^(.75){}="b" \ar@{=>} "a";"b"_{\rho(z)} %|-{X}
\\                 
{\text{\footnotesize $F(d)$}} \ar[u]^{\rho(d)}  & {\text{\footnotesize $F(c)$}}  
\ar[l]^{\widetilde{F(u)}} \ar[u]|-{\rho(c)\vphantom{\ul{\dot f}}}  
& {\text{\footnotesize $F(a)$}}\ar[u]_{\rho(a)}\ar[l]^{F(z)}             
}
\label{dcnatr5}
\end{equation}
\end{defi}
The conventionally defined double natural transformations do not
require a prior assignment of a folding. Further, they can be
either horizontal or vertical. The naturality condition they satisfy
mimics that of the ordinary natural transformations with arrows
replaced by arrow squares of the form \eqref{dcnatr2} and arrow composition 
replaced by the horizontal and vertical square compositions,
respectively.

If we forget the distinction between horizontal and vertical arrows of $E$ exploiting the edge symmetry of the latter,
the naturality condition can be viewed as the requirement of commutativity of the cube diagram 
\begin{equation}
\vbox{
\hbox{
\hspace{1.2cm}\xymatrix{\ar@{.>}'[dr]^{\hspace{-.4cm}\rho(y)\vphantom{\ul{\ul{\ul{\ul{\ul{g}}}}}}}[ddrr]&&
\\
&&
\\
&&
\\
&&}
}
\vspace{-3.cm}
\hbox{
\xymatrix@C=1.3pc@R=2pc{\\
&&&
\\
\ar@{.>}[rrr]^{{\bar\rho}(u)\hspace{1.5cm}}&&&
\\
&&&
}
\hspace{-1.7cm}
\xymatrix@C=1.3pc@R=2pc{
{\text{\footnotesize $G(d)$}}  && {\text{\footnotesize $G(b)$}} \ar[ll]_{G(y)} 
\ar@{}[dl]^(.25){}="a"^(.75){}="b" \ar@{} "a";"b"|-{G(X)}
\\
& {\text{\footnotesize $G(c)$}} \ar[ul]^{G(u)\hspace{-.05cm}} 
&& {\text{\footnotesize $G(a)$}}\ar[ul]_{\hspace{-.15cm}G(x)} \ar[ll]^{G(z)} 
\\
{\text{\footnotesize $F(d)$}} \ar[uu]^{\rho(d)} &&  {\text{\footnotesize $F(b)$}} \ar'[l]_{F(y)\hspace{.3cm}}[ll] \ar'[u]_{\rho(b)}[uu] 
\ar@{}[dl]^(.25){}="a"^(.75){}="b" \ar@{} "a";"b"|-{F(X)}
\\
&  {\text{\footnotesize $F(c)$}} \ar[ul]^{F(u)\hspace{-.05cm}} \ar[uu]^{\rho(c)\vphantom{\ul{\ul{\ul{\ul{\ul{\ul{g}}}}}}}} 
&& {\text{\footnotesize $F(a)$}} \ar[ul]_{\hspace{-.15cm}F(x)} \ar[ll]^{F(z)} \ar[uu]_{\rho(a)}
}
\hspace{-1.7cm}
\xymatrix@C=1.3pc@R=2pc{\\
&&&
\\
&&&\ar@{.>}[lll]_{\hspace{1.5cm}{\bar\rho}(x)}
\\
&&&
}
}
\vspace{-3.cm}
\hbox{
\hspace{4.cm}
\xymatrix{
&&
\\
&&
\\
&&
\\
&&\ar@{.>}'[ul]^{\vphantom{\Big[}\rho(z)\hspace{-.4cm}}[uull]}
}
}
\label{dcnatr6}
\end{equation} %\hspace{-.4cm}
Here, we have dropped all double arrows in order not to %excessively 
clog the diagram. 

%\ar@{}[dl]^(.25){}="a"^(.75){}="b" \ar@{=>} "a";"b"
%\ar@{}[ddll]^(.25){}="a"^(.75){}="b" \ar@{=>} "a";"b"

%\vfil\eject

\subsection{\normalsize \textcolor{blue}{Double modifications}}\label{sec:dcmod}

\hspace{.5cm} The non standard definition of double modification given
below is dictated by  the non standard notion of double natural
transformation of subapp. \ref{sec:dcnatr}. 

Let $D$, $E$ be double categories or groupoids with $E$ edge symmetric and 
folded (cf. apps. \ref{sec:dcedge}, \ref{sec:dcedsym}). 
Let $F,G:D\rightarrow E$ be  double functors
and $\rho,\sigma:F\Rightarrow G$ be double natural transformations (cf. subapps.
 \ref{sec:dcfnctr},\ref{sec:dcnatr}). 

\begin{defi} A double modification $\rho\Rrightarrow \sigma$
%$\xymatrix{\rho    \ar@3{->}[r]&\sigma}$ 
consists of a single datum.
\begin{enumerate}

\item A mapping of the set of objects of $D$ into the set of $2$--cells of $V\!D$, 
\begin{equation}
\vbox{
\xymatrix{{\text{\footnotesize $a$}}\hspace{.3cm}\ar@{|->}[r]
&
}
\vspace{-.75cm}}
\xymatrix{
{\text{\footnotesize $G(a)$}}  && {\text{\footnotesize $G(a)$}}\ar[ll]_{\id_{hG(a)}}  
\ar@{}[dll]^(.25){}="a"^(.75){}="b" \ar@{=>} "a";"b"_{T(a)} %|-{X}
\\                 
{\text{\footnotesize $F(a)$}} \ar[u]^{\sigma(a)} && {\text{\footnotesize $F(a)$}}\ar[u]_{\rho(a)}\ar[ll]^{\id_{hF(a)}}             
}
\label{dcmod1}
\end{equation}

\end{enumerate}
This must satisfy the modification axioms. For any
horizontal arrow of $D$
\begin{equation}
\xymatrix{{\text{\footnotesize $b$}} & \ar[l]_x {\text{\footnotesize $a$}} & }
\label{dcmod2}
\end{equation}
one has 
\begin{equation}
\xymatrix@C=2.8pc{
{\text{\footnotesize $G(b)$}} & {\text{\footnotesize $G(b)$}} 
\ar[l]_{\id_{hG(b)}} \ar@{}[dl]^(.25){}="a"^(.75){}="b" \ar@{=>} "a";"b"_{T(b)}  
& {\text{\footnotesize $G(a)$}}\ar[l]_{G(x)}  \ar@{}[dl]^(.25){}="a"^(.75){}="b" \ar@{=>} "a";"b"_{\rho(x)} %|-{X}
\\                 
{\text{\footnotesize $F(b)$}} \ar[u]^{\sigma(b)}  & {\text{\footnotesize $F(b)$}}  \ar[l]^{\id_{hF(b)}} \ar[u]|-{\rho(b)\vphantom{\ul{\dot f}}} 
& {\text{\footnotesize $F(a)$}}\ar[u]_{\rho(a)}\ar[l]^{F(x)}             
}
\hspace{.25cm}
\xymatrix{~\ar@{}[d]|-{\text{\normalsize $=\vphantom{\dot h}$}}
\\
~}
\hspace{.25cm}
\xymatrix@C=2.8pc{
{\text{\footnotesize $G(b)$}} & {\text{\footnotesize $G(a)$}} \ar[l]_{G(x)} 
\ar@{}[dl]^(.25){}="a"^(.75){}="b" \ar@{=>} "a";"b"_{\sigma(x)}
& {\text{\footnotesize $G(a)$}}\ar[l]_{\id_{hG(a)}}
\ar@{}[dl]^(.25){}="a"^(.75){}="b" \ar@{=>} "a";"b"_{T(a)} %|-{X}
\\                 
{\text{\footnotesize $F(b)$}} \ar[u]^{\sigma(b)}  & {\text{\footnotesize $F(a)$}}  \ar[l]^{F(x)} \ar[u]|-{\sigma(a)\vphantom{\ul{\dot f}}} 
& {\text{\footnotesize $F(a)$}}\ar[u]_{\rho(a)}\ar[l]^{\id_{hF(a)}}             
}
\label{dcmod3}
\end{equation}
For any vertical arrow of $D$ \hphantom{xxxxxxxxxxxxxxxxx}
\begin{equation}
\xymatrix{{\text{\footnotesize $b$}} \\ 
\ar[u]_x {\text{\footnotesize $a$}} & }
\label{dcmod4}
\end{equation}
one has 
\begin{equation}
\xymatrix@C=2.8pc{
{\text{\footnotesize $G(b)$}} & {\text{\footnotesize $G(b)$}} 
\ar[l]_{\id_{hG(b)}} \ar@{}[dl]^(.25){}="a"^(.75){}="b" \ar@{=>} "a";"b"_{T(b)}  
& {\text{\footnotesize $G(a)$}}\ar[l]_{\widetilde{G(x)}}  \ar@{}[dl]^(.25){}="a"^(.75){}="b" \ar@{=>} "a";"b"_{{\bar\rho}(x)} %|-{X}
\\                 
{\text{\footnotesize $F(b)$}} \ar[u]^{\sigma(b)}  & {\text{\footnotesize $F(b)$}}  \ar[l]^{\id_{hF(b)}} \ar[u]|-{\rho(b)\vphantom{\ul{\dot f}}} 
& {\text{\footnotesize $F(a)$}}\ar[u]_{\rho(a)}\ar[l]^{\widetilde{F(x)}}             
}
\hspace{.25cm}
\xymatrix{~\ar@{}[d]|-{\text{\normalsize $=\vphantom{\dot h}$}}
\\
~}
\hspace{.25cm}
\xymatrix@C=2.8pc{
{\text{\footnotesize $G(b)$}} & {\text{\footnotesize $G(a)$}} \ar[l]_{\widetilde{G(x)}} 
\ar@{}[dl]^(.25){}="a"^(.75){}="b" \ar@{=>} "a";"b"_{{\bar\sigma}(x)}
& {\text{\footnotesize $G(a)$}}\ar[l]_{\id_{hG(a)}}
\ar@{}[dl]^(.25){}="a"^(.75){}="b" \ar@{=>} "a";"b"_{T(a)} %|-{X}
\\                 
{\text{\footnotesize $F(b)$}} \ar[u]^{\sigma(b)}  & {\text{\footnotesize $F(a)$}}  \ar[l]^{\widetilde{F(x)}} \ar[u]|-{\sigma(a)\vphantom{\ul{\dot f}}} 
& {\text{\footnotesize $F(a)$}}\ar[u]_{\rho(a)}\ar[l]^{\id_{hF(a)}}             
}
\label{dcmod5}
\end{equation}
\end{defi}

The axioms can be interpreted as the commutativity condition of the following cylinder diagrams
\hphantom{xxxxxxxxxxxxxxxxx}
\begin{subequations}
\begin{equation}
\xymatrix@C=.4pc@R=.9pc@=.6pc{
& {\text{\scriptsize $\rho(b)$}}\ar@/_.7pc/[dl]& & &{\text{\scriptsize $\rho(a)$}}\ar@/_.7pc/[dl] 
\ar@{.}[lll]|-{\rho(x)}&
\\
{\text{\footnotesize $G(b)$}}& & \text{$\hphantom{x}$} \ar[ll]_{G(x)}& {\text{\footnotesize  $G(a)$}} \ar@{-}[l] &&
\\
&& {\text{\footnotesize $F(b)$}} \ar@{-}@/_.95pc/[uul] \ar@{}[ull]|-{T(b)}
& &&\ar@{-}@/_.95pc/[uul]\ar[lll]_{F(x)}{\text{\footnotesize  $F(a)$}} \ar@{}[ull]|-{T(a)}
\\
&{\text{\scriptsize $\sigma(b)$}} \ar@/^1pc/[uul]\ar@{-}@/_.7pc/[ur]&& 
&{\text{\scriptsize $\sigma(a)$}}\ar@/^1pc/[uul]|->>>>>{~\text{$\vphantom{\ul{\ul{\ul{\ul{\ul{f}}}}}}$}}
\ar@{-}@/_.7pc/[ur] \ar@{.}[lll]|-{\sigma(x)}&
}
\label{dcmod6}
\end{equation}
\vspace{0mm}
\begin{equation}
\xymatrix@C=.4pc@R=.9pc@=.6pc{
& {\text{\scriptsize $\rho(b)$}}\ar@/_.7pc/[dl]& & &{\text{\scriptsize $\rho(a)$}}\ar@/_.7pc/[dl] 
\ar@{.}[lll]|-{{\bar\rho}(x)}&
\\
{\text{\footnotesize $G(b)$}}& & \text{$\hphantom{x}$} \ar[ll]_{G(x)}& {\text{\footnotesize  $G(a)$}} \ar@{-}[l] &&
\\
&& {\text{\footnotesize $F(b)$}} \ar@{-}@/_.95pc/[uul] \ar@{}[ull]|-{T(b)}
& &&\ar@{-}@/_.95pc/[uul]\ar[lll]_{F(x)}{\text{\footnotesize  $F(a)$}} \ar@{}[ull]|-{T(a)}
\\
&{\text{\scriptsize $\sigma(b)$}} \ar@/^1pc/[uul]\ar@{-}@/_.7pc/[ur]&& 
&{\text{\scriptsize $\sigma(a)$}}\ar@/^1pc/[uul]|->>>>>{~\text{$\vphantom{\ul{\ul{\ul{\ul{\ul{f}}}}}}$}}
\ar@{-}@/_.7pc/[ur] \ar@{.}[lll]|-{{\bar\sigma}(x)}&
}
\label{dcmod7}
\end{equation}
\end{subequations}
\vskip4mm\noindent
Above all double arrows have been dropped. Further the identity morphisms 
of the modification arrow squares have been collapsed.

%%\vfil\eject

\subsection{\normalsize \textcolor{blue}{The double groupoid of plane rectangles}}\label{sec:dcplane}

\hspace{.5cm} 
Rectangles in $\mathbb{R}^2$ can be organized in a double groupoid.

\begin{prop}
There is a double groupoid $\mathbb{GR}^2$ defined as follows.
\begin{enumerate}

\item For each $x,y\in\mathbb{R}$, there is an object $(x,y)$ of $\mathbb{GR}^2$.

\item For each $x,x',y\in\mathbb{R}$ there is a unique horizontal arrow 
\vspace{-.2cm}
\begin{equation}
\vbox{
\xymatrix{{\text{\footnotesize $(x',y)$}}&{\text{\footnotesize $(x,y)$}}\ar[l]}}
\label{dcplane1}
\end{equation}
\vspace{-.2cm}
For each $x,y,y'\in\mathbb{R}$ there is a unique vertical arrow 
%\vspace{-.3cm}
\begin{equation}
\xymatrix{{\text{\footnotesize $(x,y')$}}
\\
{\text{\footnotesize $(x,y)$}}\ar[u]}
\label{dcplane2}
\end{equation}
%\vspace{-.3cm}

\item For each quadruple $x,x',y,y'\in\mathbb{R}$ there is a unique arrow square
\begin{equation}
\xymatrix{
{\text{\footnotesize $(x',y')$}}  & {\text{\footnotesize $(x,y')$}}\ar[l] \ar@{}[dl]^(.25){}="a"^(.75){}="b" \ar@{=>} "a";"b" %|-{X}
\\                 
{\text{\footnotesize $(x',y)$}} \ar[u] & {\text{\footnotesize $(x,y)$}}\ar[u] \ar[l]            
}
\label{dcplane3}
\end{equation}

\end{enumerate}
The horizontal and vertical composition of arrows and arrow squares are codified in the diagrams
\begin{equation}
\xymatrix{
{\text{\footnotesize $(x'',y')$}} & {\text{\footnotesize $(x',y')$}} \ar[l] \ar@{}[dl]^(.25){}="a"^(.75){}="b" \ar@{=>} "a";"b"  
& {\text{\footnotesize $(x,y')$}}\ar[l]  \ar@{}[dl]^(.25){}="a"^(.75){}="b" \ar@{=>} "a";"b" %|-{X}
\\                 
{\text{\footnotesize $(x'',y)$}} \ar[u]  & {\text{\footnotesize $(x',y)$}} \ar[l] \ar[u]
& {\text{\footnotesize $(x,y)$}} \ar[u]\ar[l]          
}
\quad
\xymatrix{~\ar@{}[d]|-{\text{\normalsize $=\vphantom{\dot h}$}}
\\
~}
\quad
\xymatrix{
{\text{\footnotesize $(x'',y')$}}  & {\text{\footnotesize $(x,y')$}} \ar[l]
\ar@{}[dl]^(.25){}="a"^(.75){}="b" \ar@{=>} "a";"b" %|-{X}
\\                 
{\text{\footnotesize $(x'',y)$}} \ar[u] & {\text{\footnotesize $(x,y)$}}\ar[u]\ar[l]            
}
\label{dcplane4}
\end{equation}
\begin{equation}
\hspace{.85cm}\xymatrix{
{\text{\footnotesize $(x',y'')$}} & {\text{\footnotesize $(x,y'')$}} \ar[l]
\ar@{}[dl]^(.25){}="a"^(.75){}="b" \ar@{=>} "a";"b" %|-{X}
\\
{\text{\footnotesize $(x',y')$}} \ar[u] & {\text{\footnotesize $(x,y')$}} \ar[l] \ar[u]
\ar@{}[dl]^(.25){}="a"^(.75){}="b" \ar@{=>} "a";"b" %|-{X}
\\                 
{\text{\footnotesize $(x',y)$}} \ar[u] & {\text{\footnotesize $(x,y)$}} \ar[u]\ar[l]            
}
\quad
\xymatrix{~\ar@{}[d]|-{\text{\normalsize $=\vphantom{\dot h}$}}
\\
~}
\quad
\xymatrix{
{\text{\footnotesize $(x',y'')$}}  & {\text{\footnotesize $(x,y'')$}} \ar[l]  
\ar@{}[dl]^(.25){}="a"^(.75){}="b" \ar@{=>} "a";"b" %|-{X}
\\                 
{\text{\footnotesize $(x',y)$}} \ar[u] & {\text{\footnotesize $(x,y)$}}\ar[u]\ar[l]            
}
\nonumber%\label{dcplane5}
\end{equation}
\vfil\eject
\noindent
respectively. 
The horizontal and vertical composition identity arrows and arrow squares are similarly encoded in the diagrams
\begin{equation}
\xymatrix{
{\text{\footnotesize $(x,y')$}}  & {\text{\footnotesize $(x,y')$}}\ar[l] 
\ar@{}[dl]^(.25){}="a"^(.75){}="b" \ar@{=>} "a";"b" %|-{X}
\\                 
{\text{\footnotesize $(x,y)$}} \ar[u] & {\text{\footnotesize $(x,y)$}}\ar[u]\ar[l]            
}
\qquad 
\xymatrix{
{\text{\footnotesize $(x',y)$}}  & {\text{\footnotesize $(x,y)$}}\ar[l] 
\ar@{}[dl]^(.25){}="a"^(.75){}="b" \ar@{=>} "a";"b" %|-{X}
\\                 
{\text{\footnotesize $(x',y)$}} \ar[u] & {\text{\footnotesize $(x,y)$}}\ar[u]\ar[l]            
}
\label{dcplane6}
\end{equation}
respectively. Finally, the horizontal and vertical inverses of arrows and arrow squares in \eqref{dcplane3} are 
\begin{equation}
\xymatrix{
{\text{\footnotesize $(x,y')$}}  & {\text{\footnotesize $(x',y')$}}\ar[l] 
\ar@{}[dl]^(.25){}="a"^(.75){}="b" \ar@{=>} "a";"b" %|-{X}
\\                 
{\text{\footnotesize $(x,y)$}} \ar[u] & {\text{\footnotesize $(x',y)$}}\ar[u]\ar[l]            
}
\qquad
\xymatrix{
{\text{\footnotesize $(x',y)$}}  & {\text{\footnotesize $(x,y)$}}\ar[l]
\ar@{}[dl]^(.25){}="a"^(.75){}="b" \ar@{=>} "a";"b" %|-{X}
\\                 
{\text{\footnotesize $(x',y')$}} \ar[u] & {\text{\footnotesize $(x,y')$}}\ar[u]\ar[l]           
}
\label{dcplane7}
\end{equation}
\end{prop}

%\vfil\eject

\subsection{\normalsize \textcolor{blue}{The double groupoid of a crossed module}}\label{sec:dccrossed}

\hspace{.5cm} 
Let $(G,H)$ be a crossed module with target map $t:H\rightarrow G$ and $G$ action
$m:G\times H\rightarrow H$. 

\begin{prop}
There is a double groupoid $B(G,H)$ defined as follows.
\begin{enumerate}

\item There is a unique object $*$. 

\item For each element $x\in G$, there is one horizontal and one vertical arrow, 
\vspace{-.3cm}
\begin{equation}
\vbox{
\xymatrix{{\text{\footnotesize $*$}}&{\text{\footnotesize $*$}}\ar[l]_x}
\vspace{-.75cm}}
\qquad
\xymatrix{{\text{\footnotesize $*$}}
\\
{\text{\footnotesize $*$}}\ar[u]_x}
\label{dccrossed1}
\end{equation}
\vspace{-.3cm}

\item For each quadruple $x,y,u,v\in G$ and each $X\in H$ satisfying the target matching condition 
\hphantom{xxxxxxxxxx}
\begin{equation}
vy=uxt(X)
\label{dccrossed2}
\end{equation}
there is one arrow square \hphantom{xxxxxxxxxx}
\begin{equation}
\xymatrix{
{\text{\footnotesize $*$}}  & {\text{\footnotesize $*$}}\ar[l]_u  \ar@{}[dl]^(.25){}="a"^(.75){}="b" \ar@{=>} "a";"b"_X %|-{X}
\\                 
{\text{\footnotesize $*$}} \ar[u]^v & {\text{\footnotesize $*$}}\ar[u]_x\ar[l]^y             
}
\label{dccrossed3}
\end{equation}

\end{enumerate}
The horizontal and vertical composition of arrows and arrow squares are codified in the diagrams
\begin{equation}
\xymatrix{
{\text{\footnotesize $*$}} & {\text{\footnotesize $*$}} \ar[l]_v \ar@{}[dl]^(.25){}="a"^(.75){}="b" \ar@{=>} "a";"b"_Y  
& {\text{\footnotesize $*$}}\ar[l]_u  \ar@{}[dl]^(.25){}="a"^(.75){}="b" \ar@{=>} "a";"b"_X %|-{X}
\\                 
{\text{\footnotesize $*$}} \ar[u]^t  & {\text{\footnotesize $*$}} \ar[l]^y \ar[u]|-{s\vphantom{\ul{\dot f}}} 
& {\text{\footnotesize $*$}} \ar[u]_r\ar[l]^x             
}
\quad
\xymatrix{~\ar@{}[d]|-{\text{\normalsize $=\vphantom{\dot h}$}}
\\
~}
\quad
\xymatrix@C=3.9pc{
{\text{\footnotesize $*$}}  && {\text{\footnotesize $*$}} \ar[ll]_{vu}  
\ar@{}[dll]^(.25){}="a"^(.75){}="b" \ar@{=>} "a";"b"_{Xm(x^{-1})(Y)\hspace{.75cm}} %|-{X}
\\                 
{\text{\footnotesize $*$}} \ar[u]^t && {\text{\footnotesize $*$}}\ar[u]_r\ar[ll]^{yx}             
}
\label{dccrossed4}
\end{equation}
\begin{equation}
\hspace{.85cm}\xymatrix{
{\text{\footnotesize $*$}} & {\text{\footnotesize $*$}} \ar[l]_z 
\ar@{}[dl]^(.25){}="a"^(.75){}="b" \ar@{=>} "a";"b"_Y %|-{X}
\\
{\text{\footnotesize $*$}} \ar[u]^t & {\text{\footnotesize $*$}} \ar[l]|-{\,\,y\,\,} \ar[u]_r 
\ar@{}[dl]^(.25){}="a"^(.75){}="b" \ar@{=>} "a";"b"_X %|-{X}
\\                 
{\text{\footnotesize $*$}} \ar[u]^s & {\text{\footnotesize $*$}} \ar[u]_q\ar[l]^x             
}
\quad
\xymatrix{~\ar@{}[d]|-{\text{\normalsize $=\vphantom{\dot h}$}}
\\
~}
\qquad\quad
\xymatrix@C=3.9pc{
{\text{\footnotesize $*$}}  && {\text{\footnotesize $*$}} \ar[ll]_z  
\ar@{}[dll]^(.25){}="a"^(.75){}="b" \ar@{=>} "a";"b"_{m(q^{-1})(Y)X\hspace{.75cm}} %|-{X}
\\                 
{\text{\footnotesize $*$}} \ar[u]^{ts} && {\text{\footnotesize $*$}}\ar[u]_{rq}\ar[ll]^x             
}
\nonumber%\label{dccrossed5}
\end{equation}
%%\vfil\eject\noindent
\vskip3truemm\noindent
respectively. The horizontal and vertical identity arrows and arrow squares are similarly encoded in the diagrams
\begin{equation}
\xymatrix{
{\text{\footnotesize $*$}}  && {\text{\footnotesize $*$}}\ar[ll]_{1_G}  
\ar@{}[dll]^(.25){}="a"^(.75){}="b" \ar@{=>} "a";"b"_{1_H} %|-{X}
\\                 
{\text{\footnotesize $*$}} \ar[u]^x && {\text{\footnotesize $*$}}\ar[u]_x\ar[ll]^{1_G}             
}
\quad 
\xymatrix{
{\text{\footnotesize $*$}}  && {\text{\footnotesize $*$}}\ar[ll]_x  
\ar@{}[dll]^(.25){}="a"^(.75){}="b" \ar@{=>} "a";"b"_{1_H} %|-{X}
\\                 
{\text{\footnotesize $*$}} \ar[u]^{1_G} && {\text{\footnotesize $*$}}\ar[u]_{1_G}\ar[ll]^x             
}
\label{dccrossed6}
\end{equation}
respectively. Finally the horizontal and vertical inverses of arrows and arrow squares in \eqref{dccrossed3} are 
\begin{equation}
\xymatrix@C=3.4pc{
{\text{\footnotesize $*$}}  && {\text{\footnotesize $*$}}\ar[ll]_{u^{-1}}  
\ar@{}[dll]^(.25){}="a"^(.75){}="b" \ar@{=>} "a";"b"_{m(y)(X^{-1})\hspace{.6cm}} %|-{X}
\\                 
{\text{\footnotesize $*$}} \ar[u]^x && {\text{\footnotesize $*$}}\ar[u]_v\ar[ll]^{y^{-1}}             
}
\qquad\quad
\xymatrix@C=3.4pc{
{\text{\footnotesize $*$}}  && {\text{\footnotesize $*$}}\ar[ll]_y  
\ar@{}[dll]^(.25){}="a"^(.75){}="b" \ar@{=>} "a";"b"_{m(x)(X^{-1})\hspace{.6cm}} %|-{X}
\\                 
{\text{\footnotesize $*$}} \ar[u]^{v^{-1}} && {\text{\footnotesize $*$}}\ar[u]_{x^{-1}}\ar[ll]^u             
}
\label{dccrossed7}
\end{equation}
\end{prop} 
We remark that the target matching condition \eqref{dccrossed2} is essential for the exchange law
\eqref{dcdef6} to be satisfied.

\begin{prop}
The double groupoid $B(G,H)$ is edge symmetric. 
\end{prop}
The invertible functor $V\!B(G,H)\rightarrow H\!B(G,H)$ 
implementing edge symmetry is defined as 
\begin{equation}
\xymatrix{
{\text{\footnotesize $*$}}  & {\text{\footnotesize $*$}}\ar[l]_{1_G}  \ar@{}[dl]^(.25){}="a"^(.75){}="b" \ar@{=>} "a";"b"_X %|-{X}
\\                 
{\text{\footnotesize $*$}} \ar[u]^y & {\text{\footnotesize $*$}}\ar[u]_x\ar[l]^{1_G}             
}
\hspace{.4cm}
\vbox{
\xymatrix{\ar@{|->}[r]
&
}
\vspace{-.75cm}}
\hspace{.4cm}
\xymatrix@C=3pc{
{\text{\footnotesize $*$}}  & {\text{\footnotesize $*$}}\ar[l]_y  
\ar@{}[dl]^(.25){}="a"^(.75){}="b" \ar@{=>} "a";"b"_{X^{-1}} %|-{X}
\\                 
{\text{\footnotesize $*$}} \ar[u]^{1_G} & {\text{\footnotesize $*$}}\ar[u]_{1_G}\ar[l]^x           
}
\label{dccrossed8}
\end{equation}

\begin{prop} The mapping 
\begin{equation}
\xymatrix{
{\text{\footnotesize $*$}}  & {\text{\footnotesize $*$}}\ar[l]_u  \ar@{}[dl]^(.25){}="a"^(.75){}="b" \ar@{=>} "a";"b"_X %|-{X}
\\                 
{\text{\footnotesize $*$}} \ar[u]^v & {\text{\footnotesize $*$}}\ar[u]_x\ar[l]^y             
}
\hspace{.4cm}
\vbox{
\xymatrix{\ar@{|->}[r]
&
}
\vspace{-.75cm}}
\hspace{.4cm}
\xymatrix@C=3pc{
{\text{\footnotesize $*$}}  & {\text{\footnotesize $*$}}\ar[l]_{ux}  
\ar@{}[dl]^(.25){}="a"^(.75){}="b" \ar@{=>} "a";"b"_{X} %|-{X}
\\                 
{\text{\footnotesize $*$}} \ar[u]^{1_G} & {\text{\footnotesize $*$}}\ar[u]_{1_G}\ar[l]^{vy}            
}
\label{dccrossed9}
\end{equation}
defines a folding of $B(G,H)$. 
\end{prop}

\vfil\eject

\end{document}